\documentclass[journal]{IEEEtran}
\usepackage{cite}

\ifCLASSINFOpdf
\else
\fi
\usepackage{bbding}
\usepackage{tabularx}
\usepackage{graphicx}
\usepackage{hyperref}
\usepackage{tabularray}
\usepackage{booktabs}
\usepackage{makecell}
\usepackage{multirow}
\usepackage{float}
\hypersetup{hidelinks,
	colorlinks=true,
	allcolors=black,
	pdfstartview=Fit,
	breaklinks=true}
\usepackage{longtable}
\usepackage{ltxtable}

\begin{document}
\newcolumntype{C}[1]{>{\centering\arraybackslash}p{#1}}
\newcommand{\tabitem}{~~\llap{\textbullet}~~}
%
\title{Foundation Model for Advancing Healthcare: Challenges, Opportunities and Future Directions}
%
%
%

\author{Yuting He,
        Fuxiang Huang,
        Xinrui Jiang,
        Yuxiang Nie,
        Minghao Wang,
        Jiguang Wang,
        Hao Chen\IEEEauthorrefmark{1},~\IEEEmembership{Senior Member,~IEEE,} 
\thanks{\textit{\IEEEauthorrefmark{1}Corresponding author: H. Chen. (e-mail: jhc@cse.ust.hk)}}
\thanks{H. Chen is with the Department of Computer Science and Engineering, Department of Chemical and Biological Engineering, and Division of Life Science, The Hong Kong University of Science and Technology, Hong Kong, China.}
\thanks{Y. He, F. Huang, X. Jiang, Y. Nie are with the Department of Computer Science and Engineering, The Hong Kong University of Science and Technology, Hong Kong, China.}
\thanks{M. Wang is with the Department of Chemical and Biological Engineering, The Hong Kong University of Science and Technology, Hong Kong, China.}
\thanks{J. Wang is with the Department of Chemical and Biological Engineering, Division of Life Science, State Key Laboratory of Molecular Neuroscience, Hong Kong University of Science and Technology, Hong Kong, China. SIAT-HKUST Joint Laboratory of Cell Evolution and Digital Health, Shenzhen-Hong Kong Collaborative Innovation Research Institute, Futian, Shenzhen, Guangdong 518045, China. Hong Kong Center for Neurodegenerative Diseases, InnoHK, Hong Kong, China.}}

%
%

\markboth{Journal of \LaTeX\ Class Files,~Vol.~14, No.~8, August~2015}%
{Shell \MakeLowercase{\textit{et al.}}: Bare Demo of IEEEtran.cls for IEEE Journals}
%



\maketitle

\begin{abstract}
Foundation model, which is pre-trained on broad data and is able to adapt to a wide range of tasks, is advancing healthcare. It promotes the development of healthcare artificial intelligence (AI) models, breaking the contradiction between limited AI models and diverse healthcare practices. Much more widespread healthcare scenarios will benefit from the development of a healthcare foundation model (HFM), improving their advanced intelligent healthcare services. Despite the impending widespread deployment of HFMs, there is currently a lack of clear understanding about how they work in the healthcare field, their current challenges, and where they are headed in the future. To answer these questions, a comprehensive and deep survey of the challenges, opportunities, and future directions of HFMs is presented in this survey. It first conducted a comprehensive overview of the HFM including the methods, data, and applications for a quick grasp of the current progress. Then, it made an in-depth exploration of the challenges present in data, algorithms, and computing infrastructures for constructing and widespread application of foundation models in healthcare. This survey also identifies emerging and promising directions in this field for future development. We believe that this survey will enhance the community's comprehension of the current progress of HFM and serve as a valuable source of guidance for future development in this field. The latest HFM papers and related resources are maintained on our \href{https://github.com/YutingHe-list/Awesome-Foundation-Models-for-Advancing-Healthcare}{\color{magenta}website}.

\end{abstract}


\IEEEpeerreviewmaketitle

\section{Introduction}
\begin{figure*}
    \centering
    \includegraphics[width=\linewidth]{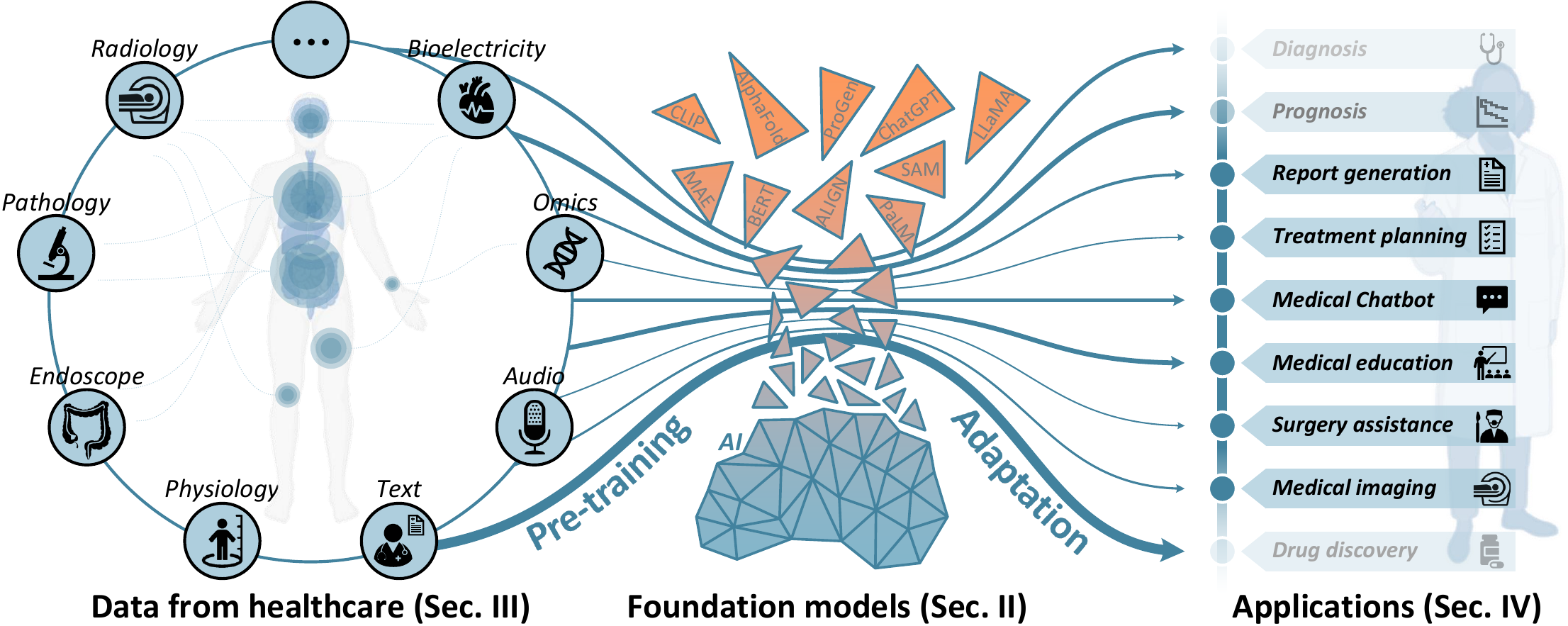}
    \caption{The pipeline of the healthcare foundation models (HFMs) including the methods (Sec.\ref{sec:methods}), datasets (Sec.\ref{sec:datasets}), and applications (Sec.\ref{sec:applications}).}
    \label{fig:overview}
\end{figure*}
\IEEEPARstart{I}{n} the past decade, with the development of artificial intelligence (AI) \cite{nilsson1982principles}, especially deep learning (DL) \cite{lecun2015deep}, healthcare techniques have undergone subversive advancement \cite{gu2023beyond,jiang2017artificial,rajpurkar2022ai}. Benefiting from the learning of healthcare data, AI models are able to unlock relevant information inner the data, and in turn, assist healthcare practices. In some influential clinical diseases, including pancreatic cancer \cite{46-PANDA}, retinal disease \cite{de2018clinically}, skin cancer \cite{esteva2017dermatologist}, etc., AI models have acquired the ability of specialists showing professional performance in the diagnosis or treatment, showing a promising future. However, before that, there is still a large contradiction between the specialist AI models that are implemented for specific healthcare tasks and the diverse healthcare scenarios and requirements, hindering their applications in widespread healthcare practices \cite{rajpurkar2022ai}. Therefore, there is an open question: ``\textit{Can we construct AI models to benefit a variety of healthcare tasks?}"

As shown in Fig.\ref{fig:overview}, the recent research of foundation models has enabled the AI models to learn general abilities and be applied to wide healthcare scenarios, giving a promising answer to this question \cite{bommasani2021opportunities,moor2023foundation,azad2023foundational,qiu2023large}. In the related sub-fields of healthcare AI, including language, vision, bioinformatics, and multi-modality, the healthcare foundation model (HFM) has shown impressive success. \textbf{a)} Language foundation model (LFM) or named large language model (LLM) \cite{thirunavukarasu2023large,he2023survey} has caused excitement and concern for the benefit of patients and clinicians \cite{thirunavukarasu2023large}. It learned large-scale medical language data, and has shown extraordinary performance in medical text processing \cite{yang2022large}  and dialogue \cite{singhal2023large} tasks. \textbf{b)} Vision foundation model (VFM) has demonstrated remarkable potential in medical images. Modality \cite{li2023d,4-Endo-FM}, organ \cite{3-RETFound}, task \cite{Kirillov2023ICCV,rombach2022high} -specific VFMs have shown their adaptability and general performance to potential medical scenarios. \textbf{c)} Bioinformatics foundation model (BFM) has helped researchers unlock the secrets of life, endowing us with prospects for the scenarios like the protein sequences, DNA, RNA, etc \cite{shen2024omnina,dalla2023nucleotide,brandes2022proteinbert,jumper2021highly,chen2022interpretable}. \textbf{d)} Multimodal foundation model (MFM) \cite{wu2023towards,fei2022towards,zhang2023biomedgpt} has provided an effective way for generalist HFMs \cite{acosta2022multimodal,moor2023foundation,tu2023towards}. It integrates the information from multiple modalities thus achieving the ability to interpret various medical modalities and perform multiple modality-dependent tasks \cite{tu2023towards,shrestha2023medical,azad2023foundational}. Therefore, these models have provided a foundation to address complex clinical issues and improve the efficiency and effectiveness of healthcare practices, thus advancing the healthcare field \cite{azad2023foundational}.

The emergence of the HFMs comes from the continuous accumulation of healthcare data, the development of AI algorithms, and the improvement of computing infrastructure \cite{bommasani2021opportunities,qiu2023large}. However the current lack of development in data, algorithms, and computing infrastructures is still the root of various challenges in HFMs. The ethics, diversity, heterogeneity, and cost of healthcare data make it extremely challenging to construct a large enough dataset to train a generalizable HFM \cite{qiu2023large,willemink2020preparing} in wide healthcare practices. The demand of adaptability, capacity, reliability, and responsibility in AI algorithms further makes it difficult to be applied to real scenarios \cite{hatherley2020limits,markus2021role}. Due to the high dimension and large size of healthcare data (e.g., 3D CT images, whole slide images (WSI), etc.), the demand for computing infrastructure is much larger than that of other fields which is extremely expensive in terms of consumption \cite{moor2023foundation,qiu2023large} and environment \cite{wu2022sustainable}. 

In general, foundation models for advancing healthcare are showing us a new future with both opportunities and challenges. In this survey, we have raised the following questions in current HFMs with a comprehensive perspective: \textbf{1)} Although the foundation models have achieved remarkable success, \textit{what are their current progresses in healthcare?} \textbf{2)} With the development of the foundation models, \textit{what challenges are they facing?} \textbf{3)} For further development of HFMs, \textit{what potential future directions deserve our attention and exploration?} The answers to the above questions will construct an overview for the current situation of the HFMs and provide a clear vision for their future development. Due to the emergence of the HFM, it has spawned hundreds of papers in recent years. Therefore, it is challenging to review all of them and all aspects in a limited paper space. In this article, we focus on the current progress in language, vision, bioinformatics, and multimodal foundation models in the healthcare field from 2018 (the beginning of foundation model era \cite{bommasani2021opportunities}) to 2024, and the challenges and future directions of the HFMs. We hope this survey will assist researchers in quickly grasping the development of HFMs and ignite a spark of creativity to further push the boundaries in healthcare.

\subsection{Brief History of Foundation Models in Healthcare}
Following the definition from Bommasani \textit{et al.} \cite{bommasani2021opportunities}, the term ``foundation model" in this survey is any model that is \textit{pre-trained} on broad data and has the ability to \textit{adapt} to a wide range of tasks. Another sociological feature \cite{bommasani2021opportunities} of the foundation model era is that it is widely accepted to apply a certain foundational AI model to a large number of different tasks. The representative inflection point of the foundation model era is the BERT model \cite{kenton2019bert} in natural language processing (NLP) at the end of 2018, after that, pre-trained models became a foundation in NLP and then spread to other fields.

\begin{figure}
    \centering
    \includegraphics[width=\linewidth]{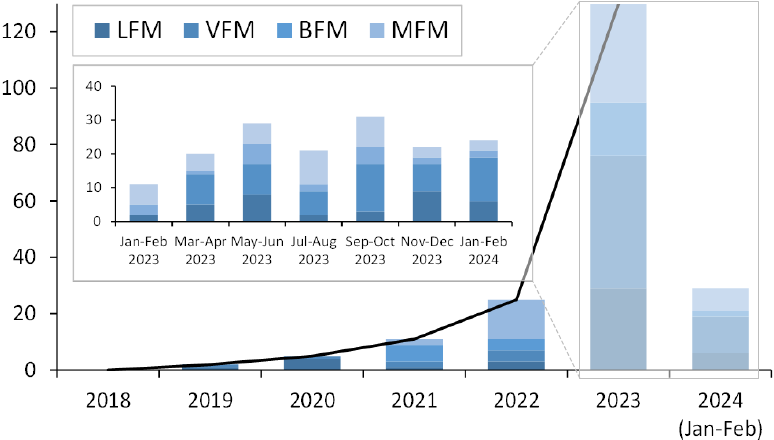}
    \caption{The number of representative papers on healthcare foundation models from 2018 to 2024 (Jan-Feb).}
    \label{fig:develop}
\end{figure}

AI in healthcare is also gradually moving from specific targets to general targets \cite{moor2023foundation} driven by the development of foundation models. BioBERT \cite{lee2020biobert} was made public after the BERT \cite{kenton2019bert} in early 2019, achieving a LFM in healthcare. At the end of 2022, ChatGPT \cite{patel2023chatgpt}, with its powerful versatility, enabled more healthcare-related practitioners to benefit from the foundation models, thus attracting their attention and further igniting the research upsurge of the HFMs. In August 2023 alone, more than 200 ChatGPT-related studies on healthcare were published \cite{qiu2023large}. For the VFMs, numerous preliminary works \cite{he2023geometric,15-MG} focused on independent pre-training or transfer learning. Owing to the extensive influence of the SAM \cite{Kirillov2023ICCV}, universal vision models \cite{22-MedSAM,mazurowski2023segment,DINOv2forrad} in healthcare have set off a research upsurge. In bioinformatics, the AlphaFold2 \cite{jumper2021highly} won the first place of CASP14 in protein structure prediction in 2020, arousing interest in BFMs and advancing the research in RNA \cite{wang2023uni}, DNA \cite{ji2021dnabert}, protein \cite{jumper2021highly}, etc. In early 2021, OpenAI constructed the CLIP \cite{radford2021learning} which constructed the large-scale learning of vision and language, achieving remarkable performance. Due to the natural multimodal property of healthcare data, this technology was quickly applied to healthcare \cite{zhao2023clip} and integrated the multimodal data from images, omics, text, etc. Until February 2024, the representative paper amount of HFM in the reviewed four sub-fields was growing exponentially (Fig.\ref{fig:develop}), in addition to the above typical technologies and events, some emerging paradigms and technologies are developing rapidly in HFM.

\subsection{Comparison of Related Surveys and Our Contributions}
In our extensive search, we discovered 17 representative surveys related to healthcare foundation models and it should be noted that existing surveys have provided insightful ideas in terms of different aspects of HFMs \cite{wang2023pre,thirunavukarasu2023large,zhou2023survey,he2023survey,yuan2023large,lee2024foundation,azad2023foundational,li2024progress,liu2024large,zhao2023clip,shrestha2023medical,qiu2023pre,qiu2023large,wang2023accelerating,moor2023foundation,ZHANG2024102996,zhang2024data}. Compared with these works, this survey has conducted a more comprehensive overview and analysis of HFM including the methods, data, and applications, and provided in-depth discussion and prospects for the challenges and future directions. Specifically, it has the following unique advantages: \textbf{1)} \textit{Systematic taxonomy and study of sub-fields in HFM.} This survey has covered four sub-fields related to HFM, including language, vision, bioinformatics, and multimodal. Compared with the existing surveys \cite{wang2023pre,thirunavukarasu2023large,he2023survey,yuan2023large,lee2024foundation,azad2023foundational,liu2024large,li2024progress,zhao2023clip,shrestha2023medical}, it provides a more comprehensive perspective on the whole HFM field. \textbf{2)} \textit{In-deep analysis of the methods in HFM.} This survey has deeply analyzed the methods in different sub-fields from pre-training to adaption which runs through the construction of a general AI model in healthcare. Compared with the existing surveys \cite{shrestha2023medical,wang2023pre,qiu2023pre,zhang2024data,zhao2023clip}, it provides a systematic summary of HFM methods. \textbf{3)} \textit{Extensive review of HFMs with different properties.} This survey has introduced the HFMs with the techniques of the whole process and is not limited to some special properties, like the ``large" \cite{qiu2023large}. Compared with the existing surveys \cite{qiu2023large,wang2023accelerating}, it provides an extensive view of the HFMs with different properties. \textbf{4)} \textit{Comprehensive and deeper exploration of different concerns in HFM.} This survey has explored comprehensive contents including the methods, data, applications, challenges, and future directions. Compared with the existing surveys \cite{moor2023foundation,ZHANG2024102996,wang2023accelerating}, it provides a complete vision for HFM so that the readers will achieve a deeper understanding.

This survey provides insight into the healthcare foundation models and our contributions are listed below:
\begin{enumerate}
    \item \textit{\textbf{Systematic Review of Methods}} (Section \ref{sec:methods}): A total of 200 technical papers from 2018 to 2024 (Jan-Feb) related to HFMs are enrolled in this survey. We presented a novel taxonomy for these papers and reviewed them in the pre-training and adaptation for language, vision, bioinformatics, and multimodal sub-fields. It provides insights into the potential technical innovations for healthcare foundation models.
    \item \textit{\textbf{Comprehensive Survey on Datasets}} (Section \ref{sec:datasets}): We surveyed 114 large-scale datasets/databases potentially available for HFM training across the four sub-fields in HFM. It identifies the current limitations in the healthcare datasets and provides data resource guidance for HFM researchers.
    \item \textit{\textbf{Thorough Overview of Applications}} (Section \ref{sec:applications}): We overview 16 potential healthcare applications in the current HFM works. It demonstrates the current development of HFM technologies in healthcare practices, providing a reference for future applications in more scenarios.
    \item \textit{\textbf{In-depth Discussion of Key Challenges}} (Section \ref{sec:challenges}): We discuss the key challenges related to data, algorithms, and computing infrastructures. It points out the current shortcomings of the HFM, providing new opportunities for researchers.
    \item \textit{\textbf{Farsighted Exploration of Emerging Future Directions}} (Section \ref{sec:directions}): We look forward to the future directions of HFM in terms of its role, implementation, application, and emphasis. It shows a transformation of healthcare AI from the conventional paradigm to the foundation model era, highlighting the future perspectives that hold promise in advancing the field.
\end{enumerate}

\section{Methods}
\label{sec:methods}
As shown in Fig.\ref{fig:overview}, HFM learns the representation for large-scale information from massive, diverse healthcare data, and then adapts to a wide range of healthcare applications. Therefore, in this section, we overview the LFM, VFM, BFM, and MFM from the perspectives of pre-training and adaptation. In this survey, we divide the pre-training paradigms as \textit{generative learning} (GL) learns a representation of data such that the model can generate meaningful information from the represented features; \textit{contrastive learning} (CL) that learns a representation of data such that similar instances are close together in the representation space, while dissimilar instances are far apart; \textit{hybrid learning} (HL) that learns a representation of data with a mixture of different learning methods; \textit{supervised learning} (SL) that use labeled data to train models to predict outcomes and recognize patterns. We divide the adaptation paradigms as \textit{fine-tuning} (FT) that adjusts the parameters inner pre-trained models; \textit{adapter tuning} (AT) that adds new parameters (adapters) into pre-trained models, and only train these additional parameters; \textit{prompt engineering} (PE) that inputs the designed or learned prompts into the pre-trained models to perform desired tasks.

\subsection{Language Foundation Models for Healthcare}
LFMs~\cite{kenton2019bert,radford2019language} have significantly advanced natural language processing (NLP) in healthcare \cite{jin2023medcpt,peng2023study,gu2021domain}. As shown in Tab.\ref{tab:lfm}, most of LFMs constructed GL methods in pre-training, and they utilized FT and prompt PE in adaptation.

\subsubsection{Pre-training}
LFM pre-training in healthcare is vital for training models on large, diverse medical textual datasets. This drives models to learn representation ability for generalizable features and achieves transferring capability for downstream tasks.

\textbf{a)} \textit{GL-based pre-training} is most widely used pre-training paradigm in LFM, which learns to generate medical text from large-scale medical corpora, learning representation ability for languages. One of the famous GL-based methods is next token prediction (NTP) \cite{peng2023study,wang2023huatuo,zhang2023huatuogpt,wu2023pmc,chen2023huatuogpt,chen2023meditron,zhang2023alpacare,chen2023bianque,li2023chatdoctor,han2023medalpaca,ye2023qilin,luo2023taiyi,wang2023gpt,xiong2023doctorglm,wang2023clinicalgpt,li2023beginner,labrak2024biomistral}, that predicts next token in a sequence via previous tokens. Representatively, GatorTronGPT \cite{peng2023study} combined medical and general text to pre-train a GPT-like model via the NTP, achieving effectiveness on multiple medical NLP tasks. Based on pre-trained LLaMA \cite{touvron2023llama}, PMC-LLaMA \cite{wu2023pmc} also constructed a data-centric knowledge injection process and learning medical language via NTP. Another widely used GL method is masked language modeling (MLM) \cite{gu2021domain,lee2020biobert,alsentzer2019publicly} that randomly masks a portion of the input tokens in a sentence and asks the model to predict the masked token.  AlphaBERT~\cite{chen2020modified}, and BEHRT~\cite{li2020behrt} are two typical methods that combined medical and general text to pre-train LFMs with MLM in BERT \cite{kenton2019bert} architectures.

\textbf{b)} \textit{Other pre-training} paradigms, including the CL and HL in healthcare LFMs, are investigating alternative techniques to capture medical linguistic structures and relationships. MedCPT \cite{jin2023medcpt}, which is a representative CL-based method, utilized PubMed search logs and learned a contrastive loss with query–document pairs and in-batch negatives, achieving new SOTA performance on six biomedical tasks. Inspired by BERT \cite{kenton2019bert}, some other methods also fused a next sentence prediction (NSP) approach, which trains the network to judge whether a sentence pair is adjacent, into the learning of MLM as HL methods. Some typical methods, i.e., PubMedBERT \cite{gu2021domain}, BioBERT \cite{lee2020biobert}, and ClinicalBERT \cite{alsentzer2019publicly}, all constructed BERT-like pre-training algorithms and learned the LFM for healthcare via a combination of MLM and NSP \cite{kenton2019bert}. Despite facing challenges in computational costs and data quality, these methods provide diverse strategies for the development of healthcare LFM.

\begin{table*}[tp]
\centering
\caption{The summary of LFM in healthcare. The abbreviations here are CL: contrastive learning, GL: generative learning, HL: hybrid learning, FT: fine-tuning, PE: prompt engineering, IR: information retrieval, NER: named entity recognition, RE: relation extraction, QA: question answering, VQA: visual question answering, DIAL: dialogue, NLI: natural language inference, TC: text classification, STS: semantic textual similarity, SUM: summarization, REC: recommendation, CLS: image classification, RG: report generation, and SEG: image segmentation.}
\begin{tabular}{lccccccc}
\toprule
Methods
& Pre-training
& Adaptation
& Backbone
& Downstream
& Year
& Code
\\
\midrule
MedCPT~\cite{jin2023medcpt}
& CL
& FT
& PubMedBERT
& IR
& 2023
& \href{https://github.com/ncbi/MedCPT}{\color{magenta}\Checkmark}
\\
AlphaBERT~\cite{chen2020modified}
& GL
& FT
& BERT
& NER, RE, QA
& 2020
& \href{https://github.com/wicebing/AlphaBERT}{\color{magenta}\Checkmark}
\\
BEHRT~\cite{li2020behrt}
& GL
& FT
& BERT
& NER, RE, QA
& 2020
& \href{https://github.com/deepmedicine/BEHRT}{\color{magenta}\Checkmark}
\\
BioBART~\cite{yuan-etal-2022-biobart}
& GL
& FT
& BART
& NER, RE, QA
& 2020
& \href{https://github.com/GanjinZero/BioBART}{\color{magenta}\Checkmark}
\\
PMC-LLaMA~\cite{wu2023pmc}
& GL
& FT
& LLaMA
& QA
& 2023
& \href{https://github.com/chaoyi-wu/PMC-LLaMA}{\color{magenta}\Checkmark}
\\
BioMistral~\cite{labrak2024biomistral}
& GL
& FT
& Mistral
& QA
& 2024
& \href{https://huggingface.co/BioMistral/BioMistral-7B}{\color{magenta}\Checkmark}
\\
Zhongjing~\cite{yang2024zhongjing}
& GL
& FT
& Ziya-LLaMA
& QA, DIAL
& 2024
& \href{https://github.com/SupritYoung/Zhongjing}{\color{magenta}\Checkmark}
\\
Me LLaMA~\cite{xie2024me}
& GL
& FT
& LLaMA2
& NER, RE, QA, NLI, SUM, CLS
& 2024
& \href{https://github.com/BIDS-Xu-Lab/Me-LLaMA}{\color{magenta}\Checkmark}
\\
OncoGPT~\cite{jia2024oncogpt}
& GL
& FT
& LLaMA
& DIAL
& 2024
& \href{https://github.com/OncoGPT1}{\color{magenta}\Checkmark}
\\
JMLR~\cite{wang2024jmlr}
& GL
& FT
& LLaMA-2
& QA
& 2024
&
\\
MEDITRON-70B~\cite{chen2023meditron}
& GL
& FT
& LLaMA-2
& QA
& 2023
& \href{https://github.com/epfLLM/meditron}{\color{magenta}\Checkmark}
\\
Qilin-Med~\cite{ye2023qilin}
& GL
& FT
& Baichuan
& QA
& 2023
& \href{https://github.com/williamliujl/Qilin-Med}{\color{magenta}\Checkmark}
\\
HuatuoGPT-II~\cite{chen2023huatuogpt}
& GL
& FT
& Baichuan2
& QA
& 2023
& \href{https://github.com/freedomintelligence/huatuogpt-ii}{\color{magenta}\Checkmark}
\\
ANTPLM-Med-10B~\cite{li2023beginner}
& GL
& FT
& AntGLM
& QA
& 2023
&
\\
GatorTronGPT~\cite{peng2023study}
& GL
& FT
& Transformer
& NER, RE, QA, NLI, STS
& 2023
& \href{https://github.com/uf-hobi-informatics-lab/gatortrongpt}{\color{magenta}\Checkmark}
\\
BioBERT~\cite{lee2020biobert}
& HL
& FT
& BERT
& NER, RE, QA
& 2019
& \href{https://github.com/dmis-lab/biobert}{\color{magenta}\Checkmark}
\\
PubMedBERT~\cite{gu2021domain}
& HL
& FT
&BERT
& NER, RE, QA, STS
& 2021
& \href{https://huggingface.co/microsoft/BiomedNLP-BiomedBERT-base-uncased-abstract-fulltext}{\color{magenta}\Checkmark}
\\
ClinicalBERT~\cite{alsentzer2019publicly}
& HL
& FT
& BERT
& NLI
& 2019
& \href{https://github.com/EmilyAlsentzer/clinicalBERT}{\color{magenta}\Checkmark}
\\
GatorTron~\cite{yang2022large}
& HL
& FT
& Transformer
& NER, RE, QA, NLI, STS
& 2022
& \href{https://github.com/uf-hobi-informatics-lab/GatorTron}{\color{magenta}\Checkmark}
\\
BenTsao~\cite{wang2023huatuo}
& -
& FT
& LLaMA
& QA
& 2023
& \href{https://github.com/SCIR-HI/Huatuo-Llama-Med-Chinese}{\color{magenta}\Checkmark}
\\
ChatDoctor~\cite{li2023chatdoctor}
& -
& FT
& LLaMA
& QA
& 2023
& \href{https://github.com/Kent0n-Li/ChatDoctor}{\color{magenta}\Checkmark}
\\
MedAlpaca~\cite{han2023medalpaca}
& -
& FT
& LLaMA
& QA
& 2023
& \href{https://github.com/kbressem/medAlpaca}{\color{magenta}\Checkmark}
\\
Alpacare~\cite{zhang2023alpacare}
& -
& FT
& LLaMA/LLaMA-2
& QA
& 2023
& \href{https://github.com/xzhang97666/alpacare}{\color{magenta}\Checkmark}
\\
MedPaLM~\cite{singhal2023publisher}
& -
& FT
& PaLM
& QA
& 2023
&
\\
MedPaLM 2~\cite{singhal2023towards}
& -
& FT
& PaLM-2
& QA
& 2023
&
\\
HuatuoGPT~\cite{zhang2023huatuogpt}
& -
& FT
& Baichuan
& QA, DIAL
& 2023
& \href{https://github.com/FreedomIntelligence/HuatuoGPT}{\color{magenta}\Checkmark}
\\
GPT-Doctor~\cite{wang2023gpt}
& -
& FT
& Baichuan2
& DIAL
& 2023
&
\\
DoctorGLM~\cite{xiong2023doctorglm}
& -
& FT
& ChatGLM
& QA
& 2023
& \href{https://github.com/xionghonglin/DoctorGLM}{\color{magenta}\Checkmark}
\\
Bianque~\cite{chen2023bianque}
& -
& FT
& ChatGLM
& QA
& 2023
& \href{https://github.com/scutcyr/BianQue}{\color{magenta}\Checkmark}
\\
Taiyi~\cite{luo2023taiyi}
& -
& FT
& Qwen
& NER, RE, TC, QA
& 2023
& \href{https://github.com/DUTIR-BioNLP/Taiyi-LLM}{\color{magenta}\Checkmark}
\\
BiMediX~\cite{pieri2024bimedix}
& -
& FT
& Mistral
& QA
& 2024
& \href{https://github.com/mbzuai-oryx/BiMediX}{\color{magenta}\Checkmark}
\\
ClinicalGPT~\cite{wang2023clinicalgpt}
& -
& FT
& BLOOM
& QA, DIAL
& 2023
&
\\
Visual Med-Alpaca \cite{shu2023visual}
& -
& FT, PE
& LLaMA
& VQA
& 2023
& \href{https://github.com/cambridgeltl/visual-med-alpaca}{\color{magenta}\Checkmark}
\\
OphGLM~\cite{gao2023ophglm}
& -
& FT, PE
& ChatGLM
& CLS, SEG
& 2023
& \href{https://github.com/ML-AILab/OphGLM}{\color{magenta}\Checkmark}
\\
ChatCAD~\cite{wang2023chatcad}
& -
& PE
& ChatGPT
& CLS, RG
& 2023
& \href{https://github.com/zhaozh10/ChatCAD}{\color{magenta}\Checkmark}
\\
ChatCAD+~\cite{zhao2023chatcad+}
& -
& PE
& ChatGPT
& CLS, RG
& 2023
& \href{https://github.com/zhaozh10/ChatCAD}{\color{magenta}\Checkmark}
\\
DeID-GPT~\cite{liu2023deid}
& -
& PE
& ChatGPT
& NER
& 2023
& \href{https://github.com/yhydhx/ChatGPT-API}{\color{magenta}\Checkmark}
\\
Dr.Knows~\cite{gao2023leveraging}
& -
& PE
& ChatGPT
& TC, SUM
& 2023
&
\\
Medprompt~\cite{nori2023can}
& -
& PE
& ChatGPT-4
& QA
& 2023
& \href{https://github.com/microsoft/promptbase}{\color{magenta}\Checkmark}
\\
HealthPrompt~\cite{sivarajkumar2022healthprompt}
& -
& PE
& ChatGPT
& TC
& 2022
&
\\
MedAgents~\cite{tang2023medagents}
& -
& PE
& ChatGPT / Flan-PaLM
& QA
& 2023
& \href{https://github.com/gersteinlab/MedAgents}{\color{magenta}\Checkmark}
\\
SPT~\cite{elfrink2023soft}
& -
& PE
& MedRoBERTa.nl
& TC
& 2023
& \href{https://bitbucket.org/aumc-kik/prompt_tuning_cancer_prediction/src/master/}{\color{magenta}\Checkmark}
\\
PBP~\cite{abaho2022position}
& -
& PE
& SciBERT
& TC
& 2022
&
\\
NapSS~\cite{lee2023clinical}
& -
& PE
& GPT-2
& REC
& 2023
& \href{https://clinical-decision-transformer.github.io/}{\color{magenta}\Checkmark}
\\
\bottomrule
\end{tabular}
\label{tab:lfm}
\end{table*}

\subsubsection{Adaptation}
Adaptation methods transfer LFMs in the general domain to specific tasks or domains using labeled data or natural language prompts, achieving their generalist applications in healthcare. As shown in Tab.\ref{tab:lfm}, with the fast development of the foundation models in the language field, a lot of works in healthcare focused on the adaptation from a pre-trained foundation language model. Most of the LFMs utilized FT and PE in adaptation.

\textbf{a)} \textit{FT-based adaptation} adjusts the parameters of pre-trained networks, adapting the LFM to downstream tasks without additional parameters. A lot of works \cite{wang2023huatuo,zhang2023huatuogpt,wu2023pmc,chen2023meditron,singhal2023large,singhal2023towards,zhang2023alpacare,chen2023bianque,pieri2024bimedix,li2023chatdoctor,han2023medalpaca} utilized full-parameter FT that directly adjusted all parameters using existing training datasets or human/LLM-generated instructions to improve their performance on target downstream tasks. For example, BenTsao~\cite{wang2023huatuo} was fine-tuned on 8K Chinese instruction data from CMeKG-8K~\cite{byambasuren2019preliminary}. HuatuoGPT~\cite{zhang2023huatuogpt} was fine-tuned on a mixture of 226k dialogue and instruction data for medical consultation. PMC-LLaMA~\cite{wu2023pmc} was further pre-trained on medical books and papers using a LLaMA~\cite{touvron2023llama} model, and then fine-tuned on the constructed instruction data collected from medical conversation~\cite{han2023medalpaca}, medical QA~\cite{jin2021disease}, and medical knowledge graph prompting~\cite{lindberg1993unified}. Another FT approaches
\cite{ye2023qilin,luo2023taiyi,wang2023gpt,xiong2023doctorglm,wang2023clinicalgpt,li2023beginner,yang2024zhongjing,xie2024me,jia2024oncogpt,wang2024jmlr}
are parameter-efficient FT that only adjust a part of parameters thus preserving a part of representation from pre-training and reducing the adapting costs. Some early LFMs \cite{10.1145/3649449} achieved the adaptation by fine-tuning a part of pre-trained parameters in the general natural language field.
Recently, low-rank adaptation (LoRA) \cite{hu2021lora} as a new parameter-efficient FT method has achieved success in healthcare LFMs. It injected trainable rank decomposition matrices into each layer of the Transformer and fused the new parameters into the original parameters in the deployment, greatly reducing the trainable parameter amount for downstream tasks without additional parameters. A lot of LFMs in healthcare, including Taiyi~\cite{luo2023taiyi}, GPT-doctor~\cite{wang2023gpt} and DoctorGLM~\cite{xiong2023doctorglm}, all utilized LoRA techniques achieving low-cost adaptation on multiple medical language tasks.

\textbf{b)} \textit{PE-based adaptation} \cite{wang2023prompt}, which design efficient prompts or instructions to guide model predictions or tuning, has also been widely applied in LFMs owing to their powerful task adaptation ability. One of the PE methods is hand-crafted prompting \cite{shu2023visual,gao2023ophglm,liu2023deid,gao2023leveraging,nori2023can,sivarajkumar2022healthprompt}. It creates natural language prompts to elicit the ability of a general LFM in the healthcare domain. DelD-GPT~\cite{liu2023deid} used ChatGPT or GPT-4 as the backbone model and employed the Chain of Thought (CoT)~\cite{wei2022chain} technique to generate prompts that can de-identify sensitive information in medical data, such as names, dates, or locations. Dr. Knows~\cite{gao2023leveraging} also used ChatGPT as the backbone model and utilized the zero-shot prompting technique to generate prompts that can answer medical questions and provide automated diagnoses based on the symptoms and conditions of the patients. Medprompt~\cite{nori2023can} utilized GPT-4 as the backbone model and incorporated a combination of the CoT and ensemble prompting techniques to generate prompts that can perform various medical tasks. HealthPrompt~\cite{sivarajkumar2022healthprompt} used six different pre-trained LFMs as the backbone models and applied a manual template zero-shot approach to generate prompts that can classify medical texts into different categories, such as diseases, drugs, or procedures. Visual Med-Alpaca~\cite{shu2023visual} and OphGLM~\cite{gao2023ophglm} integrate an LFM with specialized vision models. By doing so, they address medical tasks beyond the language modality without incurring the development costs associated with creating a vision-language foundation model. Another PE method is learnable prompting~\cite{elfrink2023soft}. It utilizes soft-prompt tuning to learn natural language prompts for specific tasks. Several works \cite{abaho2022position,beltagy2019scibert,lee2023clinical,radford2019language,verkijk2021medroberta} have used this approach for medical text classification. For instance, PBP~\cite{abaho2022position} used SciBERT~\cite{beltagy2019scibert} as the backbone model and learned natural language prompts that can classify medical texts. NapSS~\cite{lee2023clinical} used GPT-2~\cite{radford2019language} as the backbone model and learned natural language prompts that can generate personalized recommendations for clinical scenarios. MedRoBERTa.nl~\cite{verkijk2021medroberta} used soft-prompt tuning to learn natural language prompts that can classify medical texts into different categories.

\subsection{Vision Foundation Models for Healthcare}
Inspired by the revolutionary impact of LFMs, VFMs also have been explored for their generalist ability, excelling in a range of downstream tasks \cite{awais2023foundational}. As shown in Tab.\ref{tab:vfm}, they undergo pre-training on extensive labeled or unlabeled medical datasets, enabling adaptation to numerous downstream tasks.

\subsubsection{Pre-training} Different from the language, the continuity of vision information makes it challenging to separate the semantics of the content \cite{MoCo}. Therefore, except for self-supervised learning (SSL), VFMs \cite{ma2023towards,ZHANG2024102996,lee2024foundation} also utilize supervised learning (SL) for task-specific pre-training.

\textbf{a)} \textit{SL pre-training} paradigm utilizes the annotation to decouple the semantics inner the medical images, learning broad applicability for specific tasks. A previous work called Med3d \cite{chen2019med3d} pre-trained a ResNet and a multi-branch decoder from eight 3D medical image segmentation (MIS) datasets for transfer learning of downstream tasks. Recently, most approaches aim to directly pre-train a unified model with generalist ability via a specific task, i.e., segmentation. A typical work is the STU-Net ~\cite{5-STU-Net} which is pre-trained on the TotalSegmentator dataset \cite{TotalSegmentator} with 1204 CT volumes and the masks for 104 organs. Due to the high costs of the annotation, some other works mix several public annotated datasets to construct a large-scale annotated dataset. UniverSeg~\cite{71-UniverSeg} pre-trained its universal segmentation ability on 53 opened MIS datasets comprising over 22k scans. Most recently, universal and interactive medical image segmentation models have been significantly driven by the segment anything model (SAM) \cite{Kirillov2023ICCV}, which mainly involves an image encoder, a prompt encoder, and a mask decoder. For instance, SAM-Med3D~\cite{30-SAM-Med3D} and SAM-Mad2D \cite{SA-Med2D-20M} further transferred the pre-trained parameters from SAM to medical images and tuning their networks on a large-scale dataset mixed by several public MIS datasets. Although these task-specific VFMs have demonstrated remarkable performance, the high annotation costs make it extremely challenging to construct large-scale training datasets. Most of the existing supervised pre-training works are still only performed on the MIS tasks lacking task diversity.

Due to the high costs of medical image annotation, self-supervised pre-training (SSP) \cite{zhou2023unified,he2023geometric} has become a widely studied paradigm in VFMs. It constructs a pretext task to drive the learning without annotation for universal feature representations from large-scale data. Therefore, it paves the way for further development of advanced VFMs in different downstream medical image tasks, holding the promise of advancing medical image analysis and broadening its applications in various contexts.

\textbf{b)} \textit{GL-based pre-training} in VFMs, including the RETFound~\cite{3-RETFound}, VisionFM~\cite{12-VisionFM}, SegVol~\cite{33-SegVol}, DeblurringMAE~\cite{9-DeblurringMAE}, USFM \cite{1-USFM}, and Models Genesis \cite{14-MG,15-MG}, learns generic vision representations by predicting or reconstructing the original input from its corrupted counterpart. A commonly used objective is masked image modeling (MIM)~\cite{iBOT,he2022masked,xie2022simmim}, employing an encoder-decoder architecture to encode corrupted images and decode the original version. For example, RETFound \cite{3-RETFound}, VisionFM \cite{12-VisionFM} and SegVol~\cite{33-SegVol} were developed based on MIM for retinal images and ophthalmic clinical tasks. DeblurringMAE \cite{9-DeblurringMAE} introduced a deblurring task into pre-training, while USFM \cite{1-USFM} proposed a spatial-frequency dual-masked MIM approach. Models Genesis \cite{14-MG, 15-MG} used image restoration as a pretext task, effectively capturing fine-grained visual information.

\textbf{c)} \textit{CL-based pre-training} in VFMs contrasts the similarities or differences between images to learn discriminative vision representations. These works~\cite{83-CTransPath,16-C2L,18-MoCo-CXR,65-Histopathology,4-Endo-FM,47-LVM-Med,wu2024voco} learn discriminative visual features by ensuring that a query image is close to its positive samples and far from its negative samples in the embedding space. With the success of the CL in natural images, some works also utilized the MoCo~\cite{MoCo} or SimCLR~\cite{SimCLR} algorithms on medical images, achieving success in pathology \cite{83-CTransPath,65-Histopathology} and X-ray \cite{18-MoCo-CXR} images. C2L~\cite{16-C2L} constructed homogeneous and heterogeneous data pairs and compared different image representations to learn general and robust features. Endo-FM~\cite{4-Endo-FM} was pre-trained under a teacher-student scheme via spatial-temporal matching on diverse video views. Both teacher and student models process these views of a video and predict one view from another in the latent feature space. LVM-Med~\cite{47-LVM-Med} was pre-trained on 1.3 million images from 55 publicly available datasets, covering a large number of organs and modalities via a second-order graph matching. Wu \textit{et al.} \cite{wu2024voco} proposed a simple yet effective VoCo framework to leverage the contextual position priors for pre-training. Besides, MIS-FM~\cite{2-MIS-FM} introduced a pretext task based on pseudo-segmentation, where Volume Fusion (VF) was proposed to generate paired images and segmentation labels to pre-train the 3D segmentation model. Ghesu \textit{et al.} \cite{54-self} proposed a method for self-supervised learning based on CL and online feature clustering.

\textbf{d)} \textit{HL-based pre-training} combine various pre-training approaches to fuse their advantages in joint training. Virchow~\cite{10-Virchow}, UNI~\cite{11-UNI}, and RudolfV~\cite{48-RudolfV} utilized the DINOv2~\cite{DINOv2} training paradigm, which integrated the MIM and CL. BROW~\cite{6-BROW} integrated color augmentation, patch shuffling, MIM, and multi-scale input to pre-train the foundation model in a self-distillation framework. TransVW~\cite{haghighi2021transferable}  integrated self-classification and
self-restoration to train the model and learned representation from multiple sources of information. GVSL \cite{he2023geometric} learned the similarity between medical images via registration learning and the reconstruction ability via self-restoration.

\begin{table*}
\centering
\caption{The methods of VFM in healthcare. The abbreviations here are SL: supervised learning, GL: generative learning, CL: contrastive learning, HL: hybrid learning, FT: fine-tuning, PE: prompt engineering, AT: adapter tuning, CLS: classification, SEG: segmentation, DET: detection, PR: prognosis, RET: retrieval, and IE: image enhancement.}
\begin{tabularx}{\linewidth}{
p{0.17\textwidth}
C{0.085\textwidth}
C{0.08\textwidth}
C{0.14\textwidth}
C{0.14\textwidth}
C{0.12\textwidth}
C{0.04\textwidth}
cccc}
\toprule
\textbf{Methods}
& \textbf{Pre-training}
& \textbf{Adaptation}
& \textbf{Backbone}
&\textbf{Modalities}
& \textbf{Downstream}
& \textbf{Year}
& \textbf{Code}
\\
\midrule
Med3D \cite{chen2019med3d}
& SL
& FT
& ResNet
& CT, MRI
& SEG, CLS
& 2019
& \href{https://github.com/Tencent/MedicalNet}{\color{magenta}\Checkmark}
\\
STU-Net~\cite{5-STU-Net}
& SL
& FT, PE
& nnU-Net
& CT
& SEG
& 2023
& \href{https://github.com/Ziyan-Huang/STU-Net}{\color{magenta}\Checkmark}
\\
UniverSeg~\cite{71-UniverSeg}
& SL
&PE
& U-Net
& Multimodal images
& SEG
& 2023
& \href{https://universeg.csail.mit.edu/}{\color{magenta}\Checkmark}
\\
SAM-Med3D~\cite{30-SAM-Med3D}
& SL
& PE
& ViT (SAM)
& Multimodal images
& SEG
& 2023
& \href{https://github.com/uni-medical/SAM-Med3D}{\color{magenta}\Checkmark}
\\
RETFound~\cite{3-RETFound}
& GL
& FT
& ViT (MAE)
& CFP, OCT
& CLS, PR, DET
& 2023
\\
VisionFM~\cite{12-VisionFM}
& GL
& FT
& -
&Multimodal images
& CLS
& 2023
&
\\
SegVol~\cite{33-SegVol}
& GL
& FT, PE
& ViT (SAM)
& CT
& SEG
& 2023
& \href{https://github.com/BAAI-DCAI/SegVol}{\color{magenta}\Checkmark}
\\
Models Genesis~\cite{14-MG,15-MG}
& GL
& FT, AT
& U-Net
& CT, X-ray
& CLS, SEG
& 2019
& \href{https://github.com/MrGiovanni/ModelsGenesis}{\color{magenta}\Checkmark}
\\
DeblurringMAE~\cite{9-DeblurringMAE}
& GL
& FT, AT
& ViT (MAE)
& US
& CLS
& 2023
& \href{https://github.com/MembrAI/DeblurringMIM}{\color{magenta}\Checkmark}
\\
USFM~\cite{1-USFM}
& GL
& AT
& -
& US
& SEG, CLS, IE
& 2024
&
\\
C2L~\cite{16-C2L}
& CL
&FT
& ResNet/DenseNet
& X-ray
& CLS
& 2020
& \href{https://github.com/funnyzhou/C2L_MICCAI2020}{\color{magenta}\Checkmark}
\\
Endo-FM~\cite{4-Endo-FM}
& CL
& FT
& ViT
& Endoscopy
& SEG, CLS, DET
& 2023
& \href{https://github.com/med-air/Endo-FM}{\color{magenta}\Checkmark}
\\
Ciga \textit{et al.}~\cite{65-Histopathology}
& CL
& FT
& ResNet (SimCLR)
& Pathology
& CLS, SEG
& 2022
& \href{https://github.com/ozanciga/self-supervised-histopathology}{\color{magenta}\Checkmark}
\\
CTransPath~\cite{83-CTransPath}
& CL
& FT
& ViT  (MoCo v3)
& Pathology
& RET, CLS
& 2022
& \href{https://github.com/Xiyue-Wang/TransPath}{\color{magenta}\Checkmark}
\\
LVM-Med~\cite{47-LVM-Med}
& CL
&FT
& ResNet, ViT
& Multimodal images
& SEG, CLS, DET
& 2024
& \href{https://github.com/duyhominhnguyen/LVM-Med}{\color{magenta}\Checkmark}
\\
MIS-FM~\cite{2-MIS-FM}
& CL
& FT
& Swin
& CT
& SEG
& 2023
& \href{https://github.com/openmedlab/MIS-FM}{\color{magenta}\Checkmark}
\\
VoCo~\cite{wu2024voco}
& CL
&FT
& Swin
&CT
& SEG, CLS
& 2024
& \href{https://github.com/Luffy03/VoCo}{\color{magenta}\Checkmark}
\\
MoCo-CXR~\cite{18-MoCo-CXR}
& CL
& FT, AT
& ResNet, DenseNet
& X-ray
& CLS
& 2021
&
\\
TransVW~\cite{haghighi2021transferable}
& HL
& FT
& U-Net
& CT, X-ray
& CLS, SEG
& 2021
&
\\
Ghesu \textit{et al.}~\cite{54-self}
& HL
& FT
& ResNet
& X-ray, CT, MRI, US
& DET, SEG
& 2022
&
\\
UNI~\cite{11-UNI}
& HL
&FT
& ViT (DINOv2)
&Pathology
& CLS, SEG
& 2024
& \href{https://github.com/mahmoodlab/UNI}{\color{magenta}\Checkmark}
\\
BROW~\cite{6-BROW}
& HL
&FT
& ViT
&Pathology
& CLS, SEG
& 2023
&
\\
Campanella \textit{et al.}~\cite{62-Pathology}
& HL
&FT
& ViT (MAE, DINO)
&Pathology
& CLS
& 2023
&
\\
RudolfV~\cite{48-RudolfV}
& HL
&FT
& ViT (DINOv2)
& Pathology
& CLS
& 2024
&
\\
Swin UNETR~\cite{68-swinunetr}
& HL
& FT
& Swin
& CT
& SEG
& 2022
& \href{https://monai.io/research/swin-unetr}{\color{magenta}\Checkmark}
\\
GVSL \cite{he2023geometric}
& HL
& FT, AT
& U-Net
& CT, MRI
& SEG, CLS
& 2023
& \href{https://github.com/YutingHe-list/GVSL}{\color{magenta}\Checkmark}
\\
Virchow~\cite{10-Virchow}
&HL
&AT
&ViT (DINOv2)
&Pathology
& CLS
& 2023
&
\\
MA-SAM \cite{chen2023ma}
& -
& FT, AT, PE
& ViT (SAM)
& CT, MRI, Endoscopy
& SEG
& 2023
& \href{https://github.com/cchen-cc/MA-SAM}{\color{magenta}\Checkmark}
\\
Pancy \textit{et al.}~\cite{56-comprehensive}
&-
& FT, AT, PE
& YOLOv8, ViT (SAM)
& Multimodal images
& SEG
& 2023
&
\\

3DSAM-adapter~\cite{24-3DSAM-adapter}
&-
& FT, AT, PE
& ViT (SAM)
& CT
& SEG
& 2023
&\href{https://github.com/med-air/3DSAM-adapter}{\color{magenta}\Checkmark}
\\
SP-SAM~\cite{88-part}
&-
&FT, AT, PE
& ViT (SAM)
& Endoscopy
& SEG
& 2023
& \href{https://github.com/wenxi-yue/SurgicalPart-SAM}{\color{magenta}\Checkmark}
\\
Baharoon \textit{et al.}~\cite{DINOv2forrad}
& -
& FT, AT, PE
& ViT (DINOv2)
& X-ray, CT, MRI
& SEG, CLS
& 2023
& \href{https://github.com/MohammedSB/DINOv2ForRadiology}{\color{magenta}\Checkmark}
\\
MedSAM~\cite{22-MedSAM}
& -
& FT, PE
& ViT (SAM)
& Multimodal images
& SEG
& 2023
&\href{https://github.com/bowang-lab/MedSAM}{\color{magenta}\Checkmark}
\\
Skinsam~\cite{34-Skinsam}
 &-
& FT, PE
& ViT (SAM)
& Dermoscopy
& SEG
& 2023
&
\\
Polyp-SAM~\cite{35-Polyp-sam}
&-
& FT, PE
& ViT (SAM)
& Endoscopy
& SEG
& 2023
& \href{https://github.com/ricklisz/Polyp-SAM}{\color{magenta}\Checkmark}
\\
SAM-OCTA~\cite{53-SAM-OCTA}
&-
& FT, PE
& ViT (SAM)
& OCT
& SEG
& 2023
& \href{https://github.com/ShellRedia/SAM-OCTA}{\color{magenta}\Checkmark}
\\
SAMed~\cite{44-SAMed}
&-
& FT, PE
& ViT (SAM)
& CT
& SEG
& 2023
& \href{https://github.com/hitachinsk/SAMed}{\color{magenta}\Checkmark}
\\
SAM-LST~\cite{36-LFTSAM}
&-
& FT, PE
& ViT (SAM)
& CT
& SEG
& 2023
& \href{https://github.com/11yxk/SAM-LST}{\color{magenta}\Checkmark}
\\
Feng \textit{et al.}~\cite{45-cheap}
&-
& FT, PE
& ViT (SAM)
& CT, MRI
& SEG
& 2023
&
\\
SemiSAM~\cite{87-semisam}
 &-
& FT, PE
& ViT (SAM)
& MRI
& SEG
& 2023
&
\\
AFTer-SAM~\cite{28-AFTer-SAM}
&-
& AT, PE
& ViT (SAM)
&CT
& SEG
& 2024
&
\\
Mammo-SAM~\cite{82-Mammo-SAM}
 &-
& AT, PE
& ViT (SAM)
& CT
& SEG
& 2023
&
\\
ProMISe~\cite{74-promise}
 &-
& AT, PE
& ViT (SAM)
& CT
& SEG
& 2023
& \href{https://github.com/MedICL-VU/ProMISe}{\color{magenta}\Checkmark}
\\
Med-SA~\cite{23-Med-SA}
 &-
& AT, PE
& ViT (SAM)
& Multimodal images
& SEG
& 2023
& \href{https://github.com/KidsWithTokens/Medical-SAM-Adapter}{\color{magenta}\Checkmark}
\\
SAM-Med2D~\cite{26-SAM-Med2D}
 &-
& AT, PE
& ViT (SAM)
& Multimodal images
& SEG
& 2023
& \href{https://github.com/uni-medical/SAM-Med2D}{\color{magenta}\Checkmark}
\\
Adaptivesam~\cite{29-Adaptivesam}
 &-
& AT, PE
& ViT (SAM)
& Multimodal images
& SEG
& 2024
& \href{https://github.com/JayParanjape/biastuning}{\color{magenta}\Checkmark}
\\
MediViSTA-SAM~\cite{31-MediViSTA-SAM}
&-
& AT, PE
& ViT (SAM)
& US
& SEG
& 2023
& \href{https://github.com/kimsekeun/MediViSTA-SAM}{\color{magenta}\Checkmark}
\\
SAMUS~\cite{51-samus}
&-
& AT, PE
& ViT (SAM)
&US
& SEG
& 2023
& 
\\
SegmentAnyBone~\cite{91-SegmentAnyBone}
&-
& AT, PE
& ViT (SAM)
& MRI
& SEG
& 2024
& \href{https://github.com/mazurowski-lab/SegmentAnyBone}{\color{magenta}\Checkmark}
\\
Swinsam~\cite{90swinsam}
 &-
& AT, PE
& ViT (SAM)
& Endoscopy
& SEG
& 2024
& 
\\
SAMAug~\cite{SAMAug}
&-
& PE
& ViT (SAM)
& Multimodal images
& SEG
& 2023
& \href{https://github.com/yizhezhang2000/SAMAug}{\color{magenta}\Checkmark}
\\
AutoSAM~\cite{40-AutoSAM}
 &-
& PE
& ViT (SAM)
& Multimodal images
& SEG
& 2023
&
\\
DeSAM~\cite{41-DeSAM}
&-
& PE
& ViT (SAM)
& Multimodal images
& SEG
& 2023
& \href{https://github.com/yifangao112/DeSAM}{\color{magenta}\Checkmark}
\\
CellSAM~\cite{86-CellSAM}
&-
& PE
& ViT (SAM)
& Multimodal images
& SEG
& 2023
& \href{https://label-dev.deepcell.org/}{\color{magenta}\Checkmark}
\\
Sam-u~\cite{42-SAM-U}
 &-
& PE
& ViT (SAM)
& Fundus
& SEG
& 2023
&
\\
Sam-path~\cite{52-Sam-path}
&-
& PE
& ViT (SAM)
& Pathology
& SEG
& 2023
&
\\
All-in-sam~\cite{27-All}
&-
& PE
& ViT (SAM)
& Pathology
& SEG
& 2023
&
\\
SurgicalSAM~\cite{58-surgicalsam}
 &-
& PE
& ViT (SAM)
& Endoscopy
& SEG
& 2024
& \href{https://github.com/wenxi-yue/SurgicalSAM}{\color{magenta}\Checkmark}
\\
Polyp-SAM++~\cite{73-polyp}
 &-
& PE
& ViT (SAM)
& Endoscopy
& SEG
& 2023
& \href{https://github.com/RisabBiswas/Polyp-SAM++}{\color{magenta}\Checkmark}
\\
UR-SAM~\cite{39-UR-SAM}
 &-
& PE
& ViT (SAM)
& CT
& SEG
& 2023
&
\\
MedLSAM~\cite{25-MedLSAM}
&-
& PE
& ViT (SAM)
& CT
& SEG
& 2023
& \href{https://github.com/openmedlab/MedLSAM}{\color{magenta}\Checkmark}
\\
nnSAM~\cite{37-nnsam}
 &-
& PE
& ViT (SAM)
& CT
& SEG
& 2023
& \href{https://github.com/Kent0n-Li/Medical-Image-Segmentation}{\color{magenta}\Checkmark}
\\
\hline
Continue to the next page.
\end{tabularx}
\label{tab:vfm}
\end{table*}
\begin{table*}[t]
\begin{tabularx}{\linewidth}{
p{0.17\textwidth}
C{0.085\textwidth}
C{0.08\textwidth}
C{0.135\textwidth}
C{0.14\textwidth}
C{0.12\textwidth}
C{0.04\textwidth}
cccc}
\multicolumn{2}{l}{Continue from the previous page.}
\\
\hline
EviPrompt~\cite{38-EviPrompt}
 &-
& PE
& ViT (SAM)
& CT, MRI
& SEG
& 2023
&
\\
Anand \textit{et al.}~\cite{57-one}
 &-
& PE
& ViT (SAM)
& CT, MRI, US
& SEG
& 2023
&
\\
SAMM~\cite{32-SAMM}
 &-
& PE
& ViT (SAM)
& CT, MRI, US
& SEG
& 2023
&\href{https://github.com/bingogome/samm}{\color{magenta}\Checkmark}
\\
SAMPOT~\cite{79-SAMPOT}
&-
& PE
& ViT (SAM)
& X-ray
& SEG
& 2023
&
\\
PUNETR \cite{fischer2024prompt}
& -
& PE
& -
& CT
& SEG
& 2024
& \href{https://github.com/marcdcfischer/PUNETR}{\color{magenta}\Checkmark}
\\
\bottomrule
\end{tabularx}
\end{table*}

\subsubsection{Adaptation}
After pre-training, VFMs further construct adaptation methods to generalize to a wide range of tasks. In addition to the classic fine-tuning, some novel methods including adapter tuning (AT) and prompt engineering (PE) have been applied to the adaptation of VFM recently.

\textbf{a)} \textit{FT-based adaptation} methods optimize the parameters inner the pre-trained VFMs on specific datasets, adapting the models' representations to the downstream tasks. Some works fine-tune all parameters of pre-trained VFMs~\cite{56-comprehensive,22-MedSAM,34-Skinsam,35-Polyp-sam,53-SAM-OCTA,he2023geometric}, demonstrating significant improvement on specific tasks. These works are closer to data-driven initialization methods, which utilize the pre-trained weights as a better initialization to learn specific tasks. However, these methods are not only time-consuming but also prone to overfitting due to data scarcity caused by privacy issues in medical images. Other works utilized parameter-efficient fine-tuning which only adjusted a part of parameters~\cite{44-SAMed, 36-LFTSAM, 24-3DSAM-adapter, 45-cheap, 88-part, 87-semisam}, which can reduce the tuning costs, improve the calculation efficiency, and effectively maintain the representation inner the pre-trained weights. However, the fine-tuned parameter selection has to be manually designed, which limits the adaptability. Therefore, recently, inspired by the LoRA-based adaptation \cite{hu2021lora} in LFM, VFMs also utilized the low-rank methods to effectively adapt the pre-trained models to downstream tasks with low costs. For example, some VFMs~\cite{23-Med-SA,44-SAMed,45-cheap} maintain the pre-trained SAM parameters and take LoRA for efficient adaptation.

\textbf{b)} \textit{AT-based adaptation} methods add some adapters into pre-trained VFMs and only optimize these adapters to adapt the VFMs to downstream tasks. Different from the FT, it will not change original parameters, thus preserving the VFM's learned generic representations from large-scale data. A previous practice called ``Linear evaluation" is widely used to evaluate the generalization ability of pre-trained backbone \cite{14-MG,15-MG,he2023geometric,18-MoCo-CXR}. It generally adds a linear layer as the adapter at the end of the backbone and optimizes this layer when adaptation, thus evaluating the representation ability of the pre-trained weights. Recently, the adapters have been further added into the inner layers of the network for better transferring ability. A lot of practices based on SAM have demonstrated the AT's great performance on medical image segmentation~\cite{56-comprehensive,28-AFTer-SAM,82-Mammo-SAM,24-3DSAM-adapter,29-Adaptivesam,31-MediViSTA-SAM,51-samus,88-part,91-SegmentAnyBone,90swinsam,74-promise}. They keep SAM's image segmentation capabilities learned from large-scale natural images and effectively transfer them to medical images by training very few parameters in the adapters. All-in-SAM \cite{27-All} constructed a weakly supervised adaptation method that utilizes the SAM for pseudo labels via prompts and then adapts the SAM via the AT following \cite{50-samadapter}. MA-SAM \cite{chen2023ma} further embedded 3D adapters into the original 2D SAM model, constructing a 3D SAM for 3D medical images.

\textbf{c)} \textit{PE-based adaptation} methods \cite{wang2023prompt} also have achieved powerful adaptation performance in VFMs. Following the SAM, a lot of SAM-based VFMs in healthcare \cite{87-semisam,40-AutoSAM,41-DeSAM,42-SAM-U,22-MedSAM,26-SAM-Med2D,30-SAM-Med3D} utilized the point, bounding box, and text as the prompts for medical image segmentation. Baharoon \textit{et al.} \cite{DINOv2forrad} studied the prompt templates suitable for medical images on DINO v2. Besides, few-shot prompting, which provides few image-label pairs as the prompts, is also used in prompt engineering. UniverSeg \cite{71-UniverSeg} utilized support sets as prompts to segment any targets on query images. Anand \textit{et al.} \cite{57-one} proposed a one-shot localization and segmentation framework to leverage the correspondence to a template image to prompt SAM. Some VFMs \cite{41-DeSAM,40-AutoSAM,56-comprehensive,fischer2024prompt} further designed automatic prompt-generation methods. AutoSAM \cite{40-AutoSAM} embedded an auxiliary prompt encoder to generate a surrogate prompt via the features of the input images, eliminating the manual prompts. PUNETR \cite{fischer2024prompt} studied the prompt tuning method which embeds some learnable prompt tokens into the pre-trained networks to adapt the prompt for medical images.

\subsection{Bioinformatics Foundation Models in Healthcare}

\begin{table*}[tp]
\label{table:bfm}
\centering
\caption{The representative methods of BFM in healthcare. The abbreviations here are CL: contrastive learning, GL: generative learning, HL: hybrid learning, FT: fine-tuning, AT: adapter tuning, PE: prompt engineering, SA: sequence analysis, IA: interaction analysis, SFA: structure and function analysis, and DR: disease research and drug response}
\begin{tabular}{lcccccccc}
\toprule
Models
& Pre-training
& Adaptation
& Backbone
& Modalities
& Downstream
& Year
& Code
\\
\midrule
ProGen~\cite{madani2023large}
& GL
& FT
& Transformer Decoder
& Protein
& SA
& 2023
& \href{https://github.com/salesforce/progen}{\color{magenta}\Checkmark}
\\
ProGen2~\cite{Nijkamp2023progen2}
& GL
& FT
& Transformer Decoder
& Protein
& SA, SFA
& 2023
& \href{https://github.com/salesforce/progen/tree/main/progen2}{\color{magenta}\Checkmark}
\\
scBERT~\cite{Yang2022scbert}
& GL
& FT
& BERT
& scRNA-seq
& SA
& 2022
& \href{https://github.com/TencentAILabHealthcare/scBERT}{\color{magenta}\Checkmark}
\\
Geneformer~\cite{Theodoris2023geneformer}
& GL
& FT
& BERT
& scRNA-seq
& IA, DR
& 2023
& \href{https://huggingface.co/ctheodoris/Geneformer}{\color{magenta}\Checkmark}
\\
DNABERT~\cite{ji2021dnabert}
& GL
& FT
& BERT
& DNA
& SA, SFA
& 2021 & \href{https://github.com/jerryji1993/DNABERT}{\color{magenta}\Checkmark}
\\
DNABERT-2~\cite{Zhou2023dnabert2}
& GL
& FT
& BERT
& DNA
& SA, DR
& 2023 & \href{https://github.com/Zhihan1996/DNABERT_2}{\color{magenta}\Checkmark}
\\
Nucleotide Transformer~\cite{dalla2023nucleotide}
& GL
& FT
& BERT
& DNA
& SA, SFA
& 2023
& \href{https://github.com/instadeepai/nucleotide-transformer}{\color{magenta}\Checkmark}
\\
Gena-LM~\cite{Fishman2023genalm}
& GL
& FT
& BERT
& DNA
& SA
& 2023 & \href{https://github.com/AIRI-Institute/GENA_LM}{\color{magenta}\Checkmark}
\\
RNA-FM~\cite{chen2022interpretable}
& GL
& FT
& BERT
& RNA
& SA, IA, SFA
& 2022
& \href{https://github.com/ml4bio/RNA-FM}{\color{magenta}\Checkmark}
\\
RNA-MSM~\cite{Zhang2023rnamsm}
& GL
& FT
& BERT
& RNA
& SFA
& 2024
& \href{https://github.com/yikunpku/RNA-MSM}{\color{magenta}\Checkmark}
\\
SpliceBERT~\cite{Chen2023splicebert}
& GL
& FT
& BERT
& RNA
& SA, SFA
& 2023
& \href{https://github.com/biomed-AI/SpliceBERT}{\color{magenta}\Checkmark}
\\
3UTRBERT~\cite{Yang20233utr}
& GL
& FT
& BERT
& RNA
& SA
& 2023
& \href{https://github.com/yangyn533/3UTRBERT}{\color{magenta}\Checkmark}
\\
ESM-2~\cite{Lin2023esm2}
& GL
& FT
& BERT
& Protein
& SFA
& 2023
& \href{https://github.com/facebookresearch/esm}{\color{magenta}\Checkmark}
\\
ProtTrans~\cite{Elnaggar2022prottrans}
& GL
& FT
& BERT
& Protein
& SFA
& 2021
& \href{https://github.com/agemagician/ProtTrans}{\color{magenta}\Checkmark}
\\

MSA Transformer~\cite{Rao2021msa}
& GL
& FT
& BERT
& Protein
& SFA
& 2021
& \href{https://github.com/rmrao/msa-transformer}{\color{magenta}\Checkmark}
\\
ESM-1b~\cite{Rives2021esm1b}
& GL
& FT
& BERT
& Protein
& SFA
& 2021
& \href{https://github.com/facebookresearch/esm}{\color{magenta}\Checkmark}
\\
AlphaFold~\cite{jumper2021highly}
& GL
& FT
& Evoformer
& Protein
& SFA
& 2021
& \href{https://github.com/google-deepmind/alphafold}{\color{magenta}\Checkmark}
\\
HyenaDNA~\cite{Nguyen2023hyenadna}
& GL
& FT, PE
& Transformer Decoder
& DNA
& SA
& 2023 & \href{https://github.com/HazyResearch/hyena-dna}{\color{magenta}\Checkmark}
\\
scFoundation~\cite{Hao2023scfoundation}
& GL
& PE, AT
& Asymmetric Encoder-decoder
& scRNA-seq
& DR
& 2023 & \href{https://github.com/biomap-research/scFoundation}{\color{magenta}\Checkmark}
\\
UCE \cite{Rosen2023uce}
& GL
& AT
& BERT
& scRNA-seq
& SFA
& 2023 & \href{https://github.com/snap-stanford/UCE}{\color{magenta}\Checkmark}
\\
DNAGPT~\cite{Zhang2023dnagpt}
& HL
& FT
& Transformer Decoder
& DNA
& SA
& 2023
& \href{https://github.com/Xianjun-Yang/DNA-GPT}{\color{magenta}\Checkmark}
\\
scGPT~\cite{cui2023scgpt}
& HL
& FT
& Transformer Decoder
& scRNA-seq
& IA, SFA
& 2023
& \href{https://github.com/bowang-lab/scGPT}{\color{magenta}\Checkmark}
\\
RNABERT~\cite{Akiyama2022rnabert}
& HL
& FT
& BERT
& RNA
& SA, SFA
& 2022
& \href{https://github.com/mana438/RNABERT}{\color{magenta}\Checkmark}
\\
AminoBERT (RGN2)~\cite{Chowdhury2022aminobert}
& HL
& FT
& BERT
& Protein
& SFA
& 2022
& \href{https://github.com/aqlaboratory/rgn2}{\color{magenta}\Checkmark}
\\
UTR-LM~\cite{Chu2023utrlm}
& HL
& FT
& BERT
& RNA
& SA, SFA
& 2023
& \href{https://github.com/a96123155/UTR-LM}{\color{magenta}\Checkmark}
\\
CellLM~\cite{Zhao2023celllm}
& HL
& FT
& BERT
& scRNA-seq
& SFA, DR
& 2023
& \href{https://github.com/PharMolix/OpenBioMed}{\color{magenta}\Checkmark}
\\
GeneBERT~\cite{GeneBERT}
& HL
& FT
& BERT
& DNA
& SA, DR
& 2021 & \href{https://github.com/ZovcIfzm/GeneBERT}{\color{magenta}\Checkmark}
\\
CodonBERT~\cite{Li2023codonbert}
& HL
& FT
& BERT
& RNA
& SFA
& 2023
& \href{https://github.com/Sanofi-Public/CodonBERT}{\color{magenta}\Checkmark}
\\
xTrimoPGLM~\cite{Chen2024xTrimopglm}
& HL
& FT, AT
& GLM~\cite{du-etal-2022-glm}
& Protein
& SFA
& 2023
&
\\
GenePT~\cite{Chen2023genept}
& -
& PE
& GPT
& DNA
& IA, DR
& 2023
& \href{https://github.com/yiqunchen/GenePT}{\color{magenta}\Checkmark}
\\
scELMo~\cite{Liu2023scelmo}
& -
& PE
& GPT
& scRNA-seq
& SFA, DR
& 2023
& \href{https://github.com/HelloWorldLTY/scELMo}{\color{magenta}\Checkmark}
\\
\bottomrule
\end{tabular}
\end{table*}

Foundation models are also rapidly developing in the area of bioinformatics \cite{li2024progress}. As discussed in the recent review \cite{li2024progress}, with the development of high-throughput sequencing \cite{highthroughput}, the existing BFMs have achieved remarkable success on the omics including single-cell RNA sequencing (scRNA-seq) \cite{Yang2022scbert}, DNA \cite{Fishman2023genalm}, RNA \cite{Chen2023splicebert}, and protein data \cite{jumper2021highly}. As shown in Tab.\ref{table:bfm}, the methods in BFM have been greatly inspired by LFM, and most of them are constructed following the basic architectures in LFM, like the BERT and the transformer decoder. They also applied the pre-training and adaptation paradigm which is widely used in VFM and LFM for generalist ability in bioinformatics tasks.

\subsubsection{Pre-training}
Inspired by LLM, most pre-training strategies in BFM are also based on GL and CL paradigms (Tab.\ref{table:bfm}), owing to its great ability to capture the features of context dependence which is essential to understanding biological systems.

\textbf{a)} \textit{GL-based pre-training} paradigms train BFMs to learn the representation of context dependence, enabling the models to discover the potential relationships between omics.

Like the GL methods in LFMs and VFMs, masked omics modeling (MOM) \cite{Yang2022scbert}, and next token prediction (NTP) are the most popular GL pretext tasks in BFM. The MOM randomly masks the expression values or sequences in biological data and trains the models to reconstruct the masked information. For scRNA-seq data, some BFM works \cite{Yang2022scbert,cui2023scgpt,Hao2023scfoundation,Rosen2023uce} utilized MOM to encode the expression values and gene names thus extracting representative information from the high-dimensional and sparse data. Representatively, scBERT \cite{Yang2022scbert} transferred the MLM in BERT from LFM as the MOM for BFM and trained its model on 1.1 million human scRNA-seq data for an effective representation of expression values. Besides, for DNA and RNA data, MOM also learned the dependence between nucleotides inner sequences, thus modeling the relationships of genes \cite{Fishman2023genalm,Zhou2023dnabert2,Yang20233utr,Chen2023splicebert,Zhang2023rnamsm}. For example, based on the BERT \cite{kenton2019bert}, GENA-LM \cite{Fishman2023genalm} and SpliceBERT \cite{Chen2023splicebert} pre-trained the representation of human DNA sequences and RNA sequences respectively via the MOM, achieving powerful transferring capability on their objective downstream tasks. In some protein pre-training works \cite{Elnaggar2022prottrans,Rao2021msa}, MOM also has achieved success via learning to reconstruct the protein sequences or structures. The NTP in BFM learns to predict the next token or sequence based on previous tokens or sequences, achieving great success on sequence data, i.e., DNA, RNA, and protein \cite{ji2021dnabert,Nguyen2023hyenadna,madani2023large,Nijkamp2023progen2}. HyenaDNA \cite{Nguyen2023hyenadna} utilized single nucleotides as tokens and introduced full global context at each layer to predict the next nucleotide. DNABERT \cite{ji2021dnabert} trained four models using 3-mer to 6-mer tokens on up to 2.75 billion nucleotide bases. However, NTP has still not been studied in seRNA-seq data because of its requirement for the sequence properties of data.

Beyond the MOM and NTP that are originally designed in LLM or VLM, some works \cite{Chu2023utrlm,Chowdhury2022aminobert,jumper2021highly} constructed new GL-based pre-training task based on the properties in biological data. UTR-LM \cite{Chu2023utrlm} proposed a secondary structure and minimum free energy prediction pretext task, achieving an efficient RNA data pre-training. Alphafold \cite{jumper2021highly} also combined a self-distillation training with a masked multiple sequence alignment learning, leveraging unlabelled protein sequence, achieving downstream highly accurate protein structure prediction.

\textbf{b)} \textit{Other pre-training} paradigms, including CL and HL, also have been studied to capture the biology information during pre-training steps \cite{Zhao2023celllm,GeneBERT,Li2023codonbert,Chen2024xTrimopglm, Akiyama2022rnabert}. Inspired by GPTs \cite{10113601}, scGPT \cite{cui2023scgpt} combined the MOM with NTP for a single-cell multi-omics foundation model. RNABERT \cite{Akiyama2022rnabert} designed a structural alignment learning to learn the relationship between two RNA sequences for a closer embedding of bases in the same column. DNAGPT \cite{Zhang2023dnagpt} utilized three pre-training tasks, including NTP, guanine-cytosine content prediction, and sequence order prediction, to pre-train the representation of DNA sequences. Following the basic learning paradigm in BERT, GeneBERT \cite{GeneBERT} combined the two tasks (MOM and NSP) together for the construction of a DNA foundation model. CellLM \cite{Zhao2023celllm} combines MOM together with cell type discrimination and CL tasks to create the pre-training tasks. CodonBERT \cite{Li2023codonbert} constructed a homologous sequence prediction method that directly models the sequence representation and understands evolutionary relationships between mRNA sequences to facilitate the pre-training. xTrimoPGLM \cite{Chen2024xTrimopglm} trained a model via MOM and general language model (GLM) \cite{du-etal-2022-glm} pre-training tasks with over 100 billion parameters on 1 trillion tokens, becoming the current largest protein foundation model.

\subsubsection{Adaptation}
As shown in Tab.\ref{table:bfm}, BFMs also adapt their pre-trained models on downstream tasks, like the function analysis, sequence analysis, etc., for their specific bioinformatics applications.

\textbf{a)} \textit{FT-based adaptation} is the most widely used paradigm in BFMs. Like the LFMs and VFMs, the FT methods used in BFMs tune the parameters inner pre-trained models to various downstream tasks with specific targets. Those works \cite{Yang2022scbert, Theodoris2023geneformer, Zhao2023celllm, cui2023scgpt,GeneBERT,Zhang2023dnagpt,dalla2023nucleotide,Fishman2023genalm,chen2022interpretable,Akiyama2022rnabert,Zhang2023rnamsm,Chen2023splicebert,Li2023codonbert,Chu2023utrlm,Yang20233utr,jumper2021highly,Lin2023esm2,Rives2021esm1b,madani2023large,Nijkamp2023progen2,Elnaggar2022prottrans,Chowdhury2022aminobert,Rao2021msa} directly adjusted the full parameters of the network to specific downstream tasks, evaluating the generalizability of the pre-trained model and their application potential in bioinformatics. For example, scBERT \cite{Yang2022scbert} fine-tuned their pre-trained model to 9 cell type annotation tasks of unseen and user-specific scRNA-seq data, surpassing the existing advanced methods on diverse benchmarks. The parameter-efficient FT also has been studied in BFMs that try to design larger models and utilize the LoRA to adjust a part of parameters \cite{Chen2024xTrimopglm,ji2021dnabert}, thus achieving efficient adaptation. DNABERT-2 \cite{ji2021dnabert} is one such model, which introduced the LoRA and significantly reduced the computation and memory costs with ignorable performance sacrifice compared to full parameter FT.

\textbf{b)} \textit{Other adaptation} paradigms including the AT and PE also have been utilized in BFMs. Since those models have learned large-scale information during pre-training, AT-based methods will efficiently reduce computing costs \cite{Rosen2023uce,jumper2021highly,Hao2023scfoundation} just via adding and training few layers in specific positions during adaptation. One of the representative works are xTrimoPGLM \cite{jumper2021highly} and UCE \cite{Rosen2023uce} that added and trained additional MLP layers after the pre-trained backbone, thus adapting the model to downstream tasks. It provided a novel way to encode biological data, thus with the embedding from the pre-trained model, only training a simple classifier on the embedding will achieve a good performance on various tasks. PE adaptation paradigm is still new in BFM and only few works \cite{Chen2023genept,Hao2023scfoundation,Nguyen2023hyenadna, Liu2023scelmo} tried to utilize PE-based methods. For example, GenePT \cite{Chen2023genept} explored a simple method by leveraging ChatGPT embeddings of genes based on literature, and utilized a zero-shot approach to capture underlying gene functionality.

\subsection{Multimodal Foundation Models for Healthcare} \label{subsec:method:multimodal}
Healthcare data is inherently multimodal (Fig.\ref{fig:overview}), thus it is promising to integrate the multiple modalities in language, vision, bioinformatics, etc., and construct a multimodal foundation model (MFM) in healthcare practices. Unlike unimodal models, MFMs are better equipped to understand the characteristics within each modality and the interconnections among them,  enhancing the capacity of FMs to process complex scenarios in healthcare.  Due to the diversity in modalities and their combinations, the pre-training and adaptation in MFMs have their unique designs.

\begin{table*}[h!]
\label{tab:medicalVLP}
\centering
\caption{The methods of MFM in healthcare. The abbreviations here are GL: generative learning, CL: contrastive learning, HL: hybrid learning; FT: fine-tuning, AT: adapter tuning, PE: prompt engineering; CLS: classification, DET: detection, SEG: segmentation, RG: reports generation, VQA: visual question answering, CMR: cross-modal retrieval,  CMG: cross-modal generation, PG: phrase-grounding, NLI: Natural language inference, PPP: protein property prediction, TS: text summarization, GVC: genomic variant calling, GMG: gaze map generation, MPP: molecular property prediction}
\resizebox{\linewidth}{!}
{
\begin{tabular}{lcccccccc}
\toprule
\textbf{Methods}
& \textbf{Pre-training}
& \textbf{Adaptation}
& \textbf{Backbone}
& \textbf{Modalities}
& \textbf{Downstream}
& \textbf{Year}
& \textbf{Code}
\\
\midrule
MMBERT \cite{khare2021mmbert}
& GL
& FT
& ResNet+BERT
& Multimodal images, Text
& VQA
&2021
&\href{https://github.com/VirajBagal/MMBERT}{\color{magenta}\Checkmark}
\\
MRM \cite{zhou2023advancing}
& GL
& FT
& ViT+Transformer
& X-ray images, Text
& CLS, SEG
& 2023
& \href{https://github.com/RL4M/MRM-pytorch}{\color{magenta}\Checkmark}
\\
BiomedGPT \cite{zhang2023biomedgpt}
& GL
& FT, PE
& Transformer
& Multimodal images, Text
& VQA, CMG, CLS, NLI, TS
& 2023
& \href{https://github.com/taokz/BiomedGPT}{\color{magenta}\Checkmark}
\\
RadFM \cite{wu2023towards}
& GL
& FT, PE
& ViT+Transformer
& Multimodal images, Text
& VQA, RG
& 2023
&\href{https://github.com/chaoyi-wu/RadFM}{\color{magenta}\Checkmark}
\\
ConVIRT \cite{zhang2022contrastive}
& CL
& FT
& ResNet+BERT
& X-ray/Musculoskeletal images, Text
& CLS, CMR
& 2022
& \href{https://github.com/edreisMD/ConVIRT-pytorch}{\color{magenta}\Checkmark}
\\
LoVT \cite{muller2022joint}
& CL
& FT
& ResNet+BERT
& X-ray images, Text
& DET, SEG
& 2022
&
\\
UniBrain \cite{lei2023unibrain}
& CL
& FT
& ResNet+BERT
& MRI images, Text
& CLS
& 2023
& \href{https://github.com/ljy19970415/UniBrain}{\color{magenta}\Checkmark}
\\
M-FLAG \cite{liu2023m}
& CL
& FT
& ResNet+BERT
& X-ray images, Text
& CLS, DET, SEG
& 2023
&\href{https://github.com/cheliu-computation/M-FLAG-MICCAI2023}{\color{magenta}\Checkmark}
\\
MGCA \cite{wang2022multi}
& CL
& FT
& ResNet/ViT+BERT
& X-ray images, Text
& CLS, DET, SEG
& 2022
& \href{https://github.com/HKU-MedAI/MGCA}{\color{magenta}\Checkmark}
\\
MedKLIP \cite{wu2023medklip}
& CL
& FT
& ResNet/ViT+BERT
& X-ray images, Text
& CLS, SEG, PG
& 2023
& \href{https://github.com/MediaBrain-SJTU/MedKLIP}{\color{magenta}\Checkmark}
\\
ETP \cite{liu2023etp}
& CL
& FT, PE
& ResNet+BERT
& ECG signals, Text
& CLS
& 2024
& 
\\
GLoRIA \cite{huang2021gloria}
& CL
& FT, PE
& ResNet+BERT
& X-ray images, Text
& CLS, SEG, CMR
& 2021
& \href{https://github.com/marshuang80/gloria}{\color{magenta}\Checkmark}
\\
IMITATE \cite{liu2023imitate}
& CL
& FT, PE
& ResNet+BERT
& X-ray images, Text
& CLS, SEG, DET
& 2023
&
\\
MedCLIP \cite{wang2022medclip}
& CL
& FT, PE
& ResNet/ViT+BERT
& X-ray images, Text
& CLS, CMR
& 2022
& \href{https://github.com/RyanWangZf/MedCLIP}{\color{magenta}\Checkmark}
\\
Med-UniC \cite{wan2024med}
& CL
& FT, PE
& ResNet/ViT+BERT
& X-ray images, Text
& CLS, SEG, DET
& 2024
&\href{https://github.com/SUSTechBruce/Med-UniC}{\color{magenta}\Checkmark}
\\
CXR-CLIP \cite{you2023cxr}
& CL
& FT, PE
& ResNet/Swin+BERT
& X-ray images, Text
& CLS, CMR
& 2023
& \href{https://github.com/kakaobrain/cxr-clip}{\color{magenta}\Checkmark}
\\
BiomedCLIP \cite{zhang2023large}
& CL
& FT, PE
& ViT+BERT
& Multimodal images, Text
& CMR, CLS, VQA
& 2023
&\href{https://github.com/LightersWang/BiomedCLIP-LoRA}{\color{magenta}\Checkmark}
\\
UMCL \cite{wang2023unified}
& CL
& FT, PE
& Swin+BERT
& X-ray images, Text
& CLS, CMR
& 2023
&
\\
KAD \cite{zhang2023knowledge}
& CL
& FT, PE
& ResNet+BERT
& X-ray images, Text
& CLS
& 2023
&
\\
MoleculeSTM \cite{liu2023multi}
& CL
& FT, PE
& MegaMolBART/GIN+BERT
& Molecule, Text
& CMR, CMG, MPP
& 2023
&\href{https://github.com/chao1224/MoleculeSTM/tree/main}{\color{magenta}\Checkmark}
\\
CLIP-Lung \cite{lei2023clip}
& CL
& PE
& ResNet+Transformer
& CT images, Text
& CLS
& 2023
& 
\\

BFSPR \cite{seibold2022breaking}
& CL
& PE
& ResNet+Transformer
& X-ray images, Text
& CLS
& 2022
& 
\\
MI-Zero \cite{lu2023visual}
& CL
& PE
& CTransPath+BERT
& Pathology images, Text
& CLS
& 2023
&\href{https://github.com/mahmoodlab/MI-Zero}{\color{magenta}\Checkmark}
\\
Clinical-BERT \cite{yan2022clinical}
& HL
& FT
& DenseNet+BERT
& X-ray images, Text
& RG, CLS
& 2022
&
\\
M$^{3}$AE \cite{chen2022multi}
& HL
& FT
&ViT+Transformer
& Multimodal images, Text
& VQA, CLS, CMR
& 2022
& \href{https://github.com/zhjohnchan/M3AE}{\color{magenta}\Checkmark}
\\
MedViLL \cite{moon2022multi}
& HL
& FT
& ResNet+BERT
& Multimodal images, Text
& CLS, CMR, VQA, RG
& 2022
& \href{https://github.com/SuperSupermoon/MedViLL}{\color{magenta}\Checkmark}
\\
PMC-CLIP \cite{lin2023pmc}
& HL
& FT
& ResNet+BERT
& Multimodal images, Text
& VQA, CLS, CMR
& 2023
&\href{https://github.com/WeixiongLin/PMC-CLIP}{\color{magenta}\Checkmark}
\\
ARL \cite{chen2022align}
& HL
& FT
& ViT+BERT
& Multimodal images, Text
& VQA, CLS, CMR
& 2022
&\href{https://github.com/zhjohnchan/ARL}{\color{magenta}\Checkmark}
\\
MaCo\cite{huang2023enhancing}
& HL
& FT
& ViT+BERT
& X-ray images, Text
& CLS, SEG, PG
& 2023
& \href{https://github.com/SZUHvern/MaCo}{\color{magenta}\Checkmark}
\\
MUMC \cite{li2023masked}
& HL
& FT
& ViT+BERT
& Multimodal images, Text
& VQA
& 2023
& \href{https://github.com/pengfeiliHEU/MUMC}{\color{magenta}\Checkmark}
\\
T3D \cite{liu2023t3d}
& HL
& FT
& Swin+BERT
& CT images, Text
& CLS, SEG
& 2023
&
\\
GIMP \cite{jin2023gene}
& HL
& FT
& ResNet+Transformer
& Pathology images, Genomic
& CLS
& 2023
&\href{https://github.com/DeepMed-Lab-ECNU/GIMP}{\color{magenta}\Checkmark}
\\

BioViL \cite{boecking2022making}
& HL
& FT, PE
& ResNet+BERT
& X-ray images, Text
& NLI, CLS, SEG, PG
& 2022
&
\\
PIROR \cite{cheng2023prior}
& HL
& FT, PE
&ResNet+BERT
& X-ray images, Text
& CLS, SEG, DET, CMR
& 2023
& \href{https://github.com/QtacierP/PRIOR}{\color{magenta}\Checkmark}
\\
CONCH \cite{lu2023towards}
& HL
& FT, PE
& ViT+Transformer
& Pathology images, Text
& CLS, CMR, SEG, CMG
& 2024
&
\\
ProteinDT \cite{liu2023text}
& HL
& PE
& ProtBERT+SciBERT
& Protein sequences, Text
& CMG, PPP
& 2023
&
\\
PubMedCLIP \cite{eslami2021does}
& -
& FT
& CLIP
& Multimodal images, Text
& VQA
& 2023
& \href{https://github.com/sarahESL/PubMedCLIP}{\color{magenta}\Checkmark}
\\
Med-PaLMM \cite{tu2023towards}
& -
& FT
& PaLM-E
& Multimodal images, Text, Genomic
& VQA, RG, CLS, GVC
& 2023
&\href{https://github.com/kyegomez/Med-PaLM}{\color{magenta}\Checkmark}
\\
Med-Flamingo \cite{moor2023med}
& -
& FT
& Flamingo
& Multimodal images, Text
& VQA
& 2023
& \href{https://github.com/snap-stanford/med-flamingo}{\color{magenta}\Checkmark}
\\
LLaVA-Med \cite{li2023llava}
& -
& FT
& LLaVA
& Multimodal images, Text
& VQA
& 2024
& \href{https://github.com/microsoft/LLaVA-Med}{\color{magenta}\Checkmark}
\\
CheXZero \cite{tiu2022expert}
& -
& FT, PE
& CLIP
& X-ray images, Text
& CLS
& 2022
& \href{https://github.com/rajpurkarlab/CheXzero}{\color{magenta}\Checkmark}
\\
QUILTNET \cite{ikezogwo2023quilt}
& -
& FT, PE
& CLIP
& Pathology images, Text
& CLS, CMR
& 2024
& \href{https://github.com/wisdomikezogwo/quilt1m}{\color{magenta}\Checkmark}
\\
PLIP \cite{huang2023visual}
& -
& FT, PE
& CLIP
& Pathology images, Text
& CLS, CMR
& 2023
& \href{https://github.com/PathologyFoundation/plip}{\color{magenta}\Checkmark}
\\
CoOpLVT \cite{baliah2023exploring}
& -
& FT, PE
& CLIP
& Ophthalmology images, Text
& CLS
& 2023
& \href{https://github.com/Sanoojan/CLIP-DRDG}{\color{magenta}\Checkmark}
\\
RoentGen \cite{chambon2022roentgen}
& -
& FT, PE
& Stable diffusion
& X-ray images, Text
& CMR
& 2022
& 
\\
 Van Sonsbeek \textit{et al.} \cite{van2023open}
& -
& FT, AT, PE
& CLIP-ViT+GPT-2/BioMedLM/BioGPT
& X-ray images, Text
& VQA
& 2023
&\href{https://github.com/tjvsonsbeek/open-ended-medical-vqa}{\color{magenta}\Checkmark}
\\
Chambon \textit{et al.} \cite{chambon2022adapting}
& -
& FT, AT, PE
& Stable diffusion
& X-ray images, Text
& CMR
& 2022
& 
\\
Qilin-Med-VL \cite{liu2023qilin}
& -
& FT, AT, PE
& CLIP-ViT+Chinese-LLaMA
& Multimodal images, Text
& VQA
& 2023
& \href{https://github.com/williamliujl/Qilin-Med-VL}{\color{magenta}\Checkmark}
\\
PathAsst \cite{sun2023pathasst}
& -
& FT, AT
& PLIP-ViT+Vicuna
& Pathology images, Text
& CLS, DET, SEG, CMR, CMG
& 2024
&\href{https://github.com/bioinfomagic/Generative-Foundation-AI-Assistant-for-Pathology/tree/main}{\color{magenta}\Checkmark}
\\
PathChat \cite{lu2023foundational}
& -
& FT, AT
& CONCH-ViT+LLaMA-2
& Pathology images, Text
& VQA
& 2023
&
\\

Lu \textit{et al.} \cite{lu2023effectively}
& -
& FT, AT
& ResNet+GPT/OpenLLaMA
& X-ray images, Text
& RG
& 2023
& 
\\
M$^{3}$AD \cite{yu2023multi}
& -
& AT
& M$^{3}$AE
& Multimodal images, Text
& VQA
& 2023
& 
\\
I-AI \cite{pham2023decoding}
& -
& AT
& BiomedCLIP
& X-ray images, Text
& GMG, CLS
& 2024
& \href{https://github.com/UARK-AICV/IAI}{\color{magenta}\Checkmark}
\\
CITE \cite{zhang2023text}
& -
& AT, PE
& CLIP-ViT+BioLinkBERT
& Pathology images, Text
& CLS
& 2023
& \href{https://github.com/openmedlab/CITE}{\color{magenta}\Checkmark}
\\
XrayGPT \cite{thawkar2023xraygpt}
& -
& AT, PE
& MedCLIP+Vicuna
& X-ray images, Text
& VQA
& 2023
& \href{https://github.com/mbzuai-oryx/XrayGPT}{\color{magenta}\Checkmark}
\\
Xplainer \cite{pellegrini2023xplainer}
& -
& PE
& BioViL
& X-ray images, Text
& CLS
& 2023
& \href{https://github.com/ChantalMP/Xplainer}{\color{magenta}\Checkmark}
\\
Qin \textit{et al.} \cite{qin2022medical}
& -
& PE
& GLIP
& Multimodal images, Text
& DET
& 2022
& \href{https://github.com/MembrAI/MIU-VL}{\color{magenta}\Checkmark}
\\
Guo \textit{et al.} \cite{guo2023multiple}
& -
& PE
& GLIP
& Multimodal images, Text
& DET
& 2023
&
\\
\bottomrule
\end{tabular}
}
\end{table*}

\subsubsection{Pre-training}
MFM involves the learning of multiple modalities, thus the typical learning paradigms in LFM, VFM,
and BFM with different modalities are widely applied. However, multimodal pre-training has higher challenges, requiring models to not only understand unimodal data but also to process and integrate information from various modalities.
Due to differences in focus, there are three main paradigms in multimodal pre-training: GL focuses on the generative capabilities of MFMs, often employing decoders to generate data across multiple modalities. CL enhances the cross-modal understanding of MFMs by encoding data from different modalities into the same space and HL combines the advantages of the first two, aiming to comprehensively improve the model's understanding and generative abilities.


\textbf{a)} \textit{GL-based pre-training} paradigm is designed by guiding networks in predicting or reconstructing images, text, or other types of data. Therefore, masked representation modeling are used either individually or in combination across different modalities. Representatively, MMBERT \cite{khare2021mmbert} integrated image features into a BERT architecture, enhancing the comprehension of medical images and text by utilizing MLM with images as the pretext task.  MRM  \cite{zhou2023advancing} further advanced visual representation by combining MLM and MIM.  These generative pre-training provides a direct method for facilitating cross-modal interactions, enabling the reconstruction of one modality based on more generic multimodal representations. With the evolution of MFMs, there's a trend towards developing more generalist AI models that are trained on larger and more diverse multimodal datasets to handle multiple tasks within a singular architecture.
Among these methods, RadFM \cite{wu2023towards} trained a visually conditioned autoregressive language generation model for radiology, addressing a wide range of medical tasks with natural language as output.  BiomedGPT \cite{zhang2023biomedgpt} employed unimodal representation modeling and task-specific multimodal learning to pre-train a unified sequence-to-sequence model.

\textbf{b)} \textit{CL-based pre-training} paradigm in MFM utilizes contrastive loss to learn multimodal data like the CLIP \cite{radford2021learning} for image and text, achieving the alignment between different modalities. Prior to CLIP, ConVIRT \cite{zhang2022contrastive} has pioneered visual-language CL in chest X-ray and musculoskeletal images, while ETP \cite{liu2023etp}, MI-Zero \cite{lu2023visual}, BiomedCLIP \cite{zhang2023large}  and  MoleculeSTM \cite{liu2023multi} further extended this strategy into Electrocardiogram (ECG) signals, pathology images, biomedical images and molecular structure information respectively, showcasing the effectiveness of CL paradigm in medical domain. Furthermore, several subsequent studies attempted to extend vision-language alignment pre-training by improving training strategies. Considering that crucial semantic information may be concentrated in specific regions of medical data,  GLoRIA \cite{huang2021gloria}, LoVT \cite{muller2022joint}, MGCA \cite{wang2022multi},  and IMITATE \cite{liu2023imitate} have focused on exploring fine-grained semantic alignment between distinct image sub-regions and text token embeddings, showing the effectiveness of fine-grained alignment in capturing nuanced semantic information. Focusing on enhancing pre-training data efficiency,  MedCLIP \cite{wang2022medclip} extended the pretraining to include large unpaired images and texts, scaling the number of training data in a combinatorial manner. CXR-CLIP \cite{you2023cxr} and UCML \cite{wang2023unified} employed prompt templates to generate image-text pairs from image-label datasets.
Given the specialized nature of medical language, MedKLIP \cite{wu2023medklip}, KAD \cite{zhang2023knowledge}, CLIP-Lung \cite{lei2023clip} and UniBrain \cite{lei2023unibrain} leveraged domain-specific knowledge from medical datasets. BioBRIDGE \cite{wang2023biobridge} utilized knowledge graphs to learn transformations between one unimodal FM and another without fine-tuning any underlying unimodal FMs. Overall, CL methods empower the model to better comprehend the intricate relationships without the necessity for task-specific fine-tuning. 

\textbf{c)} \textit{HL-based pre-training} paradigm also has been constructed to fuse the advantages in different learning paradigms and stimulate the learned capability. Clinical-BERT \cite{yan2022clinical}, Li \textit{et al.}\cite{li2020comparison}, M$^{3}$AE \cite{chen2022multi}, MedViLL \cite{moon2022multi}, and ARL \cite{chen2022align} leveraged a combination of the masked representation modeling and image-text matching learning. PMC-CLIP \cite{lin2023pmc}, PIROR \cite{cheng2023prior}, MaCo \cite{huang2023enhancing} and BioViL \cite{boecking2022making} utilized the masked representation modeling and contrastive learning.  MUMC \cite{li2023masked}  simultaneously incorporated these three learning paradigms. Specifically, T3D \cite{liu2023t3d} was pre-trained on two text-driven pretext tasks: Text-informed Image Restoration and Text-informed Contrastive Learning. CONCH \cite{lu2023towards} utilized an equal-weighted combination of the image-text contrastive loss and the captioning loss following \cite{yu2022coca}. GIMP\cite{jin2023gene} designed a masked patch modeling paradigm and gene-induced triplet learning. ProteinDT \cite{liu2023text} combined the contrastive learning and autogressive and diffusion generative paradigm. These methods involve incorporating both inter- and intra-modality generative or constrative tasks to mutually enhance each other's effectiveness.

\subsubsection{Adaptation}
The adaptation methods in MFMs also enrolled the FT, AT, and PE methods for their widely applications in healthcare practices.

\textbf{a)} \textit{FT-based adaptation} methods in MFM are employed using domain-specific or task-specific data to tune the parameters of the pre-trained models. Several methods attempted to adapt general-domain MFMs to the healthcare domain. For instance, CheXZero \cite{tiu2022expert}, PubMedCLIP \cite{eslami2021does}, QUILTNET \cite{ikezogwo2023quilt} and PLIP \cite{huang2023visual} are fine-tuned versions of CLIP \cite{radford2021learning} tailored for chest x-ray, radiology and histopathology. RoentGen \cite{chambon2022roentgen} and \cite{chambon2022adapting} are medical domain-adapted latent diffusion models based on Stable Diffusion pipeline \cite{rombach2022high}.  LLaVA-Med \cite{li2023llava}, Med-Flamingo \cite{moor2023med} and Med-PaLMM \cite{tu2023towards} followed LLaVA \cite{liu2023visual}, Flamingo \cite{alayrac2022flamingo}, and PaLM-E \cite{driess2023palm} with paired or interleaved medical image-text data. Specifically, LLaVA-Med \cite{li2023llava} introduced a two-stage curriculum learning strategy where the model first learns to align biomedical vocabulary using the image-caption pairs and then learns open-ended conversational semantics using instruction-following data.  It has inspired the development of generalist biomedical AI models like Med-Flamingo \cite{moor2023med}, Med-PaLMM \cite{tu2023towards} and RadFM \cite{wu2023towards}.  It has also sparked interest in exploring visual condition language models, which involve fine-tuning an LLM with a visual encoder to achieve a unified biomedical AI model. For instance, PathAsst \cite{sun2023pathasst}, PathChat \cite{lu2023foundational}, Qilin-Med-VL \cite{liu2023qilin} and XrayGPT \cite{thawkar2023xraygpt} combined a strong vision encoder backbone with an open-source large language model, achieving a vision language interactive AI assistant.  Besides, \cite{liu2023utilizing} and \cite{kumar2022towards} also suggested that medical VLP, such as ConVIRT \cite{zhang2022contrastive}, GLORIA \cite{huang2021gloria}, MGCA \cite{wang2022multi} and BioViL \cite{boecking2022making}, could be further fine-tuned with higher-quality medical data.  However, foundation models usually have a large number of parameters, and fine-tuning the full model weights results in long training times, risk of overfitting, and potential domain bias. Thus, parameter-efficient fine-tuning is proposed to construct MFMs.  Van Sonsbeek \textit{et al.} \cite{van2023open} explored LoRA and prefix tuning for the language backbone of the vision language interactive model, allowing for resource- and data-efficient fine-tuning. Lu \textit{et al.} \cite{lu2023effectively} also leveraged LoRA to adapt LLM to the task of radiology report generation.

\textbf{b)} \textit{AT-based adaptation} methods often involve integrating adapters into pre-trained FMs and fine-tuning these adapters to adapt to specific domains or tasks. Adapters in multimodal models can play a unique role that serves as a bridge to convert the features in a modality to another modality, thus integrating the multimodal data with low costs.
Several approaches \cite{van2023open, lu2023effectively, liu2023qilin, sun2023pathasst, lu2023foundational, thawkar2023xraygpt} utilized simple projection layers to convert medical visual features into text embeddings, likely serving as a visual-based soft prompt for the text encoder. Specially, M$^{3}$AD \cite{yu2023multi} incorporated general adapters into two unimodal encoding networks and also employed a modality-fusion adapter to enhance multimodal interactions. In general, the adapters in MFMs convert different modalities being lightweight and economical to facilitate seamless integration.

\textbf{c)} \textit{PE-based adaptation} methods in MFMs is also flourishing. Manually crafted prompts \cite{liu2023etp,huang2021gloria,liu2023imitate,wang2022medclip,you2023cxr,seibold2022breaking,tiu2022expert,thawkar2023xraygpt} are designed to guide a pre-trained model to align with downstream tasks. Especially, regarding the design principles of prompts, BFSPR \cite{seibold2022breaking} discovered that a more detailed prompt design can enhance performance, and thus it used various combinations from a small category set to explore diverse settings. Qin \textit{et al.} \cite{qin2022medical} also showed that using essential attributes, such as color, shape, and location, can enhance domain transfer capability compared to the default category names.  Besides,  Guo \textit{et al.} \cite{guo2023multiple} utilized multiple prompts fusion to comprehensively describe information about recognized objects. Xplainer \cite{pellegrini2023xplainer} presented prompts that are initially generated using ChatGPT and then refined with a seasoned radiologist for better performance. Additionally, learnable prompting methods \cite{wang2023unified, van2023open,lei2023clip,baliah2023exploring,zhang2023text} are also introduced in the medical MFMs field, which utilizes prompt tuning to adapt pre-trained models to different downstream tasks, reducing the number of trainable parameters while improving the performance on unknown tasks. CoOpLVT \cite{baliah2023exploring} and CLIP-Lung \cite{lei2023clip} leveraged image-conditioned prompt tuning to enhance the accuracy of alignment.  CITE \cite{zhang2023text} added tuning prompt tokens to the visual inputs for more effective pathological image classification.

\subsection{Analysis of the paradigms in HFM}
\begin{figure}
    \centering
    \includegraphics[width=\linewidth]{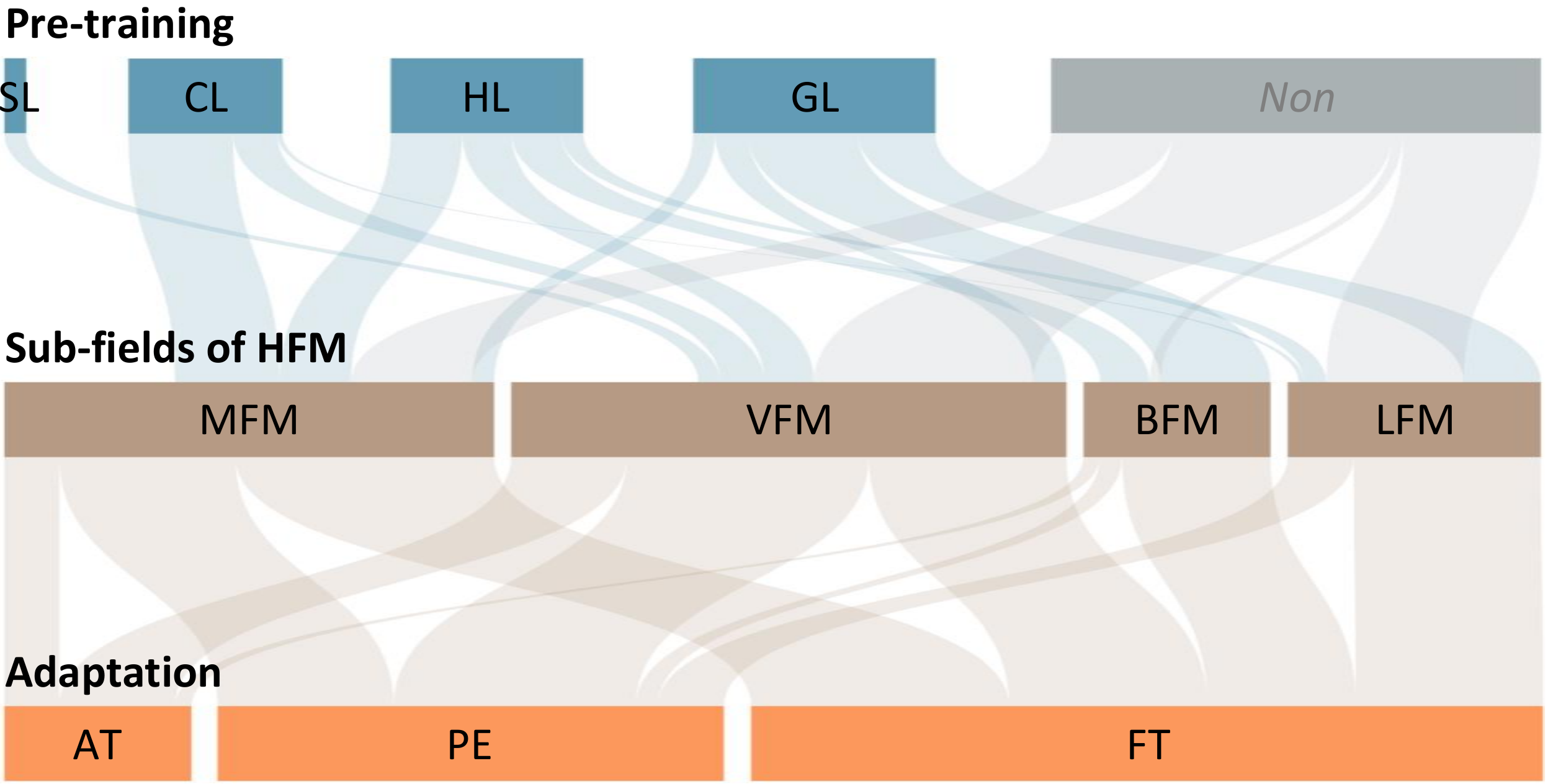}
    \caption{The Sankey diagram of healthcare foundation models demonstrates the associations between the pre-training paradigms, sub-fields of HFM, and adaptation paradigms. ``Non" means that the work directly adapted existing pre-trained models to their tasks and did not pre-train their model by themselves.}
    \label{fig:Sankey}
\end{figure}
As shown in Fig.\ref{fig:Sankey}, the Sankey diagram visualizes the paper amounts flow from the pre-training paradigms to sub-fields and then to the adaptation paradigms, demonstrating their properties and associations. From this diagram, there are five observations for the current progress of HFMs:

1) \textit{Foundation models from general fields are able to be adapt to healthcare fields.} More than 1/3 works directly adapted existing pre-trained models to their tasks in language, and multimodal fields, excepting the bioinformatics. The vision and language in healthcare and general fields are relatively unified compared with the bioinformatics, thus some successful pre-trained models in the general field also will be generalized to healthcare tasks. Bioinformatics lacks general pre-trained models for their very specific omics data, so that only two works \cite{Chen2023genept,Liu2023scelmo} studied the direct adaptation of the embedding from LFMs.

2) \textit{Most LFMs directly adapted existing pre-trained language models to their healthcare tasks.} Because language, as the data created by human beings, has strong portability. The LFM pre-trained in general field has this portability and are able to adapt to the healthcare field.

3) \textit{Most pre-training works focused on learning without annotations.} Owing to the large amount of pre-training data, supervised learning is expensive in human annotating for these data, so most works focused on self-supervised paradigms, i.e., GL, CL, or HL, to learn general representation capability with low annotation costs.

4) \textit{Supervised learning is still used in VFM pre-training.} Because the continuity of vision information makes it challenging to separate the semantics of the content \cite{MoCo} only via a self-supervised paradigm. Therefore, there are still some works tried to utilize supervised learning for a more direct optimization target, thus driving the model to decouple the semantics inner the content and learn meaningful features.

5) \textit{Fine-tuning is still widely used in adaptation.} More than 1/2 of works utilized the fine-tuning paradigm, because fine-tuning can bring a stable learning process. Some new fine-tuning techniques, e.g., LoRA, are being explored to achieve adaptation with low parameter efficiency. 

\begin{table*}[!tb]
\centering
\caption{The representative language datasets, three of them are unavailable currently. Here we use the number of language tokens or the number of data instances to express the scale of a specific dataset. The abbreviations here are LM: language modeling, DIAL: dialogue, IR: information retrieval, NRE: named entity recognition, RE: relation extraction, STS: semantic textual similarity, NLI: natural language inference, QA: question answering, and VQA: visual question answering.}
\begin{tabular}{llllccccc}
\toprule
\textbf{Dataset}
& \textbf{Text Types}
& \textbf{Scale}
& \textbf{Tasks}
& \textbf{Link}
\\
\midrule
PubMed
& Literature
& 18B tokens
& LM
& \href{https://pubmed.ncbi.nlm.nih.gov/download/}{\color{magenta}\Checkmark}
\\
MedC-I \cite{wu2023pmc}
& Literature
& 79.2B tokens
& DIAL
& \href{https://huggingface.co/datasets/axiong/pmc_llama_instructions}{\color{magenta}\Checkmark}
\\
Guidelines~\cite{chen2023meditron}
& Literature
& 47K instances
& LM
& \href{https://huggingface.co/datasets/epfl-llm/guidelines}{\color{magenta}\Checkmark}
\\
PMC-Patients~\cite{Zhao2023ALD}
& Literature
& 167K instances
& IR
& \href{https://github.com/pmc-patients/pmc-patients}{\color{magenta}\Checkmark}
\\
MIMIC-III \cite{johnson2016mimic}
& Health record 
& 122K instances
& LM
& \href{https://physionet.org/content/mimiciii/1.4/}{\color{magenta}\Checkmark}
\\
MIMIC-IV \cite{johnson2023mimic}
& Health record 
& 299K instances
& LM
& \href{https://physionet.org/content/mimiciv/2.2/}{\color{magenta}\Checkmark}
\\
eICU-CRDv2.0 \cite{pollard2018eicu}
& Health record 
& 200K instances
& LM
& \href{https://physionet.org/content/eicu-crd/2.0/}{\color{magenta}\Checkmark}
\\
EHRs \cite{yang2022large}
& Health record
& 82B tokens
& NER, RE, STS, NLI, DIAL
&
\\
MD-HER \cite{wang2023clinicalgpt}
& Health record
& 96K instances
& DIAL, QA
&
\\
IMCS-21 \cite{chen2023benchmark}
& Dialogue
& 4K instances
& DIAL
& \href{https://github.com/lemuria-wchen/imcs21}{\color{magenta}\Checkmark}
\\
Huatuo-26M \cite{li2023huatuo}
& Dialogue
& 26M instances
& QA
& \href{https://github.com/FreedomIntelligence/Huatuo-26M}{\color{magenta}\Checkmark}
\\
MedInstruct-52k \cite{zhang2023alpacare}
& Dialogue
& 52K instances
& DIAL
& \href{https://github.com/XZhang97666/AlpaCare}{\color{magenta}\Checkmark}
\\
MASH-QA \cite{zhu-etal-2020-question}
& Dialogue
& 35K instances
& QA
& \href{https://github.com/mingzhu0527/MASHQA}{\color{magenta}\Checkmark}
\\
MedQuAD \cite{ben2019question}
& Dialogue
& 47K instances
& QA
& \href{https://github.com/abachaa/MedQuAD}{\color{magenta}\Checkmark}
\\
MedDG \cite{liu2020meddg}
& Dialogue
& 17K instances
& DIAL
& \href{https://github.com/lwgkzl/MedDG}{\color{magenta}\Checkmark}
\\
CMExam \cite{liu2024benchmarking}
& Dialogue
& 68K instances
& DIAL, QA
& \href{https://github.com/williamliujl/CMExam}{\color{magenta}\Checkmark}
\\
cMedQA2 \cite{zhang2018multi}
& Dialogue
& 108K instances
& QA
& \href{https://github.com/zhangsheng93/cMedQA2}{\color{magenta}\Checkmark}
\\
CMtMedQA \cite{yang2024zhongjing}
& Dialogue
& 70K instances
& DIAL, QA
& \href{https://github.com/SupritYoung/Zhongjing}{\color{magenta}\Checkmark}
\\
CliCR \cite{suster-daelemans-2018-clicr}
& Dialogue
& 100K instances
& QA
& \href{https://github.com/clips/clicr}{\color{magenta}\Checkmark}
\\
webMedQA \cite{he2019applying}
& Dialogue
& 63K instances
& QA
& \href{https://github.com/hejunqing/webMedQA}{\color{magenta}\Checkmark}
\\
ChiMed \cite{ye2023qilin}
& Dialogue
& 1.59B tokens
& QA
& \href{https://huggingface.co/datasets/williamliu/ChiMed/tree/main}{\color{magenta}\Checkmark}
\\
MedDialog \cite{zeng2020meddialog}
& Dialogue
& 20K instances
& DIAL
& \href{https://github.com/UCSD-AI4H/Medical-Dialogue-System}{\color{magenta}\Checkmark}
\\
CMD
& Dialogue
& 882K instances
& LM
& \href{https://github.com/Toyhom/Chinese-medical-dialogue-data}{\color{magenta}\Checkmark}
\\
BianqueCorpus \cite{chen2023bianque}
& Dialogue
& 2.4M instances
& DIAL
& \href{https://github.com/scutcyr/BianQue}{\color{magenta}\Checkmark}
\\
MedQA \cite{jin2021disease}
& Dialogue
& 4K instances
& QA
& \href{https://github.com/jind11/MedQA}{\color{magenta}\Checkmark}
\\
HealthcareMagic 
& Dialogue
& 100K instancess
& DIAL
& \href{https://huggingface.co/datasets/RafaelMPereira/HealthCareMagic-100k-Chat-Format-en}{\color{magenta}\Checkmark}
\\
iCliniq
& Dialogue
& 10K instances
& DIAL
& \href{https://drive.google.com/file/d/1ZKbqgYqWc7DJHs3N9TQYQVPdDQmZaClA/view}{\color{magenta}\Checkmark}
\\
CMeKG-8K \cite{byambasuren2019preliminary}
& Dialogue
& 8K instances
& DIAL
& \href{https://github.com/WENGSYX/CMKG}{\color{magenta}\Checkmark}
\\
Hybrid SFT \cite{zhang2023huatuogpt}
& Dialogue
& 226K instances
& DIAL, QA
& \href{https://github.com/FreedomIntelligence/HuatuoGPT}{\color{magenta}\Checkmark}
\\
VariousMedQA \cite{shu2023visual} 
& Dialogue
& 54K instances
& VQA
& \href{https://github.com/cambridgeltl/visual-med-alpaca}{\color{magenta}\Checkmark}
\\
Medical Meadow \cite{han2023medalpaca}
& Dialogue
& 160K instances
& QA
& \href{https://github.com/kbressem/medAlpaca}{\color{magenta}\Checkmark}
\\
MultiMedQA \cite{singhal2023towards}
& Dialogue
& 193K instances
& QA
&
\\
BiMed1.3M \cite{pieri2024bimedix}
& Dialogue
& 250K instances
& QA
& \href{https://github.com/mbzuai-oryx/BiMediX}{\color{magenta}\Checkmark}
\\
OncoGPT \cite{jia2024oncogpt}
& Dialogue
& 180K instances
& QA
& \href{https://github.com/OncoGPT1}{\color{magenta}\Checkmark}
\\
\bottomrule
\end{tabular}
\label{tab:lfm-datasets}
\end{table*}

\section{Datasets}
\label{sec:datasets}
\subsection{Language}
The advancement of medical LFMs hinges on the diverse healthcare text datasets. Although there has been accumulated rich information inner these language data, it is still challenging in scale and specificity. A lot of LFM works combined various datasets to create a comprehensive training corpus whose detailed components are available in their articles \cite{ye2023qilin,zhang2023huatuogpt,chen2023huatuogpt,chen2023meditron,li2023beginner,wang2023gpt,wang2023clinicalgpt}. As shown in Tab.\ref{tab:lfm-datasets}, we describe the large-scale healthcare language datasets in literature, health records, and dialogues, which are crucial for the LFMs in understanding and processing of medical terms. As shown in Tab.\ref{tab:lfm-datasets}, we review relatively large datasets that are over 4K instances or 1B tokens for LFMs.

\subsubsection{Healthcare literature}
Due to the limited privacy information inner literature text and the condensation of medical knowledge, large healthcare literature datasets have been made publicly available. They are typically expansive, serving the crucial function of infusing a general domain language foundation model with a wealth of medical knowledge. PubMed \footnote{\url{https://pubmed.ncbi.nlm.nih.gov/download/}} is a large-scale database including primarily the MEDLINE database of references and abstracts on life sciences and biomedical topics. It offers a comprehensive repository of healthcare literature for the development of healthcare LFMs. MedC-I~\cite{wu2023pmc} collected more than 79B tokens from papers, books, conversations, Rationale QA, and knowledge graph. Guidelines~\cite{chen2023meditron} is composed of 47K clinical practice guidelines from 17 high-quality online medical sources. PMC-Patients~\cite{Zhao2023ALD} consists of 167K patient summaries extracted from case reports in PubMed Central.

\subsubsection{Electronic health records}
Electronic health records contain a lot of descriptions and diagnosis of diseases with a significant clinical value. These data will enable the LFMs to learn about clinical scenarios and patient outcomes, thereby enhancing the model's capability in challenging healthcare practice. Due to the large amount of privacy information included in health record data, their datasets are often much smaller than healthcare literature datasets. MIMIC-III \cite{johnson2016mimic} contains more than 122K instances of health records from forty thousand patients who stayed in critical care units, providing detailed clinical knowledge in the critical care practice, while the updated version MIMIC-IV~\cite{johnson2023mimic} contains 299K clinical records. EHRs \cite{yang2022large} encompasses more than 290M clinical notes from UF Health IDR. MD-HER \cite{wang2023clinicalgpt} contains 100k records covering a range of disease groups, and eICU-CRDv2.0~\cite{pollard2018eicu} consists of 200,859 stays at ICUs and step-down units across 208 American hospitals. However, the electronic health records datasets are still rare, and the EHRs \cite{yang2022large} and MD-HER \cite{wang2023clinicalgpt} are unavailable.

\subsubsection{Healthcare dialogue}
Healthcare dialogue data records the interactions in the healthcare setting between doctors and patients or among the doctors. These conversations are invaluable for refining communication skills and improving information retrieval processes for LFMs. There are many publicly available healthcare dialogue datasets \cite{ye2023qilin,zeng2020meddialog,chen2023bianque,jin2021disease,byambasuren2019preliminary,zhang2023huatuogpt,ben2019question,han2023medalpaca,pieri2024bimedix,jia2024oncogpt,liu2024benchmarking,li2023huatuo,10.1145/3308558.3313699,zhu-etal-2020-question,ben2019question,suster-daelemans-2018-clicr,he2019applying,yang2024zhongjing,zhang2023alpacare} and some technologies can also transform other healthcare text data or classification labels into dialogue \cite{hu2024omnimedvqa}. BianqueCorpus~\cite{chen2023bianque}, built upon several existing datasets including MedDialog~\cite{zeng2020meddialog}, IMCS-21 \cite{chen2023benchmark}, CHIP-MDCFNPC \cite{zhang2022cblue}, MedDG~\cite{liu2020meddg}, cMedQA2 \cite{zhang2018multi}, and CMD, results in a substantial Chinese medical dialogue dataset containing 2.4M instances. For English medical dialogue, Medical Meadow \cite{han2023medalpaca} also fused 11 self-created datasets and 7 external datasets, thus a large dataset with 160K instances.

\subsection{Vision}
The success of VFMs relies on large-scale medical image datasets, so many recent approaches mixed multiple publicly available or private datasets to compromise to construct a large dataset. The detailed data lists of these works are available in their articles \cite{1-USFM, 2-MIS-FM, 3-RETFound, 33-SegVol, 22-MedSAM, 26-SAM-Med2D, 30-SAM-Med3D, 71-UniverSeg, MedFMC, SA-Med2D-20M, 2-MIS-FM}. As shown in Tab.\ref{Tab:visiondatasets}, we review publicly available and relatively large datasets that are over 1K 3D medical images, 1K 2D whole slide images (WSIs), and 10k other 2D medical images/videos for VFMs.

\subsubsection{3D medical images}
The 3D medical images, including 3D CT, MRI, PET, etc., can visualize information inner the human body, being widely used in clinical practices. The Medical Segmentation Decathlon (MSD) challenge \cite{antonelli2022medical} has totally opened 1,411 3D CT and 1,222 MRI images to evaluate semantic segmentation algorithms on 10 organs or diseases. The ULS challenge \cite{FLARE22} further opened 38,842 CT volumes to evaluate the universal lesion segmentation which promotes the VFMs for lesion segmentation. Some other CT datasets including the LIDC-IDRI, TotalSegmentator \cite{TotalSegmentator, TotalSegmentatorv2}, FLARE 2022, 2023 \cite{FLARE22}, AbdomenCT-1K \cite{9497733}, CTSpine1K \cite{deng2021ctspine1k}, CTPelvic1K \cite{CTPelvic1K}, also opened more than 1K volumes for segmentation tasks. BraTS \cite{baid2021rsnaasnrmiccai,6975210, Bakas2017, labella2023asnrmiccai} challenges have opened more than 2K brain MRI volumes with multiple sequences for the brain MRI analysis. The ADNI \cite{petersen2010alzheimer} and PPMI \cite{marek2011parkinson} databases held and updated the brain MRI and other clinical data of Alzheimer's disease and Parkinson's disease, contributing to the clinical studies of these diseases. AutoPET challenges \cite{gatidis2022whole, gatidis2023autopet} opened 1,214 PET-CT pairs supporting the cross-modality image studies.

\subsubsection{Whole slide images}
WSIs are the images that visualize tissues at a microscopic level to diagnose cancer or signs of pre-cancer. Different from the other 2D medical images, WSIs have extremely high resolution (e.g., 150,000 x 85,000 pixels) \cite{greenwald2022whole} making it unable to directly analyze them on a global level. There are several WSI datasets on the TCGA \cite{TCGA} program, including the NSCLC, Lung, BRCA, GBM, KIRC, LUAD, LUSC, OV, etc., and spanning 33 cancer types. PAIP \cite{PAIP} and TissueNet \cite{greenwald2022whole} organized open challenges for pathological segmentation and diagnosis of liver cancer and cervical cancer respectively with more than 1K WSIs. Some other datasets \cite{borkowski2019lung, CRC} cropped patches, which have much smaller size, from WSIs and opened for classification.

\subsubsection{Other 2D medical images/videos}
2D medical images or videos are also widely used in medical practices. X-ray imaging is widely used in disease screening and surgical assistance accumulating a lot of data, so there are many large opened X-ray datasets \cite{wang2017chestxray, MURA} with more than 10K images. ISIC challenges \cite{rotemberg2021patient} have opened more than 30K dermoscopy images that promote the dermatosis diagnosis. The AIROG challenge \cite{de2023airogs} has opened more than 100K fundus photographs for glaucoma screening. Retinal OCT-C8 dataset \cite{9740985} fused the data from various sources and opened 24K OCT images for retinal disorders diagnosis. For ultrasound (US) images, the Ultrasound Nerve Segmentation challenge \cite{ultrasound-nerve-segmentation} has 11K images for brachial plexus segmentation. Fetal planes dataset \cite{Fetal} has opened 12,400 US images for maternal-fetal screening. The US images are used also in cardiac disease analysis, the EchoNet-Dynamic \cite{ouyang2020video} constructed a large cardiac US video dataset for cardiac function assessment. Endoscopic video is widely used in gastrointestinal disease detection and surgery, and several datasets \cite{polat2022improving,misawa2021development,smedsrud2021kvasir,ozyoruk2020endoslam, HyperKvasir,nwoye2022data} with a large number of frames have been opened in these scenarios.

\begin{table*}[h!]
\label{Tab:visiondatasets}
\centering
\caption{The publicly available vision datasets. The abbreviations here are CLS: classification, SEG: segmentation, DET: detection, REG: registration, and US: ultrasound. The ``Clinical study" means that this is a comprehensive dataset without clear task guidance.}
\begin{tabular}{llllcccc}
\toprule
\textbf{Dataset}
& \textbf{Modalities}
& \textbf{Scale}
& \textbf{Tasks}
& \textbf{Link}
\\
\midrule
LIMUC \cite{polat2022improving}
&Endoscopy	
&1043 videos (11,276 frames)
&DET
&\href{https://zenodo.org/records/5827695#.Yi8GJ3pByUk}{\color{magenta}\Checkmark}
\\
SUN \cite{misawa2021development}
&Endoscopy	
&1018 videos (158,690 frames)
&DET
&\href{http://amed8k.sundatabase.org/}{\color{magenta}\Checkmark}
\\
Kvasir-Capsule \cite{smedsrud2021kvasir}
&Endoscopy	
&117 videos (4,741,504 frames)	
&DET
&\href{https://datasets.simula.no/kvasir-capsule/}{\color{magenta}\Checkmark}
\\
EndoSLAM \cite{ozyoruk2020endoslam}
&Endoscopy	
&1020 videos (158,690 frames)
&DET, REG
&\href{https://github.com/CapsuleEndoscope/EndoSLAM}{\color{magenta}\Checkmark}
\\
LDPolypVideo \cite{ma2021ldpolypvideo}
&Endoscopy	
&263 videos (895,284 frames)
&DET
&\href{https://github.com/dashishi/LDPolypVideo-Benchmark}{\color{magenta}\Checkmark}
\\
HyperKvasir \cite{HyperKvasir}
&Endoscopy	
&374 videos (1,059,519 frames)
&DET
&\href{https://datasets.simula.no/hyper-kvasir}{\color{magenta}\Checkmark}
\\
CholecT45 \cite{nwoye2022data}
&Endoscopy	
&45 videos (90,489 frames)
& SEG, CLS
&\href{https://github.com/CAMMA-public/cholect45}{\color{magenta}\Checkmark}
\\
DeepLesion \cite{yan2018deeplesion}
& CT slices (2D)
& 32,735 images
&RET, CLS
&\href{nihcc.app.box.com}{\color{magenta}\Checkmark}
\\
LIDC-IDRI \cite{armato2011lung}
&3D CT	
&1,018 volumes	
&SEG
&\href{https://wiki.cancerimagingarchive.net/pages/viewpage.action?pageId=1966254}{\color{magenta}\Checkmark}
\\
TotalSegmentator \cite{TotalSegmentator}
&3D CT	
&1,204 volumes	
&SEG
&\href{https://pubs.rsna.org/doi/10.1148/ryai.230024}{\color{magenta}\Checkmark}
\\
TotalSegmentatorv2 \cite{TotalSegmentatorv2}
&3D CT	
&1,228 volumes	
&SEG
&\href{https://doi.org/10.5281/zenodo.6802613}{\color{magenta}\Checkmark}
\\
AutoPET \cite{gatidis2022whole,gatidis2023autopet}
&3D CT, 3D PET 	
&1,214 PET-CT pairs	
&SEG
&\href{https://wiki.cancerimagingarchive.net/pages/viewpage.action?pageId=93258287}{\color{magenta}\Checkmark}
\\
ULS
&3D CT	
&38,842 volumes	
&SEG
&\href{https://uls23.grand-challenge.org/}{\color{magenta}\Checkmark}
\\
FLARE 2022 \cite{FLARE22}
&3D CT	
&2,300 volumes	
&SEG
&\href{https://flare22.grand-challenge.org/Dataset/}{\color{magenta}\Checkmark}
\\
FLARE 2023
&3D CT	
&4,500 volumes	
&SEG
&\href{https://codalab.lisn.upsaclay.fr/competitions/12239#learn_the_details-dataset}{\color{magenta}\Checkmark}
\\
AbdomenCT-1K \cite{9497733}
&3D CT	
&1,112 volumes	
&SEG
&\href{https://github.com/JunMa11/AbdomenCT-1K}{\color{magenta}\Checkmark}
\\
CTSpine1K \cite{deng2021ctspine1k}
&3D CT	
&1,005 volumes	
&SEG
&\href{https://github.com/MIRACLE-Center/CTSpine1K}{\color{magenta}\Checkmark}
\\
CTPelvic1K \cite{CTPelvic1K}
&3D CT	
&1,184 volumes	
&SEG
&\href{https://zenodo.org/record/4588403#.YEyLq_0zaCo}{\color{magenta}\Checkmark}
\\
MSD \cite{antonelli2022medical}
&3D CT, 3D MRI	
&1,411 CT, 1,222 MRI	
&SEG
&\href{http://medicaldecathlon.com/}{\color{magenta}\Checkmark}
\\
BraTS21 \cite{baid2021rsnaasnrmiccai,6975210,Bakas2017}
&3D MRI	
&2,040 volumes	
&SEG
&\href{https://www.synapse.org/#!Synapse:syn51514105}{\color{magenta}\Checkmark}
\\
BraTS2023-MEN \cite{labella2023asnrmiccai}
&3D MRI	
&1,650 volumes	
&SEG
&\href{https://www.synapse.org/#!Synapse:syn51514106}{\color{magenta}\Checkmark}
\\
ADNI \cite{petersen2010alzheimer}
& 3D MRI
& -
& Clinical study
& \href{https://adni.loni.usc.edu/}{\color{magenta}\Checkmark}
\\
PPMI \cite{marek2011parkinson}
& 3D MRI
& -
& Clinical study
& \href{https://www.ppmi-info.org/}{\color{magenta}\Checkmark}
\\
ATLAS v2.0 \cite{liew2022large}
&3D MRI	
&1,271 volumes	
&SEG
&\href{http://fcon_1000.projects.nitrc.org/indi/retro/atlas.html}{\color{magenta}\Checkmark}
\\
PI-CAI \cite{saha2023artificial}
&3D MRI	
&1,500 volumes	
&SEG
&\href{https://zenodo.org/records/6624726}{\color{magenta}\Checkmark}
\\
MRNet \cite{bien2018deep}
&3D MRI	
&1,370 volumes	
&DET, SEG
&\href{https://stanfordmlgroup.github.io/competitions/mrnet/}{\color{magenta}\Checkmark}
\\
Retinal OCT-C8 \cite{9740985}
&2D OCT	
&24,000 imsges	
&CLS
&\href{https://www.kaggle.com/datasets/obulisainaren/retinal-oct-c8}{\color{magenta}\Checkmark}
\\

Ultrasound Nerve Segmentation \cite{ultrasound-nerve-segmentation}
& US	
&11,143 images	
&SEG
&\href{https://www.kaggle.com/c/ultrasound-nerve-segmentation/data}{\color{magenta}\Checkmark}
\\
Fetal Planes \cite{Fetal} 	
&US	
&12,400 images	
&CLS
&\href{https://zenodo.org/records/3904280}{\color{magenta}\Checkmark}
\\
EchoNet-LVH \cite{duffy2022high}
&US
&12,000 videos
&DET, Clinical study
&\href{https://echonet.github.io/lvh/}{\color{magenta}\Checkmark}
\\
EchoNet-Dynamic \cite{ouyang2020video}
&US
&10,030 videos
&Function assessment
&\href{https://echonet.github.io/dynamic/index.html}{\color{magenta}\Checkmark}
\\

AIROGS \cite{de2023airogs}
&CFP
&113,893 images	
&CLS
& \href{https://airogs.grand-challenge.org/}{\color{magenta}\Checkmark}
\\
ISIC 2020 \cite{rotemberg2021patient}
&Dermoscopy
&33,126 images	
&CLS
&\href{https://challenge2020.isic-archive.com/}{\color{magenta}\Checkmark}
\\
LC25000 \cite{borkowski2019lung}
&Pathology
&25,000 images	
&CLS
&\href{https://github.com/tampapath/lung_colon_image_set}{\color{magenta}\Checkmark}
\\

DeepLIIF \cite{ghahremani2022deep}
&Pathology	
&1,667 WSIs	
&SEG
&\href{https://deepliif.org/}{\color{magenta}\Checkmark}
\\
PAIP \cite{PAIP}	
&Pathology
&2,457 WSIs	
&SEG	
&\href{https://paip2019.grand-challenge.org/}{\color{magenta}\Checkmark}
\\
TissueNet \cite{greenwald2022whole}
&Pathology	
&1,016 WSIs	
&CLS
&\href{https://www.drivendata.org/competitions/67/competition-cervicalbiopsy/page/254/}{\color{magenta}\Checkmark}
\\
NLST \cite{national2011national}	
& 3D CT, Pathology 	
& 26,254 CT, 451 WSIs
&Clinical  study
&\href{https://www.cancerimagingarchive.net/collection/nlst/}{\color{magenta}\Checkmark}
\\
CRC \cite{CRC} 	
&Pathology	
&100k images	
&CLS
&\href{https://zenodo.org/records/1214456}{\color{magenta}\Checkmark}
\\
MURA \cite{MURA}
&X-ray
&40,895 images
&DET
& \href{https://stanfordmlgroup.github.io/competitions/mura/}{\color{magenta}\Checkmark}
\\
ChestX-ray14 \cite{wang2017chestxray}
&X-ray
&112,120 images
&DET
& \href{https://paperswithcode.com/dataset/chestx-ray14}{\color{magenta}\Checkmark}
\\
SNOW \cite{SNOW}
&Synthetic pathology
&20K image tiles
&SEG
& \href{https://zenodo.org/records/6633721#.YuE33OzMJhE}{\color{magenta}\Checkmark}
\\
\bottomrule
\end{tabular}
\end{table*}
\begin{table*}[!tb]
\label{tab:bfm-datasets}
\centering
\caption{The publicly available biological datasets. The ``Bioinformatics study" means that this is a comprehensive dataset for bioinformatics without clear task guidance.}
{
\begin{tabular}{llllc}
\toprule
 \textbf{Dataset}
& \textbf{Modalities}
& \textbf{Scale}
& \textbf{Tasks}
& \textbf{Link}
\\
\midrule
 CellxGene Corpus \cite{cellxgene}
& scRNA-seq
& over 72M scRNA-seq data
& Single cell omics study
& \href{https://cellxgene.cziscience.com}{\color{magenta}\Checkmark}
\\
 NCBI GenBank \cite{genbank}
& DNA
& 3.7B sequences
& Genomics study
& \href{https://www.ncbi.nlm.nih.gov/genbank/}{\color{magenta}\Checkmark}
\\
 SCP \cite{tarhan2023single}
& scRNA-seq
& over 40M scRNA-seq data
& Single cell omics study
& \href{https://singlecell.broadinstitute.org/single_cell}{\color{magenta}\Checkmark}
\\
 GenCode \cite{gencode}
& DNA
& -
& Genomics study
& \href{https://www.gencodegenes.org/}{\color{magenta}\Checkmark}
\\
 10x Genomics
& scRNA-seq, DNA
& -
& Single cell omics and genomics study
& \href{https://support.10xgenomics.com/single-cell-gene-expression/datasets}{\color{magenta}\Checkmark}
\\
 ABC Atlas
& scRNA-seq
& over 15M scRNA-seq data
& Single cell omics study
& \href{https://portal.brain-map.org}{\color{magenta}\Checkmark}
\\
 Human Cell Atlas \cite{HumanCellAtlas}
& scRNA-seq
& over 50M scRNA-seq data
& Single cell omics study
& \href{https://www.humancellatlas.org/}{\color{magenta}\Checkmark}
\\
 UCSC Genome Browser \cite{UCSCgenome}
& DNA
& -
& Genomics study
& \href{https://hgdownload.soe.ucsc.edu/downloads.html}{\color{magenta}\Checkmark}
\\
CPTAC \cite{doi:10.1021/pr501254j}
&DNA, RNA, protein
& -
& Genomics and proteomics study
&\href{https://gdc.cancer.gov/about-gdc/contributed-genomic-data-cancer-research/clinical-proteomic-tumor-analysis-consortium-cptac}{\color{magenta}\Checkmark}
\\
 Ensembl Project \cite{Ensembl}
& Protein
& -
& Proteomics study
& \href{https://ensembl.org/index.html}{\color{magenta}\Checkmark}
\\
 RNAcentral database \cite{RNAcentral}
& RNA
& 36M sequences
& Transcriptomics study
& \href{https://rfam.org}{\color{magenta}\Checkmark}
\\
 AlphaFold DB \cite{jumper2021highly}
& Protein
& 214M structures
& Proteomics study
& \href{https://alphafold.ebi.ac.uk/download}{\color{magenta}\Checkmark}
\\
 PDBe \cite{Armstrong2020pdbe}
& Protein
& -
& Proteomics study
& \href{https://www.ebi.ac.uk/pdbe/}{\color{magenta}\Checkmark}
\\
 UniProt \cite{Consortium2023uniprot}
& Protein
& over 250M sequences
& Proteomics study
& \href{https://www.uniprot.org}{\color{magenta}\Checkmark}
\\
 LINCS L1000 \cite{lincs1000}
& Small molecules
& 1,000 genes with 41k small molecules
& Disease research, drug response
& \href{https://lincsportal.ccs.miami.edu/dcic-portal/}{\color{magenta}\Checkmark}
\\
 GDSC \cite{gdsc}
& Small molecules
& 1,000 cancer cells with 400 compounds
& Disease research, drug response
& \href{https://www.cancerrxgene.org}{\color{magenta}\Checkmark}
\\
 CCLE \cite{CCLE}
& -
& -
& Bioinformatics study
& \href{https://sites.broadinstitute.org/ccle/}{\color{magenta}\Checkmark}
\\
\bottomrule
\end{tabular}
}
\end{table*}

\subsection{Bioinformatics}
Since high-throughput sequencing has become a fundamental technique in the biological field for over a decade \cite{highthroughput}, extensive data on DNA, RNA, protein, and scRNA-seq are large-scale scanned, providing a wealth of information for biological research. The rapid accumulation of publicly available sequencing data enables the researchers to train their BFMs. As shown in Tab.\ref{tab:bfm-datasets}, we present a list of large-scale biological datasets that contain more than millions of expression values, sequences, or structures.

\subsubsection{Genomics and single-cell omics data}
Genomics and single-cell omics data offer comprehensive insights into genetic information, gene expression patterns, and cellular functions. DNA sequence represents the genetic blueprint in organisms, while scRNA-seq profiles gene expression in individual cells, revealing functional diversity and cellular responses.
NCBI GenBank \cite{genbank} is an annotated collection of all publicly available DNA sequences. Currently, it is comprised of up to 3.7 billion sequences for set-based records. GenCode~\cite{gencode} is a scientific project aimed at annotating all evidence-based gene features in the entire human genome. Its goal is to identify all gene features, including protein-coding sequences, non-coding RNAs, pseudogenes, and their variants. The genome sequence is also included in the dataset. CellxGene Corpus \cite{cellxgene} is a broad single-cell corpus that contains 789 cell types from 1,219 datasets. The overall cell expression number reaches over 72 million. UK Biobank \cite{bycroft2018uk} provides a summary of all the information gathered by UK Biobank on 500,000 participants which contain imaging, genetics, health linkages, biomarkers, activity monitor, online questionnaires, repeat baseline assessments, etc. The SCP \cite{tarhan2023single} aggregates over 40 million cells from 645 distinct studies. This platform is composed of a diverse range of biological data, including 14 species, 83 diseases, 104 organs, and 160 different cell types. This collection offers invaluable insights into cellular behavior across various biological contexts and conditions. Some other datasets include Human Cell Atlas \cite{HumanCellAtlas}, 10x Genomics, Allen Brain Cell Atlas, etc., and also offer abundant biological genome data for scientific use.

\subsubsection{Transcriptomics and proteomics data}
Transcriptomics and proteomics together elucidate the journey from genetic information to functional proteins, uncovering complex cellular processes. Transcriptomics focuses on RNA sequences to understand gene expression, while proteomics analyzes protein structures, revealing intricate cellular mechanisms. The Ensembl project \href{https://ensembl.org/index.html}{(https://ensembl.org/index.html)} provides a comprehensive source of automatic annotation of the human genome sequence, as well as other species of biomedical interest, with confirmed gene predictions that have been integrated with external data sources \cite{Ensembl}. Genome sequence and corresponding protein sequences can be found here. RNAcentral \href{https://rnacentral.org/}{(https://rnacentral.org/)} is a database of non-coding RNA (ncRNA) sequences that aggregates data from specialized ncRNA resources and provides a single entry point for accessing ncRNA sequences of all ncRNA types from all organisms \cite{RNAcentral}. It currently contains over 36 million ncRNA sequences integrated from 53 databases. The AlphaFold Protein Structure Database \cite{jumper2021highly} provides open access to over 200 million protein structure predictions. Protein Data Bank in Europe (PDBe) \cite{Armstrong2020pdbe} and UniProt \cite{Consortium2023uniprot} also provide millions of protein structures.

\subsubsection{Other large-scale biological databases}
There are also other large-scale biological databases that are maintained for disease research drug response purposes. LINCS L1000 dataset \cite{lincs1000}, which includes information on how different types of human cells respond to various perturbations, such as exposure to drugs, toxins, or genetic modifications. It measures the expression levels of approximately 1,000 landmark genes, which are carefully selected to represent the entire human genome, on over 41k small molecules. Genomics of Drug Sensitivity in Cancer (GDSC) dataset \cite{gdsc}, which contains 1000 human cancer cell lines and screened them with around 400 compounds. There are also some other databases (Tab.\ref{Tab:databases}) like Cancer Cell Line Encyclopedia (CCLE) \cite{CCLE} and The Cancer Genome Atlas Program (TCGA) \cite{TCGA}, which can be used for validating cancer targets and for defining drug efficacy, The Chinese Glioma Genome Atlas (CGGA) \cite{Zhao2021cgga} for glioma-based disease research. The UK Biobank \cite{bycroft2018uk} is a large-scale biomedical database and research resource, containing in-depth genetic and health information from half a million UK participants. The unique and rich data resource, including genetic, lifestyle, and health information, offers an unprecedented opportunity to examine complex interactions between genetics, environment, and lifestyle in determining health outcomes.

\subsection{Multimodal}
\begin{table*}[tb]
\label{Tab:multimodaldatasets}
\centering
\caption{The publicly available multi-modal datasets. The abbreviations here are QA: question answering, VQA: visual question answering. The ``Multimodal learning" means that this is a comprehensive dataset without clear task guidance.}
\resizebox{\linewidth}{!}
{
\begin{tabular}{llllcccc}
\toprule
 \textbf{Dataset}
& \textbf{Modalities}
& \textbf{Scale}
& \textbf{Tasks}
& \textbf{Link}
\\
\midrule
 MIMIC-CXR \cite{johnson2019mimic}
& X-ray images, Medical report
& 377K images, 227K texts
& Vision-Language learning
& \href{https://physionet.org/content/mimic-cxr/2.0.0/}{\color{magenta}\Checkmark}
\\
 PadChest \cite{bustos2020padchest}
& X-ray images, Medical report
& 160K images, 109K texts
& Vision-Language learning
& \href{http://bimcv.cipf.es/bimcv-projects/padchest/}{\color{magenta}\Checkmark}
\\
 CheXpert \cite{irvin2019chexpert}
& X-ray images, Medical report
& 224K images, 224K texts
& Vision-Language learning
& \href{https://stanfordmlgroup.github.io/competitions/chexpert}{\color{magenta}\Checkmark}
\\
ImageCLEF2018 \cite{garcia2018overview}
& Multimodal images, Captions
& 232K images, 232K texts
& Image captioning
& \href{https://www.imageclef.org/2018/caption}{\color{magenta}\Checkmark}
\\
OpenPath \cite{huang2023visual}
& Pathology images, Tweets
& 208K images, 208K texts
& Vision-Language learning
& \href{https://huggingface.co/spaces/vinid/webplip}{\color{magenta}\Checkmark}
\\
 PathVQA \cite{he2020pathvqa}
& Pathology images, QA
& 4K images, 32K QA pairs
& VQA
& \href{https://github.com/UCSD-AI4H/PathVQA}{\color{magenta}\Checkmark}
\\
 Quilt-1M \cite{ikezogwo2023quilt}
& Pathology images, Mixed-source text
& 1M images, 1M texts
& Vision-Language learning
& \href{https://quilt1m.github.io/}{\color{magenta}\Checkmark}
\\
 PatchGastricADC22 \cite{tsuneki2022inference}
& Pathology images, Captions
& 991 WSIs, 991 texts
& Image captioning
& \href{https://github.com/masatsuneki/histopathology-image-caption}{\color{magenta}\Checkmark}
\\
 PTB-XL \cite{wagner2020ptb}
& ECG signals, Medical report
& 21K records, 21K texts
& Vision-Language learning
& \href{https://physionet.org/content/ptb-xl/1.0.1/}{\color{magenta}\Checkmark}
\\
 ROCO \cite{pelka2018radiology}
& Multimodal images, Captions
& 87K images, 87K texts
& Vision-Language learning
& \href{https://github.com/razorx89/roco-dataset}{\color{magenta}\Checkmark}
\\
 MedICaT \cite{subramanian2020medicat}
& Multimodal images, Captions
& 217K images, 217K texts
& Vision-Language learning
& \href{https://github.com/allenai/medicat}{\color{magenta}\Checkmark}
\\
 PMC-OA \cite{lin2023pmc}
& Multimodal images, Captions
& 1.6M images, 1.6M texts
& Vision-Language learning
& \href{https://huggingface.co/datasets/axiong/pmc_oa}{\color{magenta}\Checkmark}
\\
 ChiMed-VL \cite{liu2023qilin}
& Multimodal images, Medical report
& 580K images, 580K texts
& Vision-Language learning
& \href{https://github.com/williamliujl/Qilin-Med-VL?tab=readme-ov-file}{\color{magenta}\Checkmark}
\\
 PMC-VQA \cite{zhang2023pmc}
& Multimodal images, QA
& 149K images, 227K QA pairs
& VQA
& \href{https://huggingface.co/datasets/xmcmic/PMC-VQA}{\color{magenta}\Checkmark}
\\
SwissProtCLAP\cite{liu2023text}
&Protein Sequence, Text
&441K protein sequence, 441K texts
& Protein-Language learning
&\href{https://huggingface.co/datasets/chao1224/ProteinDT/tree/main}{\color{magenta}\Checkmark}
\\
Duke Breast Cancer MRI \cite{saha2018machine}
&Genomic, MRI images, Clinical data
& 922 patients
& Multimodal learning
&\href{https://sites.duke.edu/mazurowski/resources/breast-cancer-mri-dataset/}{\color{magenta}\Checkmark}
\\
I-SPY2 \cite{li2022ispy2}
&MRI images, Clinical data
& 719 patients
& Multimodal learning
&\href{https://www.cancerimagingarchive.net/collection/ispy2/}{\color{magenta}\Checkmark}
\\
\bottomrule
\end{tabular}
}
\end{table*}
The accumulation of multimodal healthcare data, and the arrangement and construction of large-scale multimodal datasets are the basis for the success of healthcare MFMs. However, due to the existing accessibility of healthcare image and text data, most of current multimodal healthcare datasets are still in vision and language that is limited by modal diversity. As shown in Tab.\ref{Tab:multimodaldatasets}, we provide a summary of public available and relatively large healthcare multimodal datasets for MFMs, including images, textual description, electroencephalography (EGG) signals, protein and molecular information. Here, we discuss the visual-language datasets and the multimodal datasets beyond vision and language.

\subsubsection{Visual-language data}
Due to the varied properties of different medical image modalities, the scale and composition of the existing healthcare visual-language datasets are also diverse. For X-ray imaging, MIMIC-CXR \cite{johnson2019mimic} is the most commonly used pre-training dataset for MFMs. It contains 377,110 chest X-ray images paired with 227,835 corresponding medical reports. PadChest \cite{bustos2020padchest} and CheXpert \cite{irvin2019chexpert} also comprise chest X-ray images and corresponding medical reports, enriching the variety and quantity of chest X-ray images.  For pathology imaging, OpenPath \cite{huang2023visual} is sourced from Twitter's medical knowledge-sharing platform, containing more than 200K pathology images with descriptions by medical professionals. PathVQA \cite{he2020pathvqa} is a pathology VQA dataset, which includes 4,998 images and 32,799 QA pairs. Quilt-1M \cite{Gamper_2021_CVPR} is a large histopathology dataset, which consists of 1M image-text pairs. PatchGastricADC22 \cite{tsuneki2022inference} includes 262,777 image patches extracted from 991 WSIs with the relevant diagnostic captions. Besides, there are also some visual-language datasets with multiple medical image modalities, including ROCO\cite{pelka2018radiology}, MedICaT\cite{subramanian2020medicat}, PMC-OA \cite{lin2023pmc}, ChiMed-VL\cite{liu2023qilin}. These datasets offer a diverse array of image modalities, spanning radiology and histology domains. Specially, PMC-VQA\cite{zhang2023pmc} is a multimodal medical visual question-answering dataset, containing a total of 227k VQA pairs of 149k images.

\subsubsection{Beyond visual-language data}
Besides visual-language data, there are still some other healthcare multimodal datasets that are publicly available. SwissProtCLAP \cite{liu2023text} consists of 441,000 protein-text sequence pairs, including 327,577 genes and 13,339 organisms and their corresponding texts. It is relatively smaller in scale compared to image-text pairs in the vision-language domain. PTB-X\cite{liu2023etp} includes 21,837 EGG signals paired with its corresponding medical report. Duke Breast Cancer MRI \cite{saha2018machine} includes multi-sequence MRI images and pathology, clinical treatment, and genomic data from 922 biopsy-confirmed breast cancer patients. I-SPY2 \cite{saha2018machine} comprises over 4TB of MRI and clinical data from 719 breast cancer patients. Besides, there are also some large-scale comprehensive databases (Tab.\ref{Tab:databases}) that contain numerous healthcare data in different modalities. TCGA \cite{TCGA} program is a landmark cancer genomics initiative, encompassing 2.5PB of genomic, pathology image, pathology report, and other multimodal data.

\begin{table*}[tb]
\label{Tab:databases}
\centering
\caption{Large-scale comprehensive database which contains healthcare data across multiple sub-fields.}
{
\begin{tabularx}{\linewidth}{lXc}
\toprule
 \textbf{Database}
& \textbf{Description}
& \textbf{Link}
\\
\midrule
 CGGA \cite{Zhao2021cgga}
& Chinese Glioma Genome Atlas (CGGA) database contains clinical and sequencing data of over 2,000 brain tumor samples from Chinese cohorts.
& \href{http://www.cgga.org.cn}{\color{magenta}\Checkmark}
\\
 UK Biobank \cite{bycroft2018uk}
& UK Biobank is a large-scale biomedical database and research resource containing de-identified genetic, lifestyle and health information and biological samples from half a million UK participants.
& \href{https://www.ukbiobank.ac.uk}{\color{magenta}\Checkmark}
\\
TCGA \cite{TCGA}
&The Cancer Genome Atlas program (TCGA) molecularly characterizes over 20,000 primary cancer, matches normal samples spanning 33 cancer types, and generates over 2.5 petabytes of genomic, epigenomic, transcriptomic, and proteomic data.
&\href{https://www.cancer.gov/ccg/research/genome-sequencing/tcga}{\color{magenta}\Checkmark}
\\
TCIA \cite{clark2013cancer}
&The Cancer Imaging Archive (TCIA) is a service which de-identifies and hosts a large publicly available archive of medical images of cancer.
&\href{https://www.cancerimagingarchive.net/}{\color{magenta}\Checkmark}
\\
\bottomrule
\end{tabularx}
}
\end{table*}

\section{Applications}
\label{sec:applications}
\subsection{Language}
Due to the widespread use of texts in healthcare practices, LFMs have achieved significant applications in diagnosis, education, consultation, etc. Especially, with the application of LLMs represented as ChatGPT \cite{patel2023chatgpt}, their clinical application potential has been further explored, and some general healthcare language models, like BianQue \cite{chen2023bianque} and Med-PaLMM \cite{tu2023towards}, have achieved success in healthcare scenarios.

\subsubsection{Medical diagnosis}
Medical diagnosis via LFM predicts the most likely disease based on medical tests and patient descriptions and is crucial for timely treatment and preventing complications \cite{balogh2015improving}. Recently, LFMs have been employed to enhance medical diagnosis and demonstrated generalist ability on different diseases \cite{zhang2023huatuogpt,wang2023clinicalgpt,xiong2023doctorglm,gao2023ophglm}. Ueda \textit{et al.}  \cite{ueda2023diagnostic} utilized patient history and imaging findings to diagnose please quizzes via ChatGPT. Wu \textit{et al.} \cite{wu2024collaborative} also evaluated three LFMs on the diagnosis of thyroid nodules, demonstrating the application potential of LFMs in enhancing diagnostic medical imaging. Although LFMs have shown diagnostic ability, clinicians need to trace and comprehend the logic behind each diagnostic decision and the lack of transparency is still one of the large challenges (discussed in Sec.\ref{sec:challenges}).

\subsubsection{Report generation}
LFMs have demonstrated their potential in generating medical reports, including radiology reports~\cite{singhal2023towards}, discharge summaries~\cite{patel2023chatgpt}, and referral letters~\cite{ali2023using}. These models excel at synthesizing information from diverse sources, such as EHRs, medical literature, and clinical guidelines, producing coherent and informative reports. Doctors often find writing medical reports to be tedious and time-consuming, so utilizing medical language models can alleviate their workload. One approach is by inputting diagnosis results into a language model, which then serves as a summarization tool to generate the report~\cite{wang2023chatcad}. Therefore, reasonable reports will be generated without manual editing. Radiologists also will benefit from LFMs via inputting some image descriptions, enabling it to diagnose \cite{wu2024collaborative} and create reports. Examples of such models include ChatCAD~\cite{wang2023chatcad}, ChatCAD+\cite{zhao2023chatcad+}, Visual Med-Alpaca \cite{shu2023visual}, and MedAgents \cite{tang2023medagents}.

\subsubsection{Healthcare education}
LFMs also play a significant role in healthcare education \cite{abd2023large} for both practitioners and the general public. For medical students, these models are able to generate medical questions, enhancing their understanding of medical knowledge~\cite{karabacak2023advent}. They can also play the role of medical teachers and give the students professional answers for their clinical questions. Kung \textit{et al.} \cite{kung2023performance} have evaluated that the ChatGPT has the potential to assist with medical education. Several models have been developed for medical education purposes, such as HuatuoGPT-II~\cite{chen2023huatuogpt} which focuses on Chinese medical examination. For the general public, LFMs also can translate complex medical terms into easily understandable language, facilitating public healthcare education~\cite{karabacak2023advent}. A study \cite{cocskun2024integration} has explored the integration of ChatGPT and e-Health literacy, illustrating the significant potential of LFMs for substantially enhancing the accessibility and quality of health services.

\subsubsection{Medical consultation}
LFMs can improve medical consultation \cite{lee2023benefits}, a vital aspect of healthcare. These models can use both their internal knowledge and information from medical websites, such as health forums and textbooks, to provide patients with medical information for self-diagnosis or other purposes. Furthermore, these models can also serve as chatbots that can offer mental health support to patients \cite{chen2023soulchat}, which can enhance their well-being and lessen the burden on mental health professionals. Several models have been developed for medical consultation, including BenTsao~\cite{wang2023huatuo}, MedPaLM~\cite{singhal2023large}, MedPaLM 2~\cite{singhal2023towards}, etc.~\cite{zhang2023alpacare,wang2023gpt,wang2023gpt,chen2023bianque,wang2023clinicalgpt,li2023chatdoctor}, showcasing the feasibility and effectiveness of using LFMs to improve the quality and efficiency of medical consultation and related services.

\subsection{Vision}
VFMs also have achieved success in the segmentation, classification, detection, etc., tasks, demonstrating their promising application in empowering radiologists, surgeons, or clinicians, and assisting the workflows in diagnosis, prognosis, surgery, or other healthcare practices.

\subsubsection{Medical diagnosis}
VFMs also have demonstrated their application potential in diagnosis \cite{wu2023towards} on medical images. They enable automatic disease screening on some low-risk images and assist in the detection and identification of unclear target anatomies, thus reducing the workload of radiologists and improving their diagnosis accuracy. Segmentation and detection VFMs provide the position information inner medical images, including organs~\cite{1-USFM,2-MIS-FM,26-SAM-Med2D,39-UR-SAM,45-cheap,47-LVM-Med}, tumors~\cite{24-3DSAM-adapter,22-MedSAM}, and lesions~\cite{51-samus,3-RETFound,47-LVM-Med}, assisting the radiologists to decouple the images for semantic regions and discover the interests. The classification VFMs also promote the automatic disease diagnosis~\cite{1-USFM,14-MG,15-MG,18-MoCo-CXR,16-C2L,3-RETFound,48-RudolfV} via directly predicting the categories of the input images, thus effectively reducing the cost of some low-risk images like the physical examination screening. However, owing to the limitations in trustworthiness, it is still challenging for some high-risk diagnosis applications like tumor grading.

\subsubsection{Disease prognosis}
Some VFMs also have achieved promising results in disease prognosis, which is able to provide some biomarkers to predict the likelihood or expected development of a disease. Therefore, the clinicians or radiologists will make intervention plans for patients according to the prognosis. Some large-scale pre-training VFMs, such as RETFound~\cite{3-RETFound} and VisionFM~\cite{12-VisionFM}, are able to extract the representative features, that are related to ophthalmic diseases, from retinal images, so that these features will potentially represent the progress of the disease as biomarkers. Some segmentation or detection VFMs \cite{26-SAM-Med2D,22-MedSAM} also can provide the shape, size, and position of lesions, e.g., tumors, which are also potential biomarkers for their progress. However, it is still challenging to directly contrast a foundation model for prognosis applications like survival prediction, because these practices require large-scale follow-up data which is rare clinically.

\subsubsection{Surgery planning and assistance}
Surgery is another potential application scenario of VFM which constructs plug-and-play medical imaging processing tools for surgery planning or surgery assistance without additional data collection and model training in conventional paradigms. For surgery planning, the surgeons will be able to segment the 3D objects from the medical images like CT and MRI via some 3D segmentation VFMs, like the SAM-Med-3D \cite{30-SAM-Med3D}, thus visualizing the interested objects for planning. During the surgery, VFMs like the SP-SAM \cite{88-part} (a segmentation VFM) also could segment tools or interested regions in endoscope view, thus assisting the operation and enhancing surgical outcomes. However, there is still a challenge for the interaction of VFMs when the hands of the surgeons can't operate the machine during the surgery.

\subsubsection{Other VFM applications}
VFMs encompass versatile applications beyond diagnosis, prognosis, and surgery applications in healthcare. For instance, CTransPath~\cite{83-CTransPath} is pivotal in retrieval and prototyping, facilitating the efficient retrieval of relevant medical images and aiding in the development of prototypes for medical devices and systems. USFM~\cite{1-USFM} contributes to image enhancement techniques, enhancing the quality, clarity, and interpretability of medical images, thereby assisting in accurate diagnosis and treatment planning~\cite{1-USFM}. Through their diverse applications, VFMs play a pivotal role in advancing medical imaging technologies and enhancing patient care across various clinical domains.

\subsection{Bioinformatics}
Biological foundation models serve for various downstream tasks. We can categorize them into the following levels: sequence analysis, interaction analysis, structure and function analysis, and disease and drug research. These types are instrumental in combining computational power and biological insight to unravel the complexity of life.

\subsubsection{Sequence analysis}
BFMs on sequence analysis have advanced our understanding in genomics and transcriptomics, unraveling intricate biological processes and molecular interactions. Researchers utilize BFM on this area with different targets \cite{ji2021dnabert,GeneBERT,Zhou2023dnabert2,Fishman2023genalm,dalla2023nucleotide,Nguyen2023hyenadna,Zhang2023dnagpt,Chu2023utrlm,Yang20233utr,madani2023large,Nijkamp2023progen2,Akiyama2022rnabert,Chen2023splicebert}. Several works use BFM on (core) promoter detection task \cite{ji2021dnabert,GeneBERT,Zhou2023dnabert2,Fishman2023genalm,dalla2023nucleotide,Nguyen2023hyenadna}, which focus on the identification of promoter regions in the DNA sequence. These promoters are crucial elements that initiate the transcription of genes, acting as key regulatory sequences that control gene expression.

\subsubsection{Interaction analysis}
Interaction analysis is another potential biological research application for BFMs. It gives an understanding of the complex interplay and regulatory mechanisms within cellular systems. The interactions between genes \cite{Chen2023genept, cui2023scgpt, Theodoris2023geneformer}, between protein and RNAs \cite{chen2022interpretable}, and between proteins \cite{Chen2023genept} have been effectively analyzed by current BFMs. For example, utilizing RNA-FM \cite{chen2022interpretable} embeddings with sequences achieve the best performance on three streamlines (sequence only, sequence with real secondary structure, and sequence with RNA-FM) on nearly half of the subsets, which are adopted with RNA in the HeLa cell and divided according to different corresponding RBPs. Their performance is even comparable to the real secondary structure with sequences, suggesting that embeddings from RNA-FM provide sufficient information as real secondary structures.

\subsubsection{Structure and function analysis}
BFMs have achieved application in structure and function analysis, that decipher the complex relationship between molecular structure and biological function, enhancing our comprehension of cellular behavior and genetic variability. Numerous works on genomics \cite{ji2021dnabert,dalla2023nucleotide}, transcriptomics \cite{chen2022interpretable,Chu2023utrlm,Li2023codonbert,Chen2023splicebert,Rao2021msa,Akiyama2022rnabert,Zhang2023rnamsm}, proteomics \cite{jumper2021highly,Lin2023esm2,Rives2021esm1b,Chowdhury2022aminobert,Chen2024xTrimopglm,Elnaggar2022prottrans,Nijkamp2023progen2}, and single cell omics \cite{Rosen2023uce,cui2023scgpt,Zhao2023celllm,Yang2022scbert} have achieved remarkable success. Protein structure prediction is one of the most popular tasks in proteomics. The famous Alphafold \cite{jumper2021highly} pre-trained on multiple sequence alignment datasets that have indicated functional, structural, or evolutionary relationships among the sequences. The xTrimoPGLM \cite{Chen2024xTrimopglm} model also achieved protein structure prediction and generation, significantly outperforming other advanced baselines in large-scale parameters. 

\subsubsection{Disease research and drug response}
The significance of disease research and drug response lies in their critical role in advancing medical knowledge, enabling the development of innovative therapies and cures that improve human health and extend life spans. BFM can perform drug sensitivity prediction \cite{Theodoris2023geneformer, Zhao2023celllm, Hao2023scfoundation}, transcription factor dosage sensitivity prediction \cite{Theodoris2023geneformer,Chen2023genept}, drug response prediction \cite{Hao2023scfoundation}, disease risk estimation \cite{GeneBERT}, cellular perturbation response prediction \cite{Hao2023scfoundation}, and covid variants classification \cite{Zhou2023dnabert2}, offering insights into the complex interplay between genetic factors and therapeutic outcomes, thus facilitating personalized medicine and advancing our understanding of disease mechanisms. For example, scGPT \cite{cui2023scgpt} leverages the knowledge gained from cellular responses in known experiments and extrapolates them to predict unknown responses. The utilization of self-attention mechanisms over the gene dimension enables the encoding of intricate interactions between perturbed genes and the responses of other genes to deal with the vast combinatorial space of potential gene perturbations \cite{cui2023scgpt}.

\subsection{Multimodal}
MFMs are able to fuse the information from different modalities, thus improving the performance of some applications based on a single modality (e.g., diagnosis) and also achieving cross-modality generation for some inter-modality applications (e.g., report generation).

\subsubsection{Medical diagnosis}
Although LFM and VFM have demonstrated promising abilities in medical diagnosis, MFMs further integrate multiple data sources from patients, leveraging AI models' capabilities in diagnosis and assisting doctors with more accurate diagnostic decisions. Particularly, some pre-trained MFMs \cite{huang2021gloria, liu2023etp, tiu2022expert, huang2023visual} based on CLIP are able to achieve zero-shot classification through prompting, suitable for open-ended disease diagnosis environments. Compared to VFMs, the contextual information from medical texts further enriches feature representation, especially in cases where visual clues are ambiguous. For medical images in different modalities, MFMs (e.g., RadFM \cite{wu2023towards}) also have achieved the ability to fuse their information, improving the diagnosis according to the features from different imaging conditions. The vision-language MFMs like Qilin-Med-VL \cite{liu2023qilin}, Med-Flamingo \cite{moor2023med}, LLaVA-Med \cite{li2023llava} have healthcare conversational capabilities that predict disease descriptions such as disease types and locations on medical images by asking questions. Therefore, human experts will make diagnostic decisions based on deeper insights from AI experts.

\subsubsection{Report generation}
Different from the medical report generation in LFM, visual-language MFM generates radiology reports from medical images, accelerating the efficiency of preliminary imaging diagnosis. The large-scale MFMs, such as RadFM \cite{wu2023towards}, Clinical-BERT \cite{yan2022clinical}, and MedViLL \cite{moon2022multi} leverage both image and text information, enabling a more comprehensive understanding of patient data and leading to the generation of reports from medical images. Compared with conventional manual writing, it has shown the potential to liberate radiologists from tedious report writing. Furthermore, some current MFMs also tried to support interactive report generation, enabling the creation of medical reports in specific templates and formats based on prompts \cite{thawkar2023xraygpt}. Therefore, it has achieved the advantages of both alleviating the medical professional workload and enhancing the stability of report output.

\subsubsection{Biological science}
MFMs offer a viable solution for bridging the language of life (e.g., DNA, RNA, protein, etc.) and human natural language, demonstrating the potential application in biological science. Molecule-language MFM \cite{liu2023multi} enables molecule editing with text prompts without additional molecule data and annotations, promoting science research in biochemistry. MFM (e.g., BioMedGPT-10B \cite{luo2023biomedgpt}) is also able to play the role of bioinformatics expert, and have a conversation with humans on biomedical research and development via language. Researchers can upload biological data, such as molecular structures and protein sequences, and formulate natural language queries about these data instances. Based on the conversation with the MFM, it will inspire the researchers and enhance the efficiency of discovering novel molecular structures, elucidating protein functionalities, and advancing drug research and development.

\subsubsection{Medical consultation}
Except for the professional chatbots for doctors or researchers as introduced above, MFMs also can be applied as chatbots for medical consultation of patients. Because healthcare information is often scattered and difficult to follow for people without a medical background, the MFMs will contribute to the preliminary medical consultation for patients. Some MFMs, such as Qilin-Med-VL \cite{liu2023qilin}, XrayGPT \cite{van2023open}, LLaVA-Med \cite{li2023llava} and Med-Flamingo \cite{moor2023med}, have both text and visual understanding capabilities, which can offer preliminary diagnoses and treatment advice based on patient queries and images, guiding patients to seek medical treatment. This is especially beneficial for patients with chronic illnesses, as medical chatbots can engage in ongoing conversations, helping patients understand their conditions better and make necessary adjustments to their daily routines, exercise plans, dietary habits, and other lifestyle factors. Nevertheless, the applications of medical chatbots for patients are still challenging owing to the trustworthiness that a wrong suggestion from the chatbot will be dangerous for patients who lack medical knowledge. We will discuss this challenge in Sec.\ref{sec:challenges}.

\section{Challenges}
\label{sec:challenges}
\begin{figure}
    \centering
    \includegraphics[width=1\linewidth]{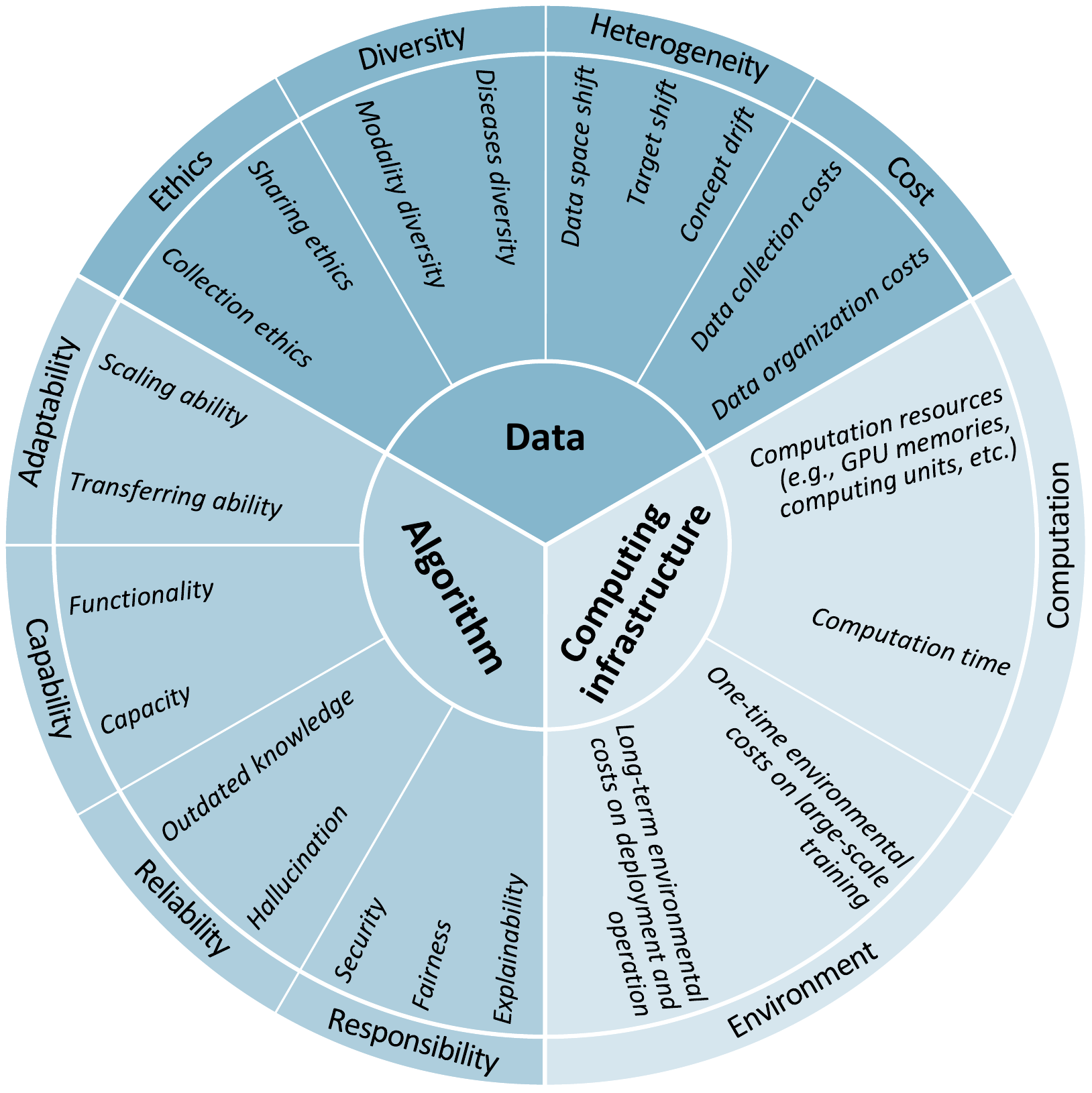}
    \caption{The challenges of the healthcare foundation model on data, algorithm, and computing infrastructure.}
    \label{fig:challenges}
\end{figure}
As shown in Fig.\ref{fig:challenges}, the data, algorithms, and computing infrastructure as three pillars of AI \cite{huawei2022general} have provided opportunities for the HFMs, but their current lack of development is still the root of various challenges. Specifically,
\subsection{Data}
The lack of data is the core challenge in HFMs. Foundation models' generalist ability relies on the learning of massive, diverse datasets \cite{bommasani2021opportunities}. However, the inherent properties including ethics, diversity, heterogeneity, high costs, etc. in healthcare data hinder large-scale dataset construction, and also pose ethical, social, and economic challenges. Therefore, ``How to construct large-scale healthcare dataset" \cite{willemink2020preparing} is the first question we have to answer for the challenges of HFMs. Specifically,

\subsubsection{Ethics}
Ethics \cite{larson2020ethics} of the healthcare data makes a critical challenge in the construction of large datasets. \textbf{a)} The acquisition of healthcare data has to meet ethical requirements. Healthcare data is scanned from the human body, while some scanning protocols or modalities will cause injuries to the human body, such as the CT imaging data \cite{salerno2019overdiagnosis}. Although these injuries may be insignificant for disease treatment, it is unethical to scan human bodies to exclusively construct the datasets for AI training. Therefore, these special data will be unable to be actively acquired for a large dataset like some existing data collection paradigms \cite{Kirillov2023ICCV}, forbidding the training for some HFM tasks. \textbf{b)} The use and sharing of healthcare data are also limited by ethics \cite{larson2020ethics}. Healthcare contains a lot of private information from the human body which is sensitive and risky, like genes. The use and sharing of these data are strictly restricted by laws and data owners. Once it is collected on a large scale in the absence of governance and used for the training of foundation models, it will be dangerous. In the further application of HFMs, the uncontrollable external environment also will enlarge this risk. For example, the language model could leak or misuse sensitive medical data, such as personal health records, testing results, genetic information, etc. Therefore, this aggravates the data challenge and makes the construction of the dataset for HFMs and their applications face stacked obstacles. Although some preliminary efforts \cite{larson2020ethics,kaur2022trustworthy} have made the community aware of it, there is still a long way.

\subsubsection{Diversity}
Due to the long-tailed distribution \cite{ZHANG2024102996} of the healthcare data, data diversity has become another significant data challenge in HFM. \textbf{a)} The refinement of healthcare applications makes the healthcare data in different modalities in a long-tailed distribution. For example, in medical images, although widely available Chest X-ray and Chest CT images serve general diagnostic purposes, other imaging modalities like optical coherence tomography (OCT), digital subtraction angiography (DSA), positron emission tomography (PET), etc., crucial for specific clinical tasks, are scarce and expensive. This makes these task-specific modalities rare in a large dataset compared with the normal modalities, restricting the generalization ability of trained HFMs. \textbf{b)} The occurrence of some diseases is also in long-tailed distribution \cite{haendel2020many}, so the images, bioinformatics data, or text records of these diseases are scarce. This means a lot of very rare disease data will not be covered or very rare in the training dataset of HFMs, limiting their generalization. Consequently, many essential yet specialized tasks remain beyond the reach of HFMs due to the limited diversity of relevant data, constraining the potential scope of AI applications and becoming a new obstacle on their way to generalists.

\subsubsection{Heterogeneity}\label{subsubsec:heter}
The features of healthcare data are varied across populations, regions, and medical centers, which makes the data heterogeneous in the real-world application of HFMs \cite{guan2021domain,gu2023beyond}. Therefore, there will be a potential distribution mismatch between the training data and the test data for the HFMs \cite{liu2024decade}. \textbf{a)} One of the urgent issues is the data space shift. The data acquisition protocol changes, sensor configuration changes, etc. will lead to variations in collected healthcare data, thus the HFMs trained for the original data will be unable to adapt to the new data form. \textbf{b)} The heterogeneity of the outcomes will occur when data are sampled from different subjects, contexts, population groups, etc., making the target shift in HFMs. For example, the incidence of different diseases would change along with personal characteristics and behaviors \cite{cassidy2007lung}, such as genetic diseases, smoking, eating habits, etc., so that it will extremely limit the HFMs' ability in the personalized and precise medical care. \textbf{c)} From a long-term perspective, the concept drift \cite{gama2014survey} is another important factor in data heterogeneity. With the development of the healthcare field, some new concepts will appear, and some wrong concepts will be corrected. Therefore, it is challenging for the HFMs to follow the changes in relationships between the input and output.

\subsubsection{Cost} \label{subsubsec:challenge:data:cost}
Data cost has been a significant challenge for healthcare AI for a long time, and the reliance of foundation models on large-scale data further amplifies this challenge in HFM \cite{moor2023foundation}. \textbf{a)} For data collection \cite{willemink2020preparing}, the acquisition of certain healthcare data modalities is exceedingly costly due to their specialized scanning methods and expensive equipment. For example, the price of a CT scan can be anywhere from \$300 to \$6,750 in the USA\footnote{\url{https://www.newchoicehealth.com/ct-scan/cost}}. Therefore, constructing a large-scale healthcare dataset, especially for some expensive modalities, for foundation model training will incur unimaginable costs, making it challenging for some institutions to implement independently. \textbf{b)} Although a lot of HFM works focus on self-supervised learning without annotation, it still costs a lot of professional manpower to organize the massive dataset further becoming another significant challenge \cite{wang2021annotation,tajbakhsh2021guest}. The specialized nature of healthcare data necessitates the involvement of skilled professionals in the filtering or annotation processes. This makes it impractical to utilize crowdsourcing \cite{Kirillov2023ICCV} for the annotation process of healthcare datasets. Therefore, it is inefficient and expensive to spend a significant amount of professional time on these repetitive tasks.

\subsection{Algorithm}
Although the algorithm has been studied for decades together with the development of AI \cite{lecun2015deep}, the unprecedented data amount, model scale, and application scope in the era of foundational models \cite{bommasani2021opportunities} have exposed new challenges in algorithms. Here, we have analyzed four of the most important algorithm challenges in healthcare, including responsibility, reliability, capability, and adaptability.

\subsubsection{Responsibility} \label{subsubsec:trust}
The responsibility of the foundation models remains a significant concern, especially owing to the close relationship between healthcare and human life, it has become particularly important \cite{sun2024trustllm} in HFM. It stems from people's skepticism about AI. \textbf{a)} One of the most important aspects is the explainability. Due to the ``blackbox" property of neural networks \cite{sokol2020one} and the much larger amount of hidden neurons, HFM is extremely more challenging to explain their behavior \cite{bommasani2023foundation}. Therefore, healthcare experts will be unable to understand the basis of the answers from the HFM, having significant concern about ethics and safety. \textbf{b)} Fairness \cite{chen2023algorithmic} is another aspect of the responsibility. Due to the distribution bias in the training dataset, HFM may be susceptible to inherent biases from the dataset, breaking the fairness of the outcome. Some work has unearthed pervasive biases and stereotypes in LFMs \cite{motoki2023more,felkner2023winoqueer}. This is dangerous, as unfair predictions may increase potential discrimination and undermine the equality of human life in healthcare, triggering potential social conflicts. \textbf{c)} Security is also a significant concern. Some LFMs have been documented to generate hate speech \cite{gehman2020realtoxicityprompts}, leading to offensive and psychologically harmful content and, in extreme cases, inciting violence. This is very dangerous for the users of HFM and become a potentially destabilizing factor in society. Some jailbreaking attacks \cite{wei2024jailbroken} even result in the output of LFMs containing private and sensitive information which is unethical in healthcare, posing a threat to data providers. Although there have been some existing studies on responsible AI \cite{baeroe2020achieve}, the unprecedented large scale and application scope of foundational models make it challenging for these technologies \cite{chen2023trustworthy}.


\subsubsection{Reliability}
Reliability is particularly critical in healthcare \cite{dwyer2023high} which poses a tremendous requirement for the reliability of HFMs, making a large challenge. It stems from the deficiencies of AI models themselves. \textbf{a)} Hallucination \cite{rawte2023survey} in foundation models is receiving increased attention. The models will output content that is not based on factual or accurate information. For example, during a medical conversation with an LFM, the model provides clinical knowledge or conclusions that contradict the facts \cite{lee2023benefits}. This raises concerns about the reliability of conclusions obtained with the assistance of HFMs. \textbf{b)} Another reliability challenge comes from outdated knowledge. As indicated in the heterogeneity of data \ref{subsubsec:heter}, the development of the healthcare field will construct some new knowledge, and correct some mistakes. Therefore, once foundational models fall behind the development of the field, they may potentially become misleading \cite{li2023task}. Although some existing efforts try to develop model editing techniques to update isolated model behavior or factual \cite{yao2023editing}, it is costly and lacks specificity, which may lead to unexpected side effects \cite{hoelscher2023detecting}. It is still a significant algorithm challenge for the long-term reliability of HFMs.

\subsubsection{Capability} \label{subsubsec:capability}
The capability of the HFMs determines their performance in applications, making it one of the most concerned challenges. \textbf{a)} Capacity \cite{raghu2017expressive} in foundational models has been the focus of researchers' efforts in recent years. It means the model's ability to represent and memorize the vast amount of accumulated knowledge \cite{bommasani2021opportunities}. Some advanced network architectures, such as ViT \cite{dosovitskiy2020image} and Swin Transformer \cite{liu2021swin}, attempt to construct a large-capacity backbone so that the foundational models will be able to learn a vast amount of knowledge from the large-scale datasets. However, enlarging the model capacity will also increase the computation and memory leading to higher costs and carbon emissions \cite{qiu2023large}, which is inefficient. Especially, for the healthcare field whose data may be very large like the 3D CT volumes, it is still a long-term problem to increase the capacity of HFMs with less computational consumption. \textbf{b)} Functionality of HFMs still seems monotonous and it is challenging to meet some complex clinical demands. For example, chronic disease management will involve information from multiple departments and various modalities \cite{zhao2022elements}, and require the implementation of multiple clinical procedures such as diagnosis, intervention, prognosis, etc. Although some generalist HFMs have shown promise in various clinical settings \cite{moor2023foundation,tu2023towards}, they are powerless to meet such complex multi-modal, multi-task requirements.

\subsubsection{Adaptability}
The adaptability of foundational models determines their ability to adapt to downstream scenarios, which remains a significant challenge in HFM. \textbf{a)} One aspect is the transferring ability for downstream tasks. The existing HFMs still struggle with data heterogeneity (\ref{subsubsec:heter}) in the real world \cite{guan2021domain} which performs poorly on some very specific domains \cite{mazurowski2023segment}. Although some methods utilize the FT or AT to apply foundational models to downstream domains, their adaptability is still limited by the original pre-trained models remaining a large cost for adaptation data \cite{22-MedSAM,23-Med-SA}. Therefore, it is still an urgent problem to efficiently and effectively reveal the hidden ability of HFMs for real-world scenarios. \textbf{b)} Another aspect is the scalability of the HFMs for downstream devices. In some potential resource-limited clinical scenarios, such as wearable medical devices \cite{chen2020deep}, initial very large foundational models will be unable to directly deploy. Therefore, HFMs require scaling methods to adapt to the potential operating environment. Although some model compression and acceleration techniques on AI models have been studied \cite{deng2020model}, scaling down such large foundational models and maintaining their generalist capabilities are still challenging.

\subsection{Computing infrastructure}
The large property (both in parameter size and data volume) of foundational models \cite{qiu2023large} makes the training and inference processes require unprecedentedly massive computing infrastructure, posing new challenges. Here, we have analyzed two of the most important computing infrastructure challenges, including computation and environment.
\subsubsection{Computation} \label{subsubsec:challenge:computing:computation}
It is excessively costly in terms of time and resources in training or even adapting the HFMs, exceeding the budgets of most researchers and organizations \cite{ding2023parameter}. \textbf{a)} The massive parameter amount of foundational models needs large computation resources in training or adaptation, like the GPU memories and the computing units, which is almost impractical in most hospitals and institutions. For example, the direct fine-tuning of GPT-3 requires approximately 175,255 billion parameters \cite{ding2023parameter}, requiring a large expense. \textbf{b)} Training on large-scale datasets will also consume a considerable amount of computation time. It had to take about 21 days to train the LLaMA with 65B parameters from 1.4T tokens on 2048 A100 GPUs with 80GB of RAM \cite{touvron2023llama}. This extends the development cycle of HFM products, and extremely enlarges the time cost of trial and error, increasing the risks in software construction. Researchers urgently require advanced GPU devices to advance the exploration of HFM, but there is a severe shortage of GPU chips currently \cite{griffith2023desperate}.
\subsubsection{Environment}
Developing such large foundation models incurs a substantial environmental cost \cite{wu2022sustainable}. \textbf{a)} Owing to their extremely large scale, their training on massive data will use immense electricity, which will consume significant one-time environmental costs. The increased carbon emissions \cite{gupta2021chasing} will negatively impact the environment. A study estimates that the carbon emissions generated from the training of a BERT-based model could only be offset by 40 trees over 10 years \cite{bommasani2021opportunities}. \textbf{b)} In the extensive deployment and operation process, directly deploying such enormous models will further entail significant long-term environmental costs \cite{henderson2020towards}. Therefore, reducing environmental impact and fostering sustainable development in foundational model construction and deployment has become an important demand. However, related technologies and policies still lag.

\section{Future Directions}
\label{sec:directions}
\begin{figure*}
    \centering
    \includegraphics[width=1\linewidth]{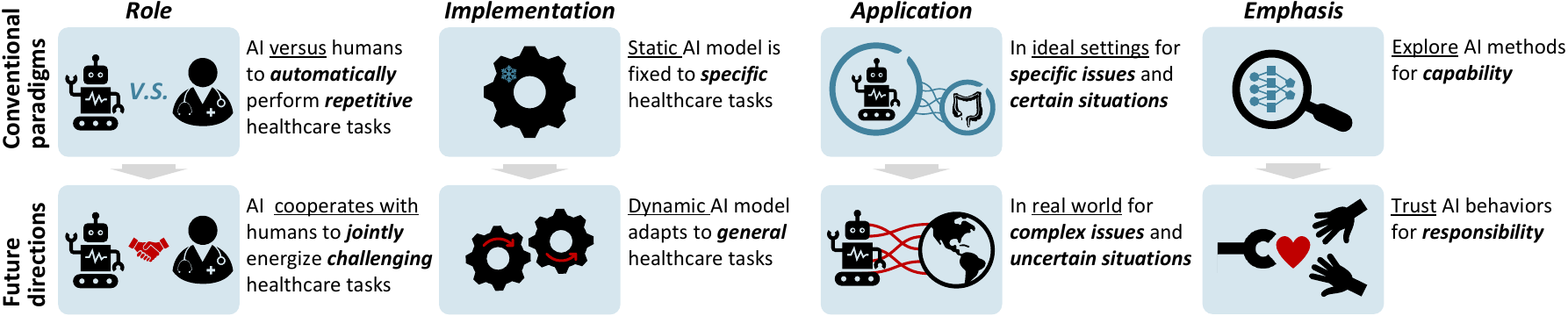}
    \caption{The future directions of the healthcare foundation model. In this paper, we discuss its transformation of role, implementation, applications, and emphasis in the future.}
    \label{fig:future}
\end{figure*}

The development of HFMs represents a progression from specific tasks to general tasks \cite{moor2023foundation}. It enables the AI with a more general capability to address a wide range of requirements and complex environments in the real world. As shown in Fig.\ref{fig:future}, we explore four future directions of HFMs in role, implementation, application, and emphasis.
\subsection{Beyond AI versus Human}
Although the conventional paradigm focuses on utilizing AI to automatically achieve some healthcare tasks instead of human manual work \cite{de2018clinically,esteva2017dermatologist}, AI-human collaboration setups \cite{rajpurkar2022ai} in HFM have demonstrated their opportunities and practical application value. In particular, there are three significant targets of AI-human collaboration in HFM.

\subsubsection{Improving healthcare capabilities}
It targets enabling AI to play a collaborative role with human doctors rather than replacing humans, energizing them to accomplish more challenging healthcare tasks with the support of AI. This creates a collaborative process where AI assists humans in quickly completing the tedious and time-consuming parts of complex tasks, while humans provide professional judgment and correct potential mistakes made by AI. Therefore, it will enable humans and AI to efficiently and accurately accomplish more challenging healthcare tasks. Compared with the conventional AI-independent paradigm and individual human experts, the AI-human collaboration has better capabilities \cite{park2019deep,steiner2018impact}, demonstrating potential in challenging healthcare problems.

\subsubsection{Meeting healthcare requirements}
It targets making AI under the supervision of human experts, thus meeting the healthcare requirements \cite{baeroe2020achieve} in real-world practices. As healthcare practices are closely tied to human life, it is very important to have accountability mechanisms for health-related incidents. Although AI has demonstrated tremendous potential in some clinical scenarios \cite{46-PANDA,esteva2017dermatologist}, even surpassing human experts, it is still difficult to be applied in real life. Because in the event of clinical incidents, independent AI models are unable to be responsible for them. Collaboration with human experts increases people's trust in AI decisions, allowing AI to have greater opportunities for application in clinical scenarios. Potential clinical incidents will also achieve accountability mechanisms, providing the patients with more legal safeguards.

\subsubsection{Optimizing collaboration}It targets improving the collaboration methods with the support of HFM. Based on the prompts from humans, HFMs have the potential to summarize broad and reasonable feedback from the vast knowledge learned from large-scale data \cite{wang2023prompt}. Therefore, it is important to design collaboration methods to effectively mine the vast knowledge inner HFMs and reduce human interaction costs. One of the important future works is the division of labor between humans and AI. Some studies find that misallocation will lead to AI-human collaboration performing worse than independent AI \cite{kim2020changes}, and compared to senior doctors, junior doctors can benefit more from AI \cite{tschandl2020human}. Therefore, it is still unclear how to divide rules in the collaboration to maximize performance in healthcare tasks. The design of prompts is another future work. There are still numerous healthcare scenarios that lack effective prompts or interactive methods. Some prompting strategies like the points or bounding boxes in SAM \cite{Kirillov2023ICCV} are challenging to apply to clinical scenarios where users have to remain focused, such as during surgeries. Therefore, it is important to design more scenario-adapted prompts for more efficient interaction.

\subsection{From Static Model to Dynamic Model}
Although static AI models have demonstrated effectiveness in specific healthcare tasks, real-world healthcare practices always need to coordinate the data with diverse modalities and different clinical requirements from multiple departments. Therefore, constructing dynamic AI models is one of the important future directions \cite{han2021dynamic} in HFM. Specifically,

\subsubsection{Representation capability} Owing to the variation of the data distribution in the real world, the inherent representation capability of HFM is crucial for their adaptation to healthcare scenarios. It is promising to construct powerful dynamic neural network structures, e.g., the recent very famous attention mechanisms (e.g., Transformer) \cite{vaswani2017attention}, mixture-of-experts (e.g., MoE) \cite{shazeer2016outrageously,you2023implicit}, selective state space model (e.g., Mamba) \cite{gu2023mamba}, etc., to represent a wider range of data distributions. Therefore, it will enable the HFMs to dynamically adapt to varied data distributions and clinical situations \cite{han2021dynamic}, boosting the HFMs for healthcare applications with sophisticated designs. What's more, to address the potential challenges in hallucination and outdated knowledge, further studies on continuous learning and model editing \cite{yao2023editing,yi2023towards} to update the representation are also very important in the lifecycle of HFMs.

\subsubsection{Task adaptation capability} To apply the model to varied healthcare scenarios and tasks, adaptation capability is important for HFMs in real-world healthcare practices. An important aspect is reducing the cost of adaptation which uses less data and computation to improve the flexibility of foundational models \cite{tajbakhsh2021guest,wang2021annotation,ding2023parameter}. Once successful, users will be easier to apply these models to their tasks which is essential for HFMs to gain wider applicability. On the other aspect, it should improve the emergence ability of HFMs \cite{bommasani2021opportunities} to stimulate the wealth of knowledge learned from large-scale data \cite{wang2023prompt}. Existing research has shown that carefully designed prompt templates will significantly improve the performance of models on target tasks without any additional training in LFM \cite{kojima2022large}. However, more powerful and flexible prompting methods in other sub-fields are still urgent.

\subsubsection{Scalability} As discussed in Sec.\ref{subsubsec:capability}, the scalability for the downstream devices is important for the deployment of HFMs in resource-limited clinical scenarios. Especially, with a plethora of expensive but computation-limited devices already being operational in medical centers, running HFMs on these devices has become a large challenge. Therefore, it is essential to develop effective scaling methods targeted for foundation models, such as the learngene \cite{wang2022learngene}, dynamically adapting the computation environments and enabling efficient inference on these devices.

\subsection{From Ideal Setting to Complex Real World}
Previous healthcare AI applications \cite{gu2023beyond,jiang2017artificial,rajpurkar2022ai} were set under ideal conditions for specific issues and certain situations that are unable to cope with the complexity and uncertainty of the real world. Therefore, exploring HFMs for real-world healthcare practices has become an important future direction.

\subsubsection{Single-domain to multi-domain} As discussed in Sec.\ref{subsubsec:heter}, the healthcare data suffers from serious heterogeneity challenges due to the variation in populations, regions, and medical centers, i.e., ``domain" \cite{guan2021domain}. Therefore, it makes the HFMs have to be able to learn and generalize to multiple domains for their wide application. The algorithms on domain adaptation \cite{guan2021domain} and domain generalization \cite{baliah2023exploring} should be further studied for their practice on foundational models to adapt to the heterogeneity. Their effectiveness in real-world healthcare applications still needs validation. Moreover, due to the privacy of healthcare data, there is a growing interest in leveraging federated learning to construct privacy-preserved large-scale cross-domain learning systems for HFMs \cite{tan2022federated,zhuang2023foundation}. It takes distributed training mechanisms to enable HFMs to learn healthcare data in various domains under protected situations without the risk of privacy breaches. However, it is extremely challenging to conduct such large-scale distributed training (large parameter and data amount) for foundational models.

\subsubsection{Single/closed-task to multi/open-task} Due to the diversity of healthcare scenarios, there is a practical requirement for the generalist models \cite{moor2023foundation} which are able to meet a wide range of healthcare tasks in the real world. Different from the conventional paradigm which is designed for single tasks, e.g., certain disease diagnoses, HFMs face open-world healthcare scenarios with multiple tasks, such as the various organs, diseases, clinical objectives, etc. Therefore, HFMs need to acquire more powerful multi-task capabilities for different healthcare scenarios. Some dynamic model techniques, such as MoE \cite{shazeer2016outrageously}, have shown their effectiveness in multi-task prediction of foundation models, showing a potential way towards the generalist AI \cite{zhu2022uni}. On the other hand, the uncertainty of the real world further introduces the open-set problem \cite{geng2020recent} for HFMs which is particularly crucial because irresponsible predictions for healthcare tasks could jeopardize human life safety. For potentially uncontrollable inputs, HFMs have to establish mechanisms to handle the requirements beyond their capability for reasonable and safe prediction results in healthcare \cite{rawte2023survey}. 

\subsubsection{Single-modality to multi-modality} In the real world, healthcare scenarios involve multiple modalities simultaneously for healthcare practices \cite{gu2023beyond} demonstrating a great potential to construct the HFMs in a multimodal uniform setting. Compared with the conventional single-modality paradigm, the multimodal setting will incorporate the representative features from different modalities, enabling the model to achieve precise and reliable results \cite{acosta2022multimodal}. As discussed in Sec.\ref{subsec:method:multimodal}, although HFMs have achieved preliminary success in the multimodal setting, most efforts still focus on language and vision modalities. Therefore, it is still an open problem to integrate more modalities in real healthcare practices into the HFMs. In addition, learning algorithms for multimodal data have emerged as a topic of interest. Some existing studies tried to stimulate the learning ability between modalities via cross-modality generation \cite{luo2023biomedgpt}, cross-modality self-supervised learning \cite{zhao2023clip}, multi-modality knowledge distillation \cite{li2023scaling}, etc. However, it is a long-term problem on how to leverage the advantages and complementarities when more modalities are enrolled in large-scale HFMs' learning. The challenge in missing modality \cite{ma2022multimodal} has also raised concerns in multimodal settings. Because the real cases are in different situations, such as different diseases and treatment plans, data will be collected from different combinations of modalities. Therefore, multimodal HFMs will be unable to access complete sets of data modalities for inference, requiring adaptability for varied combinations.

\subsection{From Exploration to Trust}
As foundational models evolve the role, implementation, and application of AI in healthcare, people's emphasis will also progress from the exploration of their capability to the trust of their behaviors. As discussed in Sec.\ref{sec:challenges}, it is still challenging to confidently and deeply apply HFMs in healthcare, making an urgent future direction. Here, we discuss three important aspects.

\subsubsection{Explainable HFM} As illustrated in Sec.\ref{subsubsec:trust}, the ``blackbox" property of the neural networks makes people difficult to understand their behavior.  Therefore, explaining the intrinsic intention behind the results of HFMs is crucial for people to trust their behavior in healthcare \cite{sokol2020one}. One of the future tasks is promoting the machine learning theoretical research \cite{mohri2018foundations} on foundational models. Analyzing the machine learning properties of foundational models will reveal their unique patterns, providing the researchers with insights to design reasonable models and enhance the research and development efficiency. However, the existing theory of representation, optimization, or generalization is unable for HFMs, because their enormous parameters and data amount extremely exceed the ideal settings for these theories \cite{yuan2023power}. Another promising direction is the discovery of more effective evidence for the explanation. Although the existing works utilized some heatmaps \cite{sokol2020one}, including attention maps, class activation maps, uncertainty maps, etc., for the explanation of individual behavior, it is still urgent to further explore higher abstraction levels of explanatory evidence. In addition, further leveraging explainable HFMs for scientific exploration, such as drug discovery \cite{jimenez2020drug}, holds potential as a future direction.

\subsubsection{Secure HFM} The security of HFMs is the foundation upon which people trust and use them, making one of the important future directions \cite{qayyum2020secure}. One aspect of the security is the ability of HFMs to against external attacks. For example, some jailbreaking attacks \cite{wei2024jailbroken} will result in the output of LFMs containing private and sensitive information, posing a threat to data providers. Therefore, defense mechanisms, like adversarial attack \cite{schlarmann2023adversarial}, should be established for HFMs to deal with potential malicious users and protect the lifecycle of HFMs. Another aspect is the reliability of HFM itself. In healthcare tasks, only reliable outputs are worthy of trust owing to their close relationship with human life. Therefore, in addition to the construction of exploration \cite{sokol2020one} methods for the measurement of reliability, it is also essential to introduce more data and design more powerful methods to enhance the robustness and accuracy. Then, it should construct reasonable accountability mechanisms \cite{habli2020artificial} during the application of HFMs to enhance the caution of users and the legal security in healthcare practices.

\subsubsection{Sustainable HFM} The further in-depth development of HFMs requires a sustainable way \cite{vinuesa2020role,wu2022sustainable}. Potential energy and environmental crises are occurring alongside the development of foundational models. The large-scale training causes massive power consumption and carbon emissions resulting in the development of HFMs based on the expense of the environment unsustainable \cite{kaack2022aligning}. Therefore, there is an urgent direction to research low-power foundational model training and deployment strategies, including the construction of greener chip technologies and model architectures. With the expansion of the application scope of HFM, the cost problem is another focus to limit sustainability (discussed in Sec.\ref{subsubsec:challenge:data:cost} and Sec.\ref{subsubsec:challenge:computing:computation}). Therefore, further study of the efficient learning \cite{menghani2023efficient} algorithms and hardware facilities is important in the future development of HFM. Reducing the costs including the data collection and processing \cite{willemink2020preparing,wang2021annotation,tajbakhsh2021guest}, model training and inference \cite{ding2023parameter}, will stimulate the commercial advantages of foundational models, thereby enhancing their sustainability.

\section{Conclusion}
\label{sec:conclusion}
This survey gives a potential answer to the question that ``Can we construct AI models to benefit a variety of healthcare tasks?" More healthcare practices will benefit from the development of HFM achieving advanced intelligent healthcare services. Although HFM is gradually showing its great application value, it still lacks clear recognition about what are the challenges,  new opportunities from HFM, and potential future directions for the foundation models in healthcare practices. This paper first presented a comprehensive overview and analysis of the HFMs including the methods, data, and applications that help to understand the current progress of HFM. Then, we provide an in-depth discussion and prospects for key challenges in data, algorithms, and computing infrastructure, illustrating the current shortcomings of the HFM. Finally, we look forward to the future directions in the role, implementation, application, and emphasis, highlighting the future perspectives that hold promise in advancing the field.

\section*{Acknowledgment}
This work was supported by the Hong Kong Innovation and Technology Fund (Project No. No. MHP/002/22 and No. PRP/034/22FX), Shenzhen Science and Technology Innovation Committee Fund (Project No. SGDX20210823103201011), the Pneumoconiosis Compensation Fund Board, HKSARS (Project No. PCFB22EG01) and Research Grants Council of the Hong Kong Special Administrative Region, China (Project No. R6003-22 and C4024-22GF).

\bibliographystyle{IEEEtran}
\bibliography{bibtotal_shortjournal_shortauthors}

\begin{thebibliography}{100}
\providecommand{\url}[1]{#1}
\csname url@samestyle\endcsname
\providecommand{\newblock}{\relax}
\providecommand{\bibinfo}[2]{#2}
\providecommand{\BIBentrySTDinterwordspacing}{\spaceskip=0pt\relax}
\providecommand{\BIBentryALTinterwordstretchfactor}{4}
\providecommand{\BIBentryALTinterwordspacing}{\spaceskip=\fontdimen2\font plus
\BIBentryALTinterwordstretchfactor\fontdimen3\font minus \fontdimen4\font\relax}
\providecommand{\BIBforeignlanguage}[2]{{%
\expandafter\ifx\csname l@#1\endcsname\relax
\typeout{** WARNING: IEEEtran.bst: No hyphenation pattern has been}%
\typeout{** loaded for the language `#1'. Using the pattern for}%
\typeout{** the default language instead.}%
\else
\language=\csname l@#1\endcsname
\fi
#2}}
\providecommand{\BIBdecl}{\relax}
\BIBdecl

\bibitem{nilsson1982principles}
N.~J. Nilsson, \emph{Principles of artificial intelligence}.\hskip 1em plus 0.5em minus 0.4em\relax Springer Science \& Business Media, 1982.

\bibitem{lecun2015deep}
Y.~LeCun \emph{et~al.}, ``Deep learning,'' \emph{Nature}, vol. 521, no. 7553, pp. 436--444, 2015.

\bibitem{gu2023beyond}
X.~Gu \emph{et~al.}, ``Beyond supervised learning for pervasive healthcare,'' \emph{IEEE Rev. Biomed. Eng.}, 2023.

\bibitem{jiang2017artificial}
F.~Jiang \emph{et~al.}, ``Artificial intelligence in healthcare: past, present and future,'' \emph{Stroke and vascular neurology}, vol.~2, no.~4, 2017.

\bibitem{rajpurkar2022ai}
P.~Rajpurkar \emph{et~al.}, ``Ai in health and medicine,'' \emph{Nat. Med.}, vol.~28, no.~1, pp. 31--38, 2022.

\bibitem{46-PANDA}
K.~Cao \emph{et~al.}, ``Large-scale pancreatic cancer detection via non-contrast ct and deep learning,'' \emph{Nat. Med.}, pp. 1--11, 2023.

\bibitem{de2018clinically}
J.~De~Fauw \emph{et~al.}, ``Clinically applicable deep learning for diagnosis and referral in retinal disease,'' \emph{Nat. Med.}, vol.~24, no.~9, pp. 1342--1350, 2018.

\bibitem{esteva2017dermatologist}
A.~Esteva \emph{et~al.}, ``Dermatologist-level classification of skin cancer with deep neural networks,'' \emph{Nature}, vol. 542, no. 7639, pp. 115--118, 2017.

\bibitem{bommasani2021opportunities}
R.~Bommasani \emph{et~al.}, ``On the opportunities and risks of foundation models,'' \emph{arXiv preprint arXiv:2108.07258}, 2021.

\bibitem{moor2023foundation}
M.~Moor \emph{et~al.}, ``Foundation models for generalist medical artificial intelligence,'' \emph{Nature}, vol. 616, no. 7956, pp. 259--265, 2023.

\bibitem{azad2023foundational}
B.~Azad \emph{et~al.}, ``Foundational models in medical imaging: A comprehensive survey and future vision,'' \emph{arXiv preprint arXiv:2310.18689}, 2023.

\bibitem{qiu2023large}
J.~Qiu \emph{et~al.}, ``Large ai models in health informatics: Applications, challenges, and the future,'' \emph{IEEE J. Biomed. Health Inform.}, 2023.

\bibitem{thirunavukarasu2023large}
A.~J. Thirunavukarasu \emph{et~al.}, ``Large language models in medicine,'' \emph{Nat. Med.}, vol.~29, no.~8, pp. 1930--1940, 2023.

\bibitem{he2023survey}
K.~He \emph{et~al.}, ``A survey of large language models for healthcare: from data, technology, and applications to accountability and ethics,'' \emph{arXiv preprint arXiv:2310.05694}, 2023.

\bibitem{yang2022large}
X.~Yang \emph{et~al.}, ``A large language model for electronic health records,'' \emph{NPJ Digit. Med.}, vol.~5, no.~1, p. 194, 2022.

\bibitem{singhal2023large}
K.~Singhal \emph{et~al.}, ``Large language models encode clinical knowledge,'' \emph{Nature}, vol. 620, no. 7972, pp. 172--180, 2023.

\bibitem{li2023d}
Z.~Li \emph{et~al.}, ``D-lmbmap: a fully automated deep-learning pipeline for whole-brain profiling of neural circuitry,'' \emph{Nat. Methods}, vol.~20, no.~10, pp. 1593--1604, 2023.

\bibitem{4-Endo-FM}
Z.~Wang \emph{et~al.}, ``Foundation model for endoscopy video analysis via large-scale self-supervised pre-train,'' in \emph{Proc. Int. Conf. Med. Image Comput. Comput.-Assisted Intervention}.\hskip 1em plus 0.5em minus 0.4em\relax Springer, 2023, pp. 101--111.

\bibitem{3-RETFound}
Y.~Zhou \emph{et~al.}, ``A foundation model for generalizable disease detection from retinal images,'' \emph{Nature}, vol. 622, no. 7981, pp. 156--163, 2023.

\bibitem{Kirillov2023ICCV}
A.~Kirillov \emph{et~al.}, ``Segment anything,'' in \emph{Proc. IEEE Int. Conf. Comput. Vis.}, October 2023, pp. 4015--4026.

\bibitem{rombach2022high}
R.~Rombach \emph{et~al.}, ``High-resolution image synthesis with latent diffusion models,'' in \emph{Proc. IEEE Conf. Comput. Vis. Pattern Recognit.}, 2022, pp. 10\,684--10\,695.

\bibitem{shen2024omnina}
X.~Shen and X.~Li, ``Omnina: A foundation model for nucleotide sequences,'' \emph{bioRxiv}, pp. 2024--01, 2024.

\bibitem{dalla2023nucleotide}
H.~Dalla-Torre \emph{et~al.}, ``The nucleotide transformer: Building and evaluating robust foundation models for human genomics,'' \emph{bioRxiv}, pp. 2023--01, 2023.

\bibitem{brandes2022proteinbert}
N.~Brandes \emph{et~al.}, ``Proteinbert: a universal deep-learning model of protein sequence and function,'' \emph{Bioinformatics}, vol.~38, no.~8, pp. 2102--2110, 2022.

\bibitem{jumper2021highly}
\BIBentryALTinterwordspacing
J.~Jumper \emph{et~al.}, ``Highly accurate protein structure prediction with alphafold,'' \emph{Nature}, vol. 596, no. 7873, pp. 583--589, 2021. [Online]. Available: \url{https://doi.org/10.1038/s41586-021-03819-2}
\BIBentrySTDinterwordspacing

\bibitem{chen2022interpretable}
J.~Chen \emph{et~al.}, ``Interpretable rna foundation model from unannotated data for highly accurate rna structure and function predictions,'' \emph{bioRxiv}, pp. 2022--08, 2022.

\bibitem{wu2023towards}
C.~Wu \emph{et~al.}, ``Towards generalist foundation model for radiology,'' \emph{arXiv preprint arXiv:2308.02463}, 2023.

\bibitem{fei2022towards}
N.~Fei \emph{et~al.}, ``Towards artificial general intelligence via a multimodal foundation model,'' \emph{Nat. Commun.}, vol.~13, no.~1, p. 3094, 2022.

\bibitem{zhang2023biomedgpt}
K.~Zhang \emph{et~al.}, ``Biomedgpt: A unified and generalist biomedical generative pre-trained transformer for vision, language, and multimodal tasks,'' \emph{arXiv preprint arXiv:2305.17100}, 2023.

\bibitem{acosta2022multimodal}
J.~N. Acosta \emph{et~al.}, ``Multimodal biomedical ai,'' \emph{Nat. Med.}, vol.~28, no.~9, pp. 1773--1784, 2022.

\bibitem{tu2023towards}
T.~Tu \emph{et~al.}, ``Towards generalist biomedical ai,'' \emph{NEJM AI}, vol.~1, no.~3, p. AIoa2300138, 2024.

\bibitem{shrestha2023medical}
P.~Shrestha \emph{et~al.}, ``Medical vision language pretraining: A survey,'' \emph{arXiv preprint arXiv:2312.06224}, 2023.

\bibitem{willemink2020preparing}
M.~J. Willemink \emph{et~al.}, ``Preparing medical imaging data for machine learning,'' \emph{Radiology}, vol. 295, no.~1, pp. 4--15, 2020.

\bibitem{hatherley2020limits}
J.~J. Hatherley, ``Limits of trust in medical ai,'' \emph{Journal of medical ethics}, 2020.

\bibitem{markus2021role}
A.~F. Markus \emph{et~al.}, ``The role of explainability in creating trustworthy artificial intelligence for health care: a comprehensive survey of the terminology, design choices, and evaluation strategies,'' \emph{Journal of biomedical informatics}, vol. 113, p. 103655, 2021.

\bibitem{wu2022sustainable}
C.-J. Wu \emph{et~al.}, ``Sustainable ai: Environmental implications, challenges and opportunities,'' \emph{Proceedings of Machine Learning and Systems}, vol.~4, pp. 795--813, 2022.

\bibitem{kenton2019bert}
J.~Devlin \emph{et~al.}, ``Bert: Pre-training of deep bidirectional transformers for language understanding,'' in \emph{Proceedings of NAACL-HLT}, 2019, pp. 4171--4186.

\bibitem{lee2020biobert}
\BIBentryALTinterwordspacing
J.~Lee \emph{et~al.}, ``Biobert: a pre-trained biomedical language representation model for biomedical text mining,'' \emph{Bioinformatics}, vol.~36, no.~4, pp. 1234--1240, 09 2019. [Online]. Available: \url{https://doi.org/10.1093/bioinformatics/btz682}
\BIBentrySTDinterwordspacing

\bibitem{patel2023chatgpt}
S.~B. Patel and K.~Lam, ``Chatgpt: the future of discharge summaries?'' \emph{The Lancet Digital Health}, vol.~5, no.~3, pp. e107--e108, 2023.

\bibitem{he2023geometric}
Y.~He \emph{et~al.}, ``Geometric visual similarity learning in 3d medical image self-supervised pre-training,'' in \emph{Proc. IEEE Conf. Comput. Vis. Pattern Recognit.}, 2023, pp. 9538--9547.

\bibitem{15-MG}
Z.~Zhou \emph{et~al.}, ``Models genesis,'' \emph{Med. Image Anal.}, vol.~67, p. 101840, 2021.

\bibitem{22-MedSAM}
J.~Ma \emph{et~al.}, ``Segment anything in medical images,'' \emph{Nat. Commun.}, vol.~15, no.~1, p. 654, 2024.

\bibitem{mazurowski2023segment}
M.~A. Mazurowski \emph{et~al.}, ``Segment anything model for medical image analysis: an experimental study,'' \emph{Med. Image Anal.}, vol.~89, p. 102918, 2023.

\bibitem{DINOv2forrad}
M.~Baharoon \emph{et~al.}, ``Towards general purpose vision foundation models for medical image analysis: An experimental study of dinov2 on radiology benchmarks,'' \emph{arXiv preprint arXiv:2312.02366}, 2023.

\bibitem{wang2023uni}
X.~Wang \emph{et~al.}, ``Uni-rna: universal pre-trained models revolutionize rna research,'' \emph{bioRxiv}, pp. 2023--07, 2023.

\bibitem{ji2021dnabert}
\BIBentryALTinterwordspacing
Y.~Ji \emph{et~al.}, ``Dnabert: pre-trained bidirectional encoder representations from transformers model for dna-language in genome,'' \emph{Bioinformatics}, vol.~37, no.~15, pp. 2112--2120, 8 2021. [Online]. Available: \url{https://doi.org/10.1093/bioinformatics/btab083}
\BIBentrySTDinterwordspacing

\bibitem{radford2021learning}
A.~Radford \emph{et~al.}, ``Learning transferable visual models from natural language supervision,'' in \emph{Proc. Int. Conf. Mach. Learn.}\hskip 1em plus 0.5em minus 0.4em\relax PMLR, 2021, pp. 8748--8763.

\bibitem{zhao2023clip}
Z.~Zhao \emph{et~al.}, ``Clip in medical imaging: A comprehensive survey,'' \emph{arXiv preprint arXiv:2312.07353}, 2023.

\bibitem{wang2023pre}
B.~Wang \emph{et~al.}, ``Pre-trained language models in biomedical domain: A systematic survey,'' \emph{ACM Computing Surveys}, vol.~56, no.~3, pp. 1--52, 2023.

\bibitem{zhou2023survey}
H.~Zhou, B.~Gu, X.~Zou, Y.~Li, S.~S. Chen, P.~Zhou, J.~Liu, Y.~Hua, C.~Mao, X.~Wu \emph{et~al.}, ``A survey of large language models in medicine: Progress, application, and challenge,'' \emph{arXiv preprint arXiv:2311.05112}, 2023.

\bibitem{yuan2023large}
M.~Yuan \emph{et~al.}, ``Large language models illuminate a progressive pathway to artificial healthcare assistant: A review,'' \emph{arXiv preprint arXiv:2311.01918}, 2023.

\bibitem{lee2024foundation}
H.~H. Lee \emph{et~al.}, ``Foundation models for biomedical image segmentation: A survey,'' \emph{arXiv preprint arXiv:2401.07654}, 2024.

\bibitem{li2024progress}
Q.~Li \emph{et~al.}, ``Progress and opportunities of foundation models in bioinformatics,'' \emph{arXiv preprint arXiv:2402.04286}, 2024.

\bibitem{liu2024large}
J.~Liu \emph{et~al.}, ``Large language models in bioinformatics: applications and perspectives,'' \emph{arXiv preprint arXiv:2401.04155}, 2024.

\bibitem{qiu2023pre}
Y.~Qiu \emph{et~al.}, ``Pre-training in medical data: A survey,'' \emph{Machine Intelligence Research}, vol.~20, no.~2, pp. 147--179, 2023.

\bibitem{wang2023accelerating}
D.-Q. Wang \emph{et~al.}, ``Accelerating the integration of chatgpt and other large-scale ai models into biomedical research and healthcare,'' \emph{MedComm--Future Medicine}, vol.~2, no.~2, p. e43, 2023.

\bibitem{ZHANG2024102996}
S.~Zhang and D.~Metaxas, ``On the challenges and perspectives of foundation models for medical image analysis,'' \emph{Med. Image Anal.}, vol.~91, p. 102996, 2024.

\bibitem{zhang2024data}
Y.~Zhang \emph{et~al.}, ``Data-centric foundation models in computational healthcare: A survey,'' \emph{arXiv preprint arXiv:2401.02458}, 2024.

\bibitem{radford2019language}
A.~Radford \emph{et~al.}, ``Language models are unsupervised multitask learners,'' \emph{OpenAI blog}, vol.~1, no.~8, p.~9, 2019.

\bibitem{jin2023medcpt}
Q.~Jin \emph{et~al.}, ``Medcpt: Contrastive pre-trained transformers with large-scale pubmed search logs for zero-shot biomedical information retrieval,'' \emph{Bioinformatics}, vol.~39, no.~11, p. btad651, 2023.

\bibitem{peng2023study}
\BIBentryALTinterwordspacing
C.~Peng \emph{et~al.}, ``A study of generative large language model for medical research and healthcare,'' \emph{NPJ Digit. Med.}, vol.~6, 2023. [Online]. Available: \url{https://doi.org/10.1038/s41746-023-00958-w}
\BIBentrySTDinterwordspacing

\bibitem{gu2021domain}
Y.~Gu \emph{et~al.}, ``Domain-specific language model pretraining for biomedical natural language processing,'' \emph{ACM Transactions on Computing for Healthcare}, vol.~3, no.~1, pp. 1--23, 2021.

\bibitem{wang2023huatuo}
H.~Wang \emph{et~al.}, ``Huatuo: Tuning llama model with chinese medical knowledge,'' \emph{arXiv preprint arXiv:2304.06975}, 2023.

\bibitem{zhang2023huatuogpt}
H.~Zhang \emph{et~al.}, ``Huatuogpt, towards taming language model to be a doctor,'' in \emph{Findings of the Association for Computational Linguistics: EMNLP 2023}, 2023, pp. 10\,859--10\,885.

\bibitem{wu2023pmc}
C.~Wu \emph{et~al.}, ``Pmc-llama: Towards building open-source language models for medicine,'' \emph{arXiv preprint arXiv:2305.10415}, vol.~6, 2023.

\bibitem{chen2023huatuogpt}
J.~Chen \emph{et~al.}, ``Huatuogpt-ii, one-stage training for medical adaption of llms,'' \emph{arXiv preprint arXiv:2311.09774}, 2023.

\bibitem{chen2023meditron}
Z.~Chen \emph{et~al.}, ``Meditron-70b: Scaling medical pretraining for large language models,'' \emph{arXiv preprint arXiv:2311.16079}, 2023.

\bibitem{zhang2023alpacare}
X.~Zhang \emph{et~al.}, ``Alpacare: Instruction-tuned large language models for medical application,'' \emph{arXiv preprint arXiv:2310.14558}, 2023.

\bibitem{chen2023bianque}
Y.~Chen \emph{et~al.}, ``Bianque: Balancing the questioning and suggestion ability of health llms with multi-turn health conversations polished by chatgpt,'' \emph{arXiv preprint arXiv:2310.15896}, 2023.

\bibitem{li2023chatdoctor}
Y.~Li \emph{et~al.}, ``Chatdoctor: A medical chat model fine-tuned on a large language model meta-ai (llama) using medical domain knowledge,'' \emph{Cureus}, vol.~15, no.~6, 2023.

\bibitem{han2023medalpaca}
T.~Han \emph{et~al.}, ``Medalpaca--an open-source collection of medical conversational ai models and training data,'' \emph{arXiv preprint arXiv:2304.08247}, 2023.

\bibitem{ye2023qilin}
Q.~Ye \emph{et~al.}, ``Qilin-med: Multi-stage knowledge injection advanced medical large language model,'' \emph{arXiv preprint arXiv:2310.09089}, 2023.

\bibitem{luo2023taiyi}
L.~Luo \emph{et~al.}, ``Taiyi: a bilingual fine-tuned large language model for diverse biomedical tasks,'' \emph{Journal of the American Medical Informatics Association}, p. ocae037, 02 2024.

\bibitem{wang2023gpt}
W.~Wang \emph{et~al.}, ``Gpt-doctor: Customizing large language models for medical consultation,'' \emph{arXiv preprint arXiv:2312.10225}, 2023.

\bibitem{xiong2023doctorglm}
H.~Xiong \emph{et~al.}, ``Doctorglm: Fine-tuning your chinese doctor is not a herculean task,'' \emph{arXiv preprint arXiv:2304.01097}, 2023.

\bibitem{wang2023clinicalgpt}
G.~Wang \emph{et~al.}, ``Clinicalgpt: Large language models finetuned with diverse medical data and comprehensive evaluation,'' \emph{arXiv preprint arXiv:2306.09968}, 2023.

\bibitem{li2023beginner}
Q.~Li \emph{et~al.}, ``From beginner to expert: Modeling medical knowledge into general llms,'' \emph{arXiv preprint arXiv:2312.01040}, 2023.

\bibitem{labrak2024biomistral}
Y.~Labrak \emph{et~al.}, ``Biomistral: A collection of open-source pretrained large language models for medical domains,'' \emph{arXiv preprint arXiv:2402.10373}, 2024.

\bibitem{touvron2023llama}
H.~Touvron \emph{et~al.}, ``Llama: Open and efficient foundation language models,'' \emph{arXiv preprint arXiv:2302.13971}, 2023.

\bibitem{alsentzer2019publicly}
E.~Alsentzer \emph{et~al.}, ``Publicly available clinical {BERT} embeddings,'' in \emph{Proceedings of the 2nd Clinical Natural Language Processing Workshop}.\hskip 1em plus 0.5em minus 0.4em\relax Minneapolis, Minnesota, USA: Association for Computational Linguistics, Jun. 2019, pp. 72--78.

\bibitem{chen2020modified}
Y.-P. Chen \emph{et~al.}, ``Modified bidirectional encoder representations from transformers extractive summarization model for hospital information systems based on character-level tokens (alphabert): development and performance evaluation,'' \emph{JMIR medical informatics}, vol.~8, no.~4, p. e17787, 2020.

\bibitem{li2020behrt}
Y.~Li \emph{et~al.}, ``Behrt: transformer for electronic health records,'' \emph{Scientific reports}, vol.~10, no.~1, p. 7155, 2020.

\bibitem{yuan-etal-2022-biobart}
\BIBentryALTinterwordspacing
H.~Yuan \emph{et~al.}, ``{B}io{BART}: Pretraining and evaluation of a biomedical generative language model,'' in \emph{Proceedings of the 21st Workshop on Biomedical Language Processing}.\hskip 1em plus 0.5em minus 0.4em\relax Dublin, Ireland: Association for Computational Linguistics, May 2022, pp. 97--109. [Online]. Available: \url{https://aclanthology.org/2022.bionlp-1.9}
\BIBentrySTDinterwordspacing

\bibitem{yang2024zhongjing}
S.~Yang \emph{et~al.}, ``Zhongjing: Enhancing the chinese medical capabilities of large language model through expert feedback and real-world multi-turn dialogue,'' in \emph{Proceedings of the AAAI Conference on Artificial Intelligence}, vol.~38, no.~17, 2024, pp. 19\,368--19\,376.

\bibitem{xie2024me}
Q.~Xie \emph{et~al.}, ``Me llama: Foundation large language models for medical applications,'' \emph{arXiv preprint arXiv:2402.12749}, 2024.

\bibitem{jia2024oncogpt}
F.~Jia \emph{et~al.}, ``Oncogpt: A medical conversational model tailored with oncology domain expertise on a large language model meta-ai (llama),'' \emph{arXiv preprint arXiv:2402.16810}, 2024.

\bibitem{wang2024jmlr}
J.~Wang \emph{et~al.}, ``Jmlr: Joint medical llm and retrieval training for enhancing reasoning and professional question answering capability,'' \emph{arXiv preprint arXiv:2402.17887}, 2024.

\bibitem{singhal2023publisher}
Singhal \emph{et~al.}, ``Large language models encode clinical knowledge,'' \emph{Nature}, vol. 620, no. 7973, p. E19, 2023.

\bibitem{singhal2023towards}
K.~Singhal \emph{et~al.}, ``Towards expert-level medical question answering with large language models,'' \emph{arXiv preprint arXiv:2305.09617}, 2023.

\bibitem{pieri2024bimedix}
S.~Pieri \emph{et~al.}, ``Bimedix: Bilingual medical mixture of experts llm,'' \emph{arXiv preprint arXiv:2402.13253}, 2024.

\bibitem{shu2023visual}
C.~Shu, B.~Chen, F.~Liu, Z.~Fu, E.~Shareghi, and N.~Collier, ``Visual med-alpaca: A parameter-efficient biomedical llm with visual capabilities,'' 2023.

\bibitem{gao2023ophglm}
W.~Gao \emph{et~al.}, ``Ophglm: Training an ophthalmology large language-and-vision assistant based on instructions and dialogue,'' \emph{arXiv preprint arXiv:2306.12174}, 2023.

\bibitem{wang2023chatcad}
S.~Wang \emph{et~al.}, ``Chatcad: Interactive computer-aided diagnosis on medical image using large language models,'' \emph{arXiv preprint arXiv:2302.07257}, 2023.

\bibitem{zhao2023chatcad+}
Z.~Zhao \emph{et~al.}, ``Chatcad+: Towards a universal and reliable interactive cad using llms,'' \emph{arXiv preprint arXiv:2305.15964}, 2023.

\bibitem{liu2023deid}
Z.~Liu \emph{et~al.}, ``Deid-gpt: Zero-shot medical text de-identification by gpt-4,'' \emph{arXiv preprint arXiv:2303.11032}, 2023.

\bibitem{gao2023leveraging}
Y.~Gao \emph{et~al.}, ``Leveraging a medical knowledge graph into large language models for diagnosis prediction,'' \emph{arXiv preprint arXiv:2308.14321}, 2023.

\bibitem{nori2023can}
H.~Nori \emph{et~al.}, ``Can generalist foundation models outcompete special-purpose tuning? case study in medicine,'' \emph{Medicine}, vol.~84, no. 88.3, pp. 77--3, 2023.

\bibitem{sivarajkumar2022healthprompt}
S.~Sivarajkumar and Y.~Wang, ``Healthprompt: A zero-shot learning paradigm for clinical natural language processing,'' in \emph{AMIA Annual Symposium Proceedings}, vol. 2022.\hskip 1em plus 0.5em minus 0.4em\relax American Medical Informatics Association, 2022, p. 972.

\bibitem{tang2023medagents}
X.~Tang \emph{et~al.}, ``Medagents: Large language models as collaborators for zero-shot medical reasoning,'' \emph{arXiv preprint arXiv:2311.10537}, 2023.

\bibitem{elfrink2023soft}
A.~Elfrink \emph{et~al.}, ``Soft-prompt tuning to predict lung cancer using primary care free-text dutch medical notes,'' in \emph{International Conference on Artificial Intelligence in Medicine}.\hskip 1em plus 0.5em minus 0.4em\relax Springer, 2023, pp. 193--198.

\bibitem{abaho2022position}
M.~Abaho \emph{et~al.}, ``Position-based prompting for health outcome generation,'' in \emph{Proceedings of the 21st Workshop on Biomedical Language Processing}, 2022, pp. 26--36.

\bibitem{lee2023clinical}
S.~Lee \emph{et~al.}, ``Clinical decision transformer: Intended treatment recommendation through goal prompting,'' \emph{arXiv preprint arXiv:2302.00612}, 2023.

\bibitem{byambasuren2019preliminary}
O.~Byambasuren \emph{et~al.}, ``Preliminary study on the construction of chinese medical knowledge graph,'' \emph{Journal of Chinese Information Processing}, vol.~33, no.~10, pp. 1--9, 2019.

\bibitem{jin2021disease}
D.~Jin \emph{et~al.}, ``What disease does this patient have? a large-scale open domain question answering dataset from medical exams,'' \emph{Applied Sciences}, vol.~11, no.~14, p. 6421, 2021.

\bibitem{lindberg1993unified}
D.~A. Lindberg \emph{et~al.}, ``The unified medical language system,'' \emph{Yearbook of medical informatics}, vol.~2, no.~01, pp. 41--51, 1993.

\bibitem{10.1145/3649449}
\BIBentryALTinterwordspacing
J.~Li \emph{et~al.}, ``Pre-trained language models for text generation: A survey,'' \emph{ACM Comput. Surv.}, mar 2024, just Accepted. [Online]. Available: \url{https://doi.org/10.1145/3649449}
\BIBentrySTDinterwordspacing

\bibitem{hu2021lora}
E.~J. Hu \emph{et~al.}, ``Lora: Low-rank adaptation of large language models,'' in \emph{Proc. Int. Conf. Learn. Represent.}, 2021.

\bibitem{wang2023prompt}
J.~Wang \emph{et~al.}, ``Prompt engineering for healthcare: Methodologies and applications,'' \emph{arXiv preprint arXiv:2304.14670}, 2023.

\bibitem{wei2022chain}
J.~Wei \emph{et~al.}, ``Chain-of-thought prompting elicits reasoning in large language models,'' \emph{Proc. Adv. Neural Inf. Process. Syst.}, vol.~35, pp. 24\,824--24\,837, 2022.

\bibitem{beltagy2019scibert}
I.~Beltagy \emph{et~al.}, ``Scibert: A pretrained language model for scientific text,'' in \emph{Proceedings of the 2019 Conference on Empirical Methods in Natural Language Processing and the 9th International Joint Conference on Natural Language Processing (EMNLP-IJCNLP)}, 2019, pp. 3615--3620.

\bibitem{verkijk2021medroberta}
S.~Verkijk and P.~Vossen, ``Medroberta. nl: a language model for dutch electronic health records,'' \emph{Computational Linguistics in the Netherlands Journal}, vol.~11, pp. 141--159, 2021.

\bibitem{awais2023foundational}
M.~Awais \emph{et~al.}, ``Foundational models defining a new era in vision: A survey and outlook,'' \emph{arXiv preprint arXiv:2307.13721}, 2023.

\bibitem{MoCo}
K.~He \emph{et~al.}, ``Momentum contrast for unsupervised visual representation learning,'' in \emph{Proc. IEEE Conf. Comput. Vis. Pattern Recognit.}, 2020, pp. 9729--9738.

\bibitem{ma2023towards}
J.~Ma and B.~Wang, ``Towards foundation models of biological image segmentation,'' \emph{Nat. Methods}, vol.~20, no.~7, pp. 953--955, 2023.

\bibitem{chen2019med3d}
S.~Chen \emph{et~al.}, ``Med3d: Transfer learning for 3d medical image analysis,'' \emph{arXiv preprint arXiv:1904.00625}, 2019.

\bibitem{5-STU-Net}
Z.~Huang \emph{et~al.}, ``Stu-net: Scalable and transferable medical image segmentation models empowered by large-scale supervised pre-training,'' \emph{arXiv preprint arXiv:2304.06716}, 2023.

\bibitem{TotalSegmentator}
J.~Wasserthal \emph{et~al.}, ``Totalsegmentator: Robust segmentation of 104 anatomic structures in ct images,'' \emph{Radiology: Artificial Intelligence}, vol.~5, no.~5, 2023.

\bibitem{71-UniverSeg}
V.~I. Butoi \emph{et~al.}, ``Universeg: Universal medical image segmentation,'' in \emph{Proceedings of the IEEE/CVF International Conference on Computer Vision}, 2023.

\bibitem{30-SAM-Med3D}
H.~Wang \emph{et~al.}, ``Sam-med3d,'' \emph{arXiv preprint arXiv:2310.15161}, 2023.

\bibitem{SA-Med2D-20M}
J.~Ye \emph{et~al.}, ``Sa-med2d-20m dataset: Segment anything in 2d medical imaging with 20 million masks,'' \emph{arXiv preprint arXiv:2311.11969}, 2023.

\bibitem{zhou2023unified}
H.-Y. Zhou \emph{et~al.}, ``A unified visual information preservation framework for self-supervised pre-training in medical image analysis,'' \emph{IEEE Trans. Pattern Anal. Mach. Intell.}, 2023.

\bibitem{12-VisionFM}
J.~Qiu \emph{et~al.}, ``Visionfm: a multi-modal multi-task vision foundation model for generalist ophthalmic artificial intelligence,'' \emph{arXiv preprint arXiv:2310.04992}, 2023.

\bibitem{33-SegVol}
Y.~Du \emph{et~al.}, ``Segvol: Universal and interactive volumetric medical image segmentation,'' \emph{arXiv preprint arXiv:2311.13385}, 2023.

\bibitem{9-DeblurringMAE}
Q.~Kang \emph{et~al.}, ``Deblurring masked autoencoder is better recipe for ultrasound image recognition,'' in \emph{International Conference on Medical Image Computing and Computer-Assisted Intervention}.\hskip 1em plus 0.5em minus 0.4em\relax Springer, 2023, pp. 352--362.

\bibitem{1-USFM}
J.~Jiao \emph{et~al.}, ``Usfm: A universal ultrasound foundation model generalized to tasks and organs towards label efficient image analysis,'' \emph{arXiv preprint arXiv:2401.00153}, 2024.

\bibitem{14-MG}
Z.~Zhou \emph{et~al.}, ``Models genesis: Generic autodidactic models for 3d medical image analysis,'' in \emph{Proc. Int. Conf. Med. Image Comput. Comput.-Assisted Intervention}.\hskip 1em plus 0.5em minus 0.4em\relax Springer, 2019, pp. 384--393.

\bibitem{iBOT}
J.~Zhou \emph{et~al.}, ``Image bert pre-training with online tokenizer,'' in \emph{International Conference on Learning Representations}, 2021.

\bibitem{he2022masked}
K.~He \emph{et~al.}, ``Masked autoencoders are scalable vision learners,'' in \emph{Proc. IEEE Conf. Comput. Vis. Pattern Recognit.}, 2022, pp. 16\,000--16\,009.

\bibitem{xie2022simmim}
Z.~Xie, Z.~Zhang, Y.~Cao, Y.~Lin, J.~Bao, Z.~Yao, Q.~Dai, and H.~Hu, ``Simmim: A simple framework for masked image modeling,'' in \emph{Proceedings of the IEEE/CVF conference on computer vision and pattern recognition}, 2022, pp. 9653--9663.

\bibitem{83-CTransPath}
X.~Wang \emph{et~al.}, ``Transformer-based unsupervised contrastive learning for histopathological image classification,'' \emph{Med. Image Anal.}, vol.~81, p. 102559, 2022.

\bibitem{16-C2L}
H.-Y. Zhou \emph{et~al.}, ``Comparing to learn: Surpassing imagenet pretraining on radiographs by comparing image representations,'' in \emph{Proc. Int. Conf. Med. Image Comput. Comput.-Assisted Intervention}.\hskip 1em plus 0.5em minus 0.4em\relax Springer, 2020, pp. 398--407.

\bibitem{18-MoCo-CXR}
H.~Sowrirajan \emph{et~al.}, ``Moco pretraining improves representation and transferability of chest x-ray models,'' in \emph{Proc. Int. Conf. Medical Imaging Deep Learn.}\hskip 1em plus 0.5em minus 0.4em\relax PMLR, 2021, pp. 728--744.

\bibitem{65-Histopathology}
O.~Ciga \emph{et~al.}, ``Self supervised contrastive learning for digital histopathology,'' \emph{Machine Learning with Applications}, vol.~7, p. 100198, 2022.

\bibitem{47-LVM-Med}
D.~M. Nguyen \emph{et~al.}, ``Lvm-med: Learning large-scale self-supervised vision models for medical imaging via second-order graph matching,'' \emph{Advances in Neural Information Processing Systems}, vol.~36, 2024.

\bibitem{wu2024voco}
L.~Wu,  \emph{et~al.}, ``Voco: A simple-yet-effective volume contrastive learning framework for 3d medical image analysis,'' in \emph{IEEE Conf. Comput. Vis. Pattern Recog.}, 2024.

\bibitem{SimCLR}
T.~Chen \emph{et~al.}, ``A simple framework for contrastive learning of visual representations,'' in \emph{Proc. Int. Conf. Mach. Learn.}\hskip 1em plus 0.5em minus 0.4em\relax PMLR, 2020, pp. 1597--1607.

\bibitem{2-MIS-FM}
G.~Wang \emph{et~al.}, ``Mis-fm: 3d medical image segmentation using foundation models pretrained on a large-scale unannotated dataset,'' \emph{arXiv preprint arXiv:2306.16925}, 2023.

\bibitem{54-self}
F.~C. Ghesu \emph{et~al.}, ``Contrastive self-supervised learning from 100 million medical images with optional supervision,'' \emph{Journal of Medical Imaging}, vol.~9, no.~6, pp. 064\,503--064\,503, 2022.

\bibitem{10-Virchow}
E.~Vorontsov \emph{et~al.}, ``Virchow: A million-slide digital pathology foundation model,'' \emph{arXiv preprint arXiv:2309.07778}, 2023.

\bibitem{11-UNI}
R.~J. Chen \emph{et~al.}, ``Towards a general-purpose foundation model for computational pathology,'' \emph{Nature Medicine}, 2024.

\bibitem{48-RudolfV}
J.~Dippel \emph{et~al.}, ``Rudolfv: A foundation model by pathologists for pathologists,'' \emph{arXiv preprint arXiv:2401.04079}, 2024.

\bibitem{DINOv2}
M.~Oquab \emph{et~al.}, ``Dinov2: Learning robust visual features without supervision,'' \emph{Transactions on Machine Learning Research}, 2023.

\bibitem{6-BROW}
Y.~Wu \emph{et~al.}, ``Brow: Better features for whole slide image based on self-distillation,'' \emph{arXiv preprint arXiv:2309.08259}, 2023.

\bibitem{haghighi2021transferable}
F.~Haghighi \emph{et~al.}, ``Transferable visual words: Exploiting the semantics of anatomical patterns for self-supervised learning,'' \emph{IEEE transactions on medical imaging}, vol.~40, no.~10, pp. 2857--2868, 2021.

\bibitem{62-Pathology}
G.~Campanella \emph{et~al.}, ``Computational pathology at health system scale--self-supervised foundation models from billions of images,'' in \emph{AAAI 2024 Spring Symposium on Clinical Foundation Models}, 2024.

\bibitem{68-swinunetr}
Y.~Tang \emph{et~al.}, ``Self-supervised pre-training of swin transformers for 3d medical image analysis,'' in \emph{Proc. IEEE Conf. Comput. Vis. Pattern Recognit.}, 2022, pp. 20\,730--20\,740.

\bibitem{chen2023ma}
C.~Chen \emph{et~al.}, ``Ma-sam: Modality-agnostic sam adaptation for 3d medical image segmentation,'' \emph{arXiv preprint arXiv:2309.08842}, 2023.

\bibitem{56-comprehensive}
S.~Pandey \emph{et~al.}, ``Comprehensive multimodal segmentation in medical imaging: Combining yolov8 with sam and hq-sam models,'' in \emph{Proc. IEEE Int. Conf. Comput. Vis.}, 2023, pp. 2592--2598.

\bibitem{24-3DSAM-adapter}
S.~Gong \emph{et~al.}, ``3dsam-adapter: Holistic adaptation of sam from 2d to 3d for promptable medical image segmentation,'' \emph{arXiv preprint arXiv:2306.13465}, 2023.

\bibitem{88-part}
W.~Yue \emph{et~al.}, ``Part to whole: Collaborative prompting for surgical instrument segmentation,'' \emph{arXiv preprint arXiv:2312.14481}, 2023.

\bibitem{34-Skinsam}
M.~Hu \emph{et~al.}, ``Skinsam: Empowering skin cancer segmentation with segment anything model,'' \emph{arXiv preprint arXiv:2304.13973}, 2023.

\bibitem{35-Polyp-sam}
Y.~Li \emph{et~al.}, ``Polyp-sam: Transfer sam for polyp segmentation,'' \emph{arXiv preprint arXiv:2305.00293}, 2023.

\bibitem{53-SAM-OCTA}
C.~Wang \emph{et~al.}, ``Sam-octa: A fine-tuning strategy for applying foundation model to octa image segmentation tasks,'' in \emph{ICASSP 2024-2024 IEEE International Conference on Acoustics, Speech and Signal Processing (ICASSP)}.\hskip 1em plus 0.5em minus 0.4em\relax IEEE, 2024, pp. 1771--1775.

\bibitem{44-SAMed}
K.~Zhang and D.~Liu, ``Customized segment anything model for medical image segmentation,'' \emph{arXiv preprint arXiv:2304.13785}, 2023.

\bibitem{36-LFTSAM}
S.~Chai \emph{et~al.}, ``Ladder fine-tuning approach for sam integrating complementary network,'' \emph{arXiv preprint arXiv:2306.12737}, 2023.

\bibitem{45-cheap}
W.~Feng \emph{et~al.}, ``Cheap lunch for medical image segmentation by fine-tuning sam on few exemplars,'' \emph{arXiv preprint arXiv:2308.14133}, 2023.

\bibitem{87-semisam}
Y.~Zhang \emph{et~al.}, ``Semisam: Exploring sam for enhancing semi-supervised medical image segmentation with extremely limited annotations,'' \emph{arXiv preprint arXiv:2312.06316}, 2023.

\bibitem{28-AFTer-SAM}
X.~Yan \emph{et~al.}, ``After-sam: Adapting sam with axial fusion transformer for medical imaging segmentation,'' in \emph{Proceedings of the IEEE/CVF Winter Conference on Applications of Computer Vision}, 2024, pp. 7975--7984.

\bibitem{82-Mammo-SAM}
X.~Xiong \emph{et~al.}, ``Mammo-sam: Adapting foundation segment anything model for automatic breast mass segmentation in whole mammograms,'' in \emph{International Workshop on Machine Learning in Medical Imaging}.\hskip 1em plus 0.5em minus 0.4em\relax Springer, 2023, pp. 176--185.

\bibitem{74-promise}
H.~Li \emph{et~al.}, ``Promise: Prompt-driven 3d medical image segmentation using pretrained image foundation models,'' \emph{arXiv preprint arXiv:2310.19721}, 2023.

\bibitem{23-Med-SA}
J.~Wu \emph{et~al.}, ``Medical sam adapter: Adapting segment anything model for medical image segmentation,'' \emph{arXiv preprint arXiv:2304.12620}, 2023.

\bibitem{26-SAM-Med2D}
J.~Cheng \emph{et~al.}, ``Sam-med2d,'' \emph{arXiv preprint arXiv:2308.16184}, 2023.

\bibitem{29-Adaptivesam}
J.~N. Paranjape \emph{et~al.}, ``Adaptivesam: Towards efficient tuning of sam for surgical scene segmentation,'' in \emph{Medical Imaging with Deep Learning}, 2024.

\bibitem{31-MediViSTA-SAM}
S.~Kim \emph{et~al.}, ``Medivista-sam: Zero-shot medical video analysis with spatio-temporal sam adaptation,'' \emph{arXiv preprint arXiv:2309.13539}, 2023.

\bibitem{51-samus}
X.~Lin \emph{et~al.}, ``Samus: Adapting segment anything model for clinically-friendly and generalizable ultrasound image segmentation,'' \emph{arXiv preprint arXiv:2309.06824}, 2023.

\bibitem{91-SegmentAnyBone}
H.~Gu \emph{et~al.}, ``Segmentanybone: A universal model that segments any bone at any location on mri,'' \emph{arXiv preprint arXiv:2401.12974}, 2024.

\bibitem{90swinsam}
Z.~Feng \emph{et~al.}, ``Swinsam: Fine-grained polyp segmentation in colonoscopy images via segment anything model integrated with a swin transformer decoder,'' \emph{Available at SSRN 4673046}.

\bibitem{SAMAug}
Y.~Zhang \emph{et~al.}, ``Input augmentation with sam: Boosting medical image segmentation with segmentation foundation model,'' in \emph{Proc. Int. Conf. Med. Image Comput. Comput.-Assisted Intervention}.\hskip 1em plus 0.5em minus 0.4em\relax Springer, 2023, pp. 129--139.

\bibitem{40-AutoSAM}
T.~Shaharabany \emph{et~al.}, ``Autosam: Adapting sam to medical images by overloading the prompt encoder,'' \emph{arXiv preprint arXiv:2306.06370}, 2023.

\bibitem{41-DeSAM}
Y.~Gao \emph{et~al.}, ``Desam: Decoupling segment anything model for generalizable medical image segmentation,'' \emph{arXiv preprint arXiv:2306.00499}, 2023.

\bibitem{86-CellSAM}
U.~Israel \emph{et~al.}, ``A foundation model for cell segmentation,'' \emph{bioRxiv}, pp. 2023--11, 2023.

\bibitem{42-SAM-U}
G.~Deng \emph{et~al.}, ``Sam-u: Multi-box prompts triggered uncertainty estimation for reliable sam in medical image,'' in \emph{International Conference on Medical Image Computing and Computer-Assisted Intervention}.\hskip 1em plus 0.5em minus 0.4em\relax Springer, 2023, pp. 368--377.

\bibitem{52-Sam-path}
J.~Zhang \emph{et~al.}, ``Sam-path: A segment anything model for semantic segmentation in digital pathology,'' in \emph{International Conference on Medical Image Computing and Computer-Assisted Intervention}.\hskip 1em plus 0.5em minus 0.4em\relax Springer, 2023, pp. 161--170.

\bibitem{27-All}
C.~Cui and R.~Deng, ``All-in-sam: from weak annotation to pixel-wise nuclei segmentation with prompt-based finetuning,'' in \emph{Asia Conference on Computers and Communications, ACCC}, 2023.

\bibitem{58-surgicalsam}
W.~Yue \emph{et~al.}, ``Surgicalsam: Efficient class promptable surgical instrument segmentation,'' in \emph{Proceedings of the AAAI Conference on Artificial Intelligence}, 2024.

\bibitem{73-polyp}
R.~Biswas, ``Polyp-sam++: Can a text guided sam perform better for polyp segmentation?'' \emph{arXiv preprint arXiv:2308.06623}, 2023.

\bibitem{39-UR-SAM}
Y.~Zhang \emph{et~al.}, ``Segment anything model with uncertainty rectification for auto-prompting medical image segmentation,'' \emph{arXiv preprint arXiv:2311.10529}, 2023.

\bibitem{25-MedLSAM}
W.~Lei \emph{et~al.}, ``Medlsam: Localize and segment anything model for 3d medical images,'' \emph{arXiv preprint arXiv:2306.14752}, 2023.

\bibitem{37-nnsam}
Y.~Li \emph{et~al.}, ``nnsam: Plug-and-play segment anything model improves nnunet performance,'' \emph{arXiv preprint arXiv:2309.16967}, 2023.

\bibitem{38-EviPrompt}
Y.~Xu \emph{et~al.}, ``Eviprompt: A training-free evidential prompt generation method for segment anything model in medical images,'' \emph{arXiv preprint arXiv:2311.06400}, 2023.

\bibitem{57-one}
D.~Anand \emph{et~al.}, ``One-shot localization and segmentation of medical images with foundation models,'' in \emph{R0-FoMo: Robustness of Few-shot and Zero-shot Learning in Large Foundation Models}, 2023.

\bibitem{32-SAMM}
Y.~Liu \emph{et~al.}, ``Samm (segment any medical model): A 3d slicer integration to sam,'' \emph{arXiv preprint arXiv:2304.05622}, 2023.

\bibitem{79-SAMPOT}
R.~Sathish \emph{et~al.}, ``Task-driven prompt evolution for foundation models,'' in \emph{Proc. Int. Conf. Med. Image Comput. Comput.-Assisted Intervention}.\hskip 1em plus 0.5em minus 0.4em\relax Springer, 2023, pp. 256--264.

\bibitem{fischer2024prompt}
M.~Fischer \emph{et~al.}, ``Prompt tuning for parameter-efficient medical image segmentation,'' \emph{Medical Image Analysis}, vol.~91, p. 103024, 2024.

\bibitem{50-samadapter}
T.~Chen \emph{et~al.}, ``Sam fails to segment anything?--sam-adapter: Adapting sam in underperformed scenes: Camouflage, shadow, and more,'' \emph{arXiv preprint arXiv:2304.09148}, 2023.

\bibitem{madani2023large}
A.~Madani \emph{et~al.}, ``Large language models generate functional protein sequences across diverse families,'' \emph{Nat. Biotechnol.}, pp. 1--8, 2023.

\bibitem{Nijkamp2023progen2}
E.~Nijkamp \emph{et~al.}, ``Progen2: Exploring the boundaries of protein language models,'' \emph{Cell Systems}, vol.~14, pp. 968--978.e3, 11 2023, doi: 10.1016/j.cels.2023.10.002.

\bibitem{Yang2022scbert}
\BIBentryALTinterwordspacing
F.~Yang \emph{et~al.}, ``scbert as a large-scale pretrained deep language model for cell type annotation of single-cell rna-seq data,'' \emph{Nat. Mach. Intell}, vol.~4, pp. 852--866, 2022. [Online]. Available: \url{https://doi.org/10.1038/s42256-022-00534-z}
\BIBentrySTDinterwordspacing

\bibitem{Theodoris2023geneformer}
\BIBentryALTinterwordspacing
C.~V. Theodoris \emph{et~al.}, ``Transfer learning enables predictions in network biology,'' \emph{Nature}, vol. 618, pp. 616--624, 2023. [Online]. Available: \url{https://doi.org/10.1038/s41586-023-06139-9}
\BIBentrySTDinterwordspacing

\bibitem{Zhou2023dnabert2}
Z.~Zhou \emph{et~al.}, ``Dnabert-2: Efficient foundation model and benchmark for multi-species genomes,'' in \emph{Proc. Int. Conf. Learn. Represent.}, 2023.

\bibitem{Fishman2023genalm}
\BIBentryALTinterwordspacing
V.~Fishman \emph{et~al.}, ``Gena-lm: A family of open-source foundational models for long dna sequences,'' \emph{bioRxiv}, p. 2023.06.12.544594, 1 2023. [Online]. Available: \url{http://biorxiv.org/content/early/2023/06/13/2023.06.12.544594.abstract}
\BIBentrySTDinterwordspacing

\bibitem{Zhang2023rnamsm}
\BIBentryALTinterwordspacing
Y.~Zhang \emph{et~al.}, ``Multiple sequence-alignment-based rna language model and its application to structural inference,'' \emph{bioRxiv}, p. 2023.03.15.532863, 1 2023. [Online]. Available: \url{http://biorxiv.org/content/early/2023/03/16/2023.03.15.532863.abstract}
\BIBentrySTDinterwordspacing

\bibitem{Chen2023splicebert}
\BIBentryALTinterwordspacing
K.~Chen \emph{et~al.}, ``Self-supervised learning on millions of pre-mrna sequences improves sequence-based rna splicing prediction,'' \emph{bioRxiv}, p. 2023.01.31.526427, 1 2023. [Online]. Available: \url{http://biorxiv.org/content/early/2023/02/03/2023.01.31.526427.abstract}
\BIBentrySTDinterwordspacing

\bibitem{Yang20233utr}
\BIBentryALTinterwordspacing
Y.~Yang \emph{et~al.}, ``Deciphering 3’ utr mediated gene regulation using interpretable deep representation learning,'' \emph{bioRxiv}, p. 2023.09.08.556883, 1 2023. [Online]. Available: \url{http://biorxiv.org/content/early/2023/09/12/2023.09.08.556883.abstract}
\BIBentrySTDinterwordspacing

\bibitem{Lin2023esm2}
Z.~Lin \emph{et~al.}, ``Evolutionary-scale prediction of atomic-level protein structure with a language model,'' \emph{Science}, vol. 379, pp. 1123--1130, 3 2023, doi: 10.1126/science.ade2574.

\bibitem{Elnaggar2022prottrans}
A.~Elnaggar \emph{et~al.}, ``Prottrans: Toward understanding the language of life through self-supervised learning,'' \emph{IEEE Trans. Pattern Anal. Mach. Intell.}, vol.~44, pp. 7112--7127, 2022.

\bibitem{Rao2021msa}
R.~M. Rao \emph{et~al.}, ``Msa transformer,'' in \emph{Proceedings of the 38th International Conference on Machine Learning}, ser. Proceedings of Machine Learning Research, M.~Meila and T.~Zhang, Eds., vol. 139.\hskip 1em plus 0.5em minus 0.4em\relax PMLR, 18--24 Jul 2021, pp. 8844--8856.

\bibitem{Rives2021esm1b}
\BIBentryALTinterwordspacing
A.~Rives \emph{et~al.}, ``Biological structure and function emerge from scaling unsupervised learning to 250 million protein sequences,'' \emph{Proc. Natl. Acad. Sci.}, vol. 118, p. e2016239118, 4 2021, doi: 10.1073/pnas.2016239118. [Online]. Available: \url{https://doi.org/10.1073/pnas.2016239118}
\BIBentrySTDinterwordspacing

\bibitem{Nguyen2023hyenadna}
E.~Nguyen \emph{et~al.}, ``Hyenadna: Long-range genomic sequence modeling at single nucleotide resolution,'' \emph{Advances in Neural Information Processing Systems}, vol.~36, 2024.

\bibitem{Hao2023scfoundation}
\BIBentryALTinterwordspacing
M.~Hao \emph{et~al.}, ``Large scale foundation model on single-cell transcriptomics,'' \emph{bioRxiv}, p. 2023.05.29.542705, 1 2023. [Online]. Available: \url{http://biorxiv.org/content/early/2023/06/15/2023.05.29.542705.abstract}
\BIBentrySTDinterwordspacing

\bibitem{Rosen2023uce}
\BIBentryALTinterwordspacing
Y.~Rosen \emph{et~al.}, ``Universal cell embeddings: A foundation model for cell biology,'' \emph{bioRxiv}, p. 2023.11.28.568918, 1 2023. [Online]. Available: \url{http://biorxiv.org/content/early/2023/11/29/2023.11.28.568918.abstract}
\BIBentrySTDinterwordspacing

\bibitem{Zhang2023dnagpt}
D.~Zhang \emph{et~al.}, ``Dnagpt: A generalized pretrained tool for multiple dna sequence analysis tasks,'' \emph{arXiv preprint arXiv:2307.05628}, 2023.

\bibitem{cui2023scgpt}
H.~Cui \emph{et~al.}, ``scgpt: towards building a foundation model for single-cell multi-omics using generative ai,'' \emph{bioRxiv}, pp. 2023--04, 2023.

\bibitem{Akiyama2022rnabert}
\BIBentryALTinterwordspacing
M.~Akiyama and Y.~Sakakibara, ``Informative rna base embedding for rna structural alignment and clustering by deep representation learning,'' \emph{NAR Genomics and Bioinformatics}, vol.~4, p. lqac012, 3 2022. [Online]. Available: \url{https://doi.org/10.1093/nargab/lqac012}
\BIBentrySTDinterwordspacing

\bibitem{Chowdhury2022aminobert}
R.~Chowdhury \emph{et~al.}, ``Single-sequence protein structure prediction using a language model and deep learning,'' \emph{Nature Biotechnol.}, vol.~40, no.~11, pp. 1617--1623, 2022.

\bibitem{Chu2023utrlm}
\BIBentryALTinterwordspacing
Y.~Chu \emph{et~al.}, ``A 5’ utr language model for decoding untranslated regions of mrna and function predictions,'' \emph{bioRxiv}, p. 2023.10.11.561938, 1 2023. [Online]. Available: \url{http://biorxiv.org/content/early/2023/10/14/2023.10.11.561938.abstract}
\BIBentrySTDinterwordspacing

\bibitem{Zhao2023celllm}
S.~Zhao \emph{et~al.}, ``Large-scale cell representation learning via divide-and-conquer contrastive learning,'' \emph{arXiv preprint arXiv:2306.04371}, 2023.

\bibitem{GeneBERT}
S.~Mo \emph{et~al.}, ``Multi-modal self-supervised pre-training for regulatory genome across cell types,'' \emph{arXiv preprint arXiv:2110.05231}, 2021.

\bibitem{Li2023codonbert}
S.~Li \emph{et~al.}, ``Codonbert: Large language models for mrna design and optimization,'' in \emph{NeurIPS 2023 Generative AI and Biology (GenBio) Workshop}, 2023.

\bibitem{Chen2024xTrimopglm}
\BIBentryALTinterwordspacing
B.~Chen \emph{et~al.}, ``xtrimopglm: Unified 100b-scale pre-trained transformer for deciphering the language of protein,'' \emph{bioRxiv}, p. 2023.07.05.547496, 1 2024. [Online]. Available: \url{http://biorxiv.org/content/early/2024/01/11/2023.07.05.547496.abstract}
\BIBentrySTDinterwordspacing

\bibitem{du-etal-2022-glm}
Z.~Du \emph{et~al.}, ``{GLM}: General language model pretraining with autoregressive blank infilling,'' in \emph{Proceedings of the 60th Annual Meeting of the Association for Computational Linguistics}.\hskip 1em plus 0.5em minus 0.4em\relax Dublin, Ireland: Association for Computational Linguistics, may 2022, pp. 320--335.

\bibitem{Chen2023genept}
\BIBentryALTinterwordspacing
Y.~T. Chen and J.~Zou, ``Genept: A simple but hard-to-beat foundation model for genes and cells built from chatgpt,'' \emph{bioRxiv}, p. 2023.10.16.562533, 1 2023. [Online]. Available: \url{http://biorxiv.org/content/early/2023/10/19/2023.10.16.562533.abstract}
\BIBentrySTDinterwordspacing

\bibitem{Liu2023scelmo}
\BIBentryALTinterwordspacing
T.~Liu \emph{et~al.}, ``scelmo: Embeddings from language models are good learners for single-cell data analysis,'' \emph{bioRxiv}, 2024. [Online]. Available: \url{https://www.biorxiv.org/content/early/2024/03/03/2023.12.07.569910}
\BIBentrySTDinterwordspacing

\bibitem{highthroughput}
\BIBentryALTinterwordspacing
B.~E. Slatko \emph{et~al.}, ``Overview of next-generation sequencing technologies,'' \emph{Current Protocols in Molecular Biology}, vol. 122, no.~1, p. e59, 2018. [Online]. Available: \url{https://currentprotocols.onlinelibrary.wiley.com/doi/abs/10.1002/cpmb.59}
\BIBentrySTDinterwordspacing

\bibitem{10113601}
T.~Wu \emph{et~al.}, ``A brief overview of chatgpt: The history, status quo and potential future development,'' \emph{IEEE/CAA J. Autom. Sin.}, vol.~10, no.~5, pp. 1122--1136, 2023.

\bibitem{khare2021mmbert}
Y.~Khare \emph{et~al.}, ``Mmbert: Multimodal bert pretraining for improved medical vqa,'' in \emph{Proc. IEEE Int. Symp. Biomed. Imaging}.\hskip 1em plus 0.5em minus 0.4em\relax IEEE, 2021, pp. 1033--1036.

\bibitem{zhou2023advancing}
H.-Y. Zhou \emph{et~al.}, ``Advancing radiograph representation learning with masked record modeling,'' \emph{The Eleventh International Conference on Learning Representations}, 2023.

\bibitem{zhang2022contrastive}
Y.~Zhang \emph{et~al.}, ``Contrastive learning of medical visual representations from paired images and text,'' in \emph{Machine Learning for Healthcare Conference}.\hskip 1em plus 0.5em minus 0.4em\relax PMLR, 2022, pp. 2--25.

\bibitem{muller2022joint}
P.~M{\"u}ller \emph{et~al.}, ``Joint learning of localized representations from medical images and reports,'' in \emph{Proc. Eur. Conf. Comput. Vis.}\hskip 1em plus 0.5em minus 0.4em\relax Springer, 2022, pp. 685--701.

\bibitem{lei2023unibrain}
J.~Lei \emph{et~al.}, ``Unibrain: Universal brain mri diagnosis with hierarchical knowledge-enhanced pre-training,'' \emph{arXiv preprint arXiv:2309.06828}, 2023.

\bibitem{liu2023m}
C.~Liu \emph{et~al.}, ``M-flag: Medical vision-language pre-training with frozen language models and latent space geometry optimization,'' in \emph{Proc. Int. Conf. Med. Image Comput. Comput.-Assisted Intervention}.\hskip 1em plus 0.5em minus 0.4em\relax Springer, 2023, pp. 637--647.

\bibitem{wang2022multi}
F.~Wang \emph{et~al.}, ``Multi-granularity cross-modal alignment for generalized medical visual representation learning,'' \emph{Proc. Adv. Neural Inf. Process. Syst.}, vol.~35, pp. 33\,536--33\,549, 2022.

\bibitem{wu2023medklip}
C.~Wu \emph{et~al.}, ``Medklip: Medical knowledge enhanced language-image pre-training,'' \emph{medRxiv}, pp. 2023--01, 2023.

\bibitem{liu2023etp}
C.~Liu \emph{et~al.}, ``Etp: Learning transferable ecg representations via ecg-text pre-training,'' in \emph{ICASSP 2024-2024 IEEE International Conference on Acoustics, Speech and Signal Processing (ICASSP)}.\hskip 1em plus 0.5em minus 0.4em\relax IEEE, 2024, pp. 8230--8234.

\bibitem{huang2021gloria}
S.-C. Huang \emph{et~al.}, ``Gloria: A multimodal global-local representation learning framework for label-efficient medical image recognition,'' in \emph{Proc. IEEE Int. Conf. Comput. Vis.}, 2021, pp. 3942--3951.

\bibitem{liu2023imitate}
C.~Liu \emph{et~al.}, ``Imitate: Clinical prior guided hierarchical vision-language pre-training,'' \emph{arXiv preprint arXiv:2310.07355}, 2023.

\bibitem{wang2022medclip}
Z.~Wang \emph{et~al.}, ``Medclip: Contrastive learning from unpaired medical images and text,'' in \emph{Proceedings of the 2022 Conference on Empirical Methods in Natural Language Processing}, 2022, pp. 3876--3887.

\bibitem{wan2024med}
Z.~Wan \emph{et~al.}, ``Med-unic: Unifying cross-lingual medical vision-language pre-training by diminishing bias,'' \emph{Advances in Neural Information Processing Systems}, vol.~36, 2024.

\bibitem{you2023cxr}
K.~You \emph{et~al.}, ``Cxr-clip: Toward large scale chest x-ray language-image pre-training,'' in \emph{Proc. Int. Conf. Med. Image Comput. Comput.-Assisted Intervention}.\hskip 1em plus 0.5em minus 0.4em\relax Springer, 2023, pp. 101--111.

\bibitem{zhang2023large}
S.~Zhang \emph{et~al.}, ``Large-scale domain-specific pretraining for biomedical vision-language processing,'' \emph{arXiv preprint arXiv:2303.00915}, 2023.

\bibitem{wang2023unified}
Y.~Wang and G.~Wang, ``Umcl: Unified medical image-text-label contrastive learning with continuous prompt,'' in \emph{2023 IEEE International Conference on Bioinformatics and Biomedicine (BIBM)}.\hskip 1em plus 0.5em minus 0.4em\relax IEEE, 2023, pp. 2285--2289.

\bibitem{zhang2023knowledge}
X.~Zhang \emph{et~al.}, ``Knowledge-enhanced visual-language pre-training on chest radiology images,'' \emph{Nat. Commun.}, vol.~14, no.~1, p. 4542, 2023.

\bibitem{liu2023multi}
S.~Liu \emph{et~al.}, ``Multi-modal molecule structure--text model for text-based retrieval and editing,'' \emph{Nature Machine Intelligence}, vol.~5, no.~12, pp. 1447--1457, 2023.

\bibitem{lei2023clip}
Y.~Lei \emph{et~al.}, ``Clip-lung: Textual knowledge-guided lung nodule malignancy prediction,'' in \emph{Medical Image Computing and Computer Assisted Intervention -- MICCAI 2023}.\hskip 1em plus 0.5em minus 0.4em\relax Cham: Springer Nature Switzerland, 2023, pp. 403--412.

\bibitem{seibold2022breaking}
C.~Seibold \emph{et~al.}, ``Breaking with fixed set pathology recognition through report-guided contrastive training,'' in \emph{Proc. Int. Conf. Med. Image Comput. Comput.-Assisted Intervention}.\hskip 1em plus 0.5em minus 0.4em\relax Springer, 2022, pp. 690--700.

\bibitem{lu2023visual}
M.~Y. Lu \emph{et~al.}, ``Visual language pretrained multiple instance zero-shot transfer for histopathology images,'' in \emph{Proc. IEEE Conf. Comput. Vis. Pattern Recognit.}, 2023, pp. 19\,764--19\,775.

\bibitem{yan2022clinical}
B.~Yan and M.~Pei, ``Clinical-bert: Vision-language pre-training for radiograph diagnosis and reports generation,'' in \emph{Proc. AAAI Conf. Artif. Intell.}, vol.~36, no.~3, 2022, pp. 2982--2990.

\bibitem{chen2022multi}
Z.~Chen \emph{et~al.}, ``Multi-modal masked autoencoders for medical vision-and-language pre-training,'' in \emph{Proc. Int. Conf. Med. Image Comput. Comput.-Assisted Intervention}.\hskip 1em plus 0.5em minus 0.4em\relax Springer, 2022, pp. 679--689.

\bibitem{moon2022multi}
J.~H. Moon \emph{et~al.}, ``Multi-modal understanding and generation for medical images and text via vision-language pre-training,'' \emph{IEEE J. Biomed. Health Inform.}, vol.~26, no.~12, pp. 6070--6080, 2022.

\bibitem{lin2023pmc}
W.~Lin and other, ``Pmc-clip: Contrastive language-image pre-training using biomedical documents,'' in \emph{Medical Image Computing and Computer Assisted Intervention -- MICCAI 2023}.\hskip 1em plus 0.5em minus 0.4em\relax Cham: Springer Nature Switzerland, 2023, pp. 525--536.

\bibitem{chen2022align}
Z.~Chen \emph{et~al.}, ``Align, reason and learn: Enhancing medical vision-and-language pre-training with knowledge,'' in \emph{Proceedings of the 30th ACM International Conference on Multimedia}, 2022, pp. 5152--5161.

\bibitem{huang2023enhancing}
W.~Huang \emph{et~al.}, ``Enhancing representation in radiography-reports foundation model: A granular alignment algorithm using masked contrastive learning,'' \emph{arXiv preprint arXiv:2309.05904}, 2023.

\bibitem{li2023masked}
P.~Li \emph{et~al.}, ``Masked vision and language pre-training with unimodal and multimodal contrastive losses for medical visual question answering,'' in \emph{Proc. Int. Conf. Med. Image Comput. Comput.-Assisted Intervention}.\hskip 1em plus 0.5em minus 0.4em\relax Springer, 2023, pp. 374--383.

\bibitem{liu2023t3d}
C.~Liu \emph{et~al.}, ``T3d: Towards 3d medical image understanding through vision-language pre-training,'' \emph{arXiv preprint arXiv:2312.01529}, 2023.

\bibitem{jin2023gene}
T.~Jin and Others, ``Gene-induced multimodal pre-training for image-omic classification,'' in \emph{International Conference on Medical Image Computing and Computer-Assisted Intervention}, 2023, pp. 508--517.

\bibitem{boecking2022making}
B.~Boecking \emph{et~al.}, ``Making the most of text semantics to improve biomedical vision--language processing,'' in \emph{Proc. Eur. Conf. Comput. Vis.}\hskip 1em plus 0.5em minus 0.4em\relax Springer, 2022, pp. 1--21.

\bibitem{cheng2023prior}
P.~Cheng \emph{et~al.}, ``Prior: Prototype representation joint learning from medical images and reports,'' in \emph{Proc. IEEE Int. Conf. Comput. Vis.}, 2023, pp. 21\,361--21\,371.

\bibitem{lu2023towards}
M.~Y. Lu \emph{et~al.}, ``A visual-language foundation model for computational pathology,'' \emph{Nature Medicine}, 2024.

\bibitem{liu2023text}
S.~Liu \emph{et~al.}, ``A text-guided protein design framework,'' \emph{arXiv preprint arXiv:2302.04611}, 2023.

\bibitem{eslami2021does}
S.~Eslami \emph{et~al.}, ``Pubmedclip: How much does clip benefit visual question answering in the medical domain?'' in \emph{Findings of the Association for Computational Linguistics: EACL 2023}, 2023, pp. 1181--1193.

\bibitem{moor2023med}
M.~Moor \emph{et~al.}, ``Med-flamingo: a multimodal medical few-shot learner,'' in \emph{Machine Learning for Health}.\hskip 1em plus 0.5em minus 0.4em\relax PMLR, 2023, pp. 353--367.

\bibitem{li2023llava}
C.~Li \emph{et~al.}, ``Llava-med: Training a large language-and-vision assistant for biomedicine in one day,'' \emph{Advances in Neural Information Processing Systems}, 2024.

\bibitem{tiu2022expert}
E.~Tiu \emph{et~al.}, ``Expert-level detection of pathologies from unannotated chest x-ray images via self-supervised learning,'' \emph{Nat. Biomed. Eng.}, vol.~6, no.~12, pp. 1399--1406, 2022.

\bibitem{ikezogwo2023quilt}
W.~Ikezogwo \emph{et~al.}, ``Quilt-1m: One million image-text pairs for histopathology,'' \emph{Advances in Neural Information Processing Systems}, 2024.

\bibitem{huang2023visual}
Z.~Huang \emph{et~al.}, ``A visual--language foundation model for pathology image analysis using medical twitter,'' \emph{Nat. Med.}, vol.~29, no.~9, pp. 2307--2316, 2023.

\bibitem{baliah2023exploring}
S.~Baliah \emph{et~al.}, ``Exploring the transfer learning capabilities of clip in domain generalization for diabetic retinopathy,'' in \emph{International Workshop on Machine Learning in Medical Imaging}.\hskip 1em plus 0.5em minus 0.4em\relax Springer, 2023, pp. 444--453.

\bibitem{chambon2022roentgen}
P.~Chambon \emph{et~al.}, ``Roentgen: vision-language foundation model for chest x-ray generation,'' \emph{arXiv preprint arXiv:2211.12737}, 2022.

\bibitem{van2023open}
T.~Van~Sonsbeek \emph{et~al.}, ``Open-ended medical visual question answering through prefix tuning of language models,'' in \emph{International Conference on Medical Image Computing and Computer-Assisted Intervention}.\hskip 1em plus 0.5em minus 0.4em\relax Springer, 2023, pp. 726--736.

\bibitem{chambon2022adapting}
P.~Chambon \emph{et~al.}, ``Adapting pretrained vision-language foundational models to medical imaging domains,'' in \emph{NeurIPS 2022 Foundation Models for Decision Making Workshop}, 2022.

\bibitem{liu2023qilin}
J.~Liu \emph{et~al.}, ``Qilin-med-vl: Towards chinese large vision-language model for general healthcare,'' \emph{arXiv preprint arXiv:2310.17956}, 2023.

\bibitem{sun2023pathasst}
Y.~Sun \emph{et~al.}, ``Pathasst: Redefining pathology through generative foundation ai assistant for pathology,'' \emph{Proc. AAAI Conf. Artif. Intell.}, 2024.

\bibitem{lu2023foundational}
M.~Y. Lu \emph{et~al.}, ``A foundational multimodal vision language ai assistant for human pathology,'' \emph{arXiv preprint arXiv:2312.07814}, 2023.

\bibitem{lu2023effectively}
Y.~Lu \emph{et~al.}, ``Effectively fine-tune to improve large multimodal models for radiology report generation,'' in \emph{Deep Generative Models for Health Workshop NeurIPS 2023}, 2023.

\bibitem{yu2023multi}
Z.~Yu \emph{et~al.}, ``Multi-modal adapter for medical vision-and-language learning,'' in \emph{International Workshop on Machine Learning in Medical Imaging}.\hskip 1em plus 0.5em minus 0.4em\relax Springer, 2023, pp. 393--402.

\bibitem{pham2023decoding}
T.~T. Pham \emph{et~al.}, ``I-ai: A controllable \& interpretable ai system for decoding radiologists' intense focus for accurate cxr diagnoses,'' in \emph{Proceedings of the IEEE/CVF Winter Conference on Applications of Computer Vision}, 2024, pp. 7850--7859.

\bibitem{zhang2023text}
Y.~Zhang \emph{et~al.}, ``Text-guided foundation model adaptation for pathological image classification,'' in \emph{Proc. Int. Conf. Med. Image Comput. Comput.-Assisted Intervention}.\hskip 1em plus 0.5em minus 0.4em\relax Springer, 2023, pp. 272--282.

\bibitem{thawkar2023xraygpt}
O.~Thawkar \emph{et~al.}, ``Xraygpt: Chest radiographs summarization using medical vision-language models,'' \emph{arXiv preprint arXiv:2306.07971}, 2023.

\bibitem{pellegrini2023xplainer}
C.~Pellegrini \emph{et~al.}, ``Xplainer: From x-ray observations to explainable zero-shot diagnosis,'' in \emph{International Conference on Medical Image Computing and Computer-Assisted Intervention}.\hskip 1em plus 0.5em minus 0.4em\relax Springer, 2023, pp. 420--429.

\bibitem{qin2022medical}
Z.~Qin \emph{et~al.}, ``Medical image understanding with pretrained vision language models: A comprehensive study,'' in \emph{The Eleventh International Conference on Learning Representations}, 2022.

\bibitem{guo2023multiple}
M.~Guo and Others, ``Multiple prompt fusion for zero-shot lesion detection using vision-language models,'' in \emph{International Conference on Medical Image Computing and Computer-Assisted Intervention}, 2023, pp. 283--292.

\bibitem{wang2023biobridge}
Z.~Wang \emph{et~al.}, ``Biobridge: Bridging biomedical foundation models via knowledge graph,'' \emph{arXiv preprint arXiv:2310.03320}, 2023.

\bibitem{li2020comparison}
Y.~Li \emph{et~al.}, ``A comparison of pre-trained vision-and-language models for multimodal representation learning across medical images and reports,'' in \emph{2020 IEEE international conference on bioinformatics and biomedicine (BIBM)}.\hskip 1em plus 0.5em minus 0.4em\relax IEEE, 2020, pp. 1999--2004.

\bibitem{yu2022coca}
J.~Yu \emph{et~al.}, ``Coca: Contrastive captioners are image-text foundation models,'' \emph{arXiv preprint arXiv:2205.01917}, 2022.

\bibitem{liu2023visual}
H.~Liu \emph{et~al.}, ``Visual instruction tuning,'' \emph{Advances in neural information processing systems}, vol.~36, 2024.

\bibitem{alayrac2022flamingo}
J.-B. Alayrac \emph{et~al.}, ``Flamingo: a visual language model for few-shot learning,'' \emph{Proc. Adv. Neural Inf. Process. Syst.}, vol.~35, pp. 23\,716--23\,736, 2022.

\bibitem{driess2023palm}
D.~Driess \emph{et~al.}, ``Palm-e: An embodied multimodal language model,'' in \emph{International Conference on Machine Learning}.\hskip 1em plus 0.5em minus 0.4em\relax PMLR, 2023, pp. 8469--8488.

\bibitem{liu2023utilizing}
C.~Liu \emph{et~al.}, ``Utilizing synthetic data for medical vision-language pre-training: Bypassing the need for real images,'' \emph{arXiv preprint arXiv:2310.07027}, 2023.

\bibitem{kumar2022towards}
B.~Kumar \emph{et~al.}, ``Towards reliable zero shot classification in self-supervised models with conformal prediction,'' \emph{arXiv preprint arXiv:2210.15805}, 2022.

\bibitem{Zhao2023ALD}
\BIBentryALTinterwordspacing
Z.~Zhao \emph{et~al.}, ``A large-scale dataset of patient summaries for retrieval-based clinical decision support systems.'' \emph{Scientific data}, vol. 10 1, p. 909, 2023. [Online]. Available: \url{https://api.semanticscholar.org/CorpusID:266360591}
\BIBentrySTDinterwordspacing

\bibitem{johnson2016mimic}
A.~E. Johnson \emph{et~al.}, ``Mimic-iii, a freely accessible critical care database,'' \emph{Sci. Data}, vol.~3, no.~1, pp. 1--9, 2016.

\bibitem{johnson2023mimic}
A.~Johnson \emph{et~al.}, ``Mimic-iv, a freely accessible electronic health record dataset,'' \emph{Scientific data}, vol.~10, no.~1, p.~1, 2023.

\bibitem{pollard2018eicu}
T.~J. Pollard \emph{et~al.}, ``The eicu collaborative research database, a freely available multi-center database for critical care research,'' \emph{Scientific data}, vol.~5, no.~1, pp. 1--13, 2018.

\bibitem{chen2023benchmark}
W.~Chen \emph{et~al.}, ``A benchmark for automatic medical consultation system: frameworks, tasks and datasets,'' \emph{Bioinformatics}, vol.~39, no.~1, p. btac817, 2023.

\bibitem{li2023huatuo}
J.~Li \emph{et~al.}, ``Huatuo-26m, a large-scale chinese medical qa dataset,'' \emph{arXiv preprint arXiv:2305.01526}, 2023.

\bibitem{zhu-etal-2020-question}
\BIBentryALTinterwordspacing
M.~Zhu \emph{et~al.}, ``Question answering with long multiple-span answers,'' in \emph{Findings of the Association for Computational Linguistics: EMNLP 2020}, T.~Cohn, Y.~He, and Y.~Liu, Eds.\hskip 1em plus 0.5em minus 0.4em\relax Online: Association for Computational Linguistics, Nov. 2020, pp. 3840--3849. [Online]. Available: \url{https://aclanthology.org/2020.findings-emnlp.342}
\BIBentrySTDinterwordspacing

\bibitem{ben2019question}
A.~Ben~Abacha and Others, ``A question-entailment approach to question answering,'' \emph{BMC bioinformatics}, vol.~20, pp. 1--23, 2019.

\bibitem{liu2020meddg}
W.~Liu \emph{et~al.}, ``Meddg: An entity-centric medical consultation dataset for entity-aware medical dialogue generation,'' in \emph{Natural Language Processing and Chinese Computing}.\hskip 1em plus 0.5em minus 0.4em\relax Cham: Springer International Publishing, 2022, pp. 447--459.

\bibitem{liu2024benchmarking}
J.~Liu \emph{et~al.}, ``Benchmarking large language models on cmexam-a comprehensive chinese medical exam dataset,'' \emph{Advances in Neural Information Processing Systems}, vol.~36, 2024.

\bibitem{zhang2018multi}
S.~Zhang \emph{et~al.}, ``Multi-scale attentive interaction networks for chinese medical question answer selection,'' \emph{IEEE Access}, vol.~6, pp. 74\,061--74\,071, 2018.

\bibitem{suster-daelemans-2018-clicr}
S.~Suster and W.~Daelemans, ``Clicr: a dataset of clinical case reports for machine reading comprehension,'' in \emph{Proceedings of the 2018 Conference of the North American Chapter of the Association for Computational Linguistics: Human Language Technologies, Volume 1 (Long Papers)}, 2018, pp. 1551--1563.

\bibitem{he2019applying}
J.~He \emph{et~al.}, ``Applying deep matching networks to chinese medical question answering: a study and a dataset,'' \emph{BMC medical informatics and decision making}, vol.~19, pp. 91--100, 2019.

\bibitem{zeng2020meddialog}
G.~Zeng \emph{et~al.}, ``Meddialog: Large-scale medical dialogue datasets,'' in \emph{Proceedings of the 2020 Conference on Empirical Methods in Natural Language Processing (EMNLP)}, 2020, pp. 9241--9250.

\bibitem{10.1145/3308558.3313699}
\BIBentryALTinterwordspacing
M.~Zhu \emph{et~al.}, ``A hierarchical attention retrieval model for healthcare question answering,'' in \emph{The World Wide Web Conference}, ser. WWW '19.\hskip 1em plus 0.5em minus 0.4em\relax New York, NY, USA: Association for Computing Machinery, 2019, p. 2472–2482. [Online]. Available: \url{https://doi.org/10.1145/3308558.3313699}
\BIBentrySTDinterwordspacing

\bibitem{hu2024omnimedvqa}
Y.~Hu \emph{et~al.}, ``Omnimedvqa: A new large-scale comprehensive evaluation benchmark for medical lvlm,'' \emph{arXiv preprint arXiv:2402.09181}, 2024.

\bibitem{zhang2022cblue}
N.~Zhang \emph{et~al.}, ``Cblue: A chinese biomedical language understanding evaluation benchmark,'' in \emph{Proceedings of the 60th Annual Meeting of the Association for Computational Linguistics (Volume 1: Long Papers)}, 2022, pp. 7888--7915.

\bibitem{MedFMC}
D.~Wang \emph{et~al.}, ``A real-world dataset and benchmark for foundation model adaptation in medical image classification,'' \emph{Scientific Data}, vol.~10, no.~1, p. 574, 2023.

\bibitem{antonelli2022medical}
M.~Antonelli \emph{et~al.}, ``The medical segmentation decathlon,'' \emph{Nat. Commun.}, vol.~13, no.~1, p. 4128, 2022.

\bibitem{FLARE22}
J.~Ma \emph{et~al.}, ``Unleashing the strengths of unlabeled data in pan-cancer abdominal organ quantification: the flare22 challenge,'' \emph{arXiv preprint arXiv:2308.05862}, 2023.

\bibitem{TotalSegmentatorv2}
J.~Wasserthal \emph{et~al.}, ``Totalsegmentator: Robust segmentation of 104 anatomic structures in ct images,'' \emph{Radiology: Artificial Intelligence}, vol.~5, no.~5, p. e230024, 2023.

\bibitem{9497733}
J.~Ma \emph{et~al.}, ``Abdomenct-1k: Is abdominal organ segmentation a solved problem?'' \emph{IEEE Trans. Pattern Anal. Mach. Intell.}, vol.~44, no.~10, pp. 6695--6714, 2022.

\bibitem{deng2021ctspine1k}
Y.~Deng \emph{et~al.}, ``Ctspine1k: A large-scale dataset for spinal vertebrae segmentation in computed tomography,'' \emph{arXiv preprint arXiv:2105.14711}, 2021.

\bibitem{CTPelvic1K}
P.~Liu \emph{et~al.}, ``Deep learning to segment pelvic bones: large-scale ct datasets and baseline models,'' \emph{International Journal of Computer Assisted Radiology and Surgery}, vol.~16, no.~5, p. 749, 2021.

\bibitem{baid2021rsnaasnrmiccai}
U.~Baid \emph{et~al.}, ``The rsna-asnr-miccai brats 2021 benchmark on brain tumor segmentation and radiogenomic classification,'' \emph{arXiv preprint arXiv:2107.02314}, 2021.

\bibitem{6975210}
B.~H. Menze \emph{et~al.}, ``The multimodal brain tumor image segmentation benchmark (brats),'' \emph{IEEE transactions on medical imaging}, vol.~34, no.~10, pp. 1993--2024, 2014.

\bibitem{Bakas2017}
S.~Bakas \emph{et~al.}, ``Advancing the cancer genome atlas glioma mri collections with expert segmentation labels and radiomic features,'' \emph{Sci. Data}, vol.~4, no.~1, p. 170117, 2017.

\bibitem{labella2023asnrmiccai}
D.~LaBella \emph{et~al.}, ``The asnr-miccai brain tumor segmentation (brats) challenge 2023: Intracranial meningioma,'' \emph{arXiv preprint arXiv:2305.07642}, 2023.

\bibitem{petersen2010alzheimer}
R.~C. Petersen \emph{et~al.}, ``Alzheimer's disease neuroimaging initiative (adni): clinical characterization,'' \emph{Neurology}, vol.~74, no.~3, pp. 201--209, 2010.

\bibitem{marek2011parkinson}
K.~Marek \emph{et~al.}, ``The parkinson progression marker initiative (ppmi),'' \emph{Progress in neurobiology}, vol.~95, no.~4, pp. 629--635, 2011.

\bibitem{gatidis2022whole}
S.~Gatidis \emph{et~al.}, ``A whole-body fdg-pet/ct dataset with manually annotated tumor lesions,'' \emph{Sci. Data}, vol.~9, no.~1, p. 601, 2022.

\bibitem{gatidis2023autopet}
------, ``The autopet challenge: Towards fully automated lesion segmentation in oncologic pet/ct imaging,'' 2023.

\bibitem{greenwald2022whole}
N.~F. Greenwald \emph{et~al.}, ``Whole-cell segmentation of tissue images with human-level performance using large-scale data annotation and deep learning,'' \emph{Nature biotechnology}, vol.~40, no.~4, pp. 555--565, 2022.

\bibitem{TCGA}
\BIBentryALTinterwordspacing
K.~Chang \emph{et~al.}, ``The cancer genome atlas pan-cancer analysis project,'' \emph{Nature Genetics}, vol.~45, pp. 1113--1120, 2013. [Online]. Available: \url{https://doi.org/10.1038/ng.2764}
\BIBentrySTDinterwordspacing

\bibitem{PAIP}
Y.~J. Kim \emph{et~al.}, ``Paip 2019: Liver cancer segmentation challenge,'' \emph{Med. Image Anal.}, vol.~67, p. 101854, 2021.

\bibitem{borkowski2019lung}
A.~A. Borkowski \emph{et~al.}, ``Lung and colon cancer histopathological image dataset (lc25000),'' \emph{arXiv preprint arXiv:1912.12142}, 2019.

\bibitem{CRC}
J.~N. Kather \emph{et~al.}, ``Predicting survival from colorectal cancer histology slides using deep learning: A retrospective multicenter study,'' \emph{PLoS Med.}, vol.~16, no.~1, p. e1002730, 2019.

\bibitem{wang2017chestxray}
X.~Wang,  \emph{et~al.}, ``Chestx-ray8: Hospital-scale chest x-ray database and benchmarks on weakly-supervised classification and localization of common thorax diseases,'' in \emph{2017 IEEE Conference on Computer Vision and Pattern Recognition (CVPR)}, 2017, pp. 3462--3471.

\bibitem{MURA}
P.~Rajpurkar \emph{et~al.}, ``Mura: Large dataset for abnormality detection in musculoskeletal radiographs,'' in \emph{Medical Imaging with Deep Learning}, 2022.

\bibitem{rotemberg2021patient}
V.~Rotemberg \emph{et~al.}, ``A patient-centric dataset of images and metadata for identifying melanomas using clinical context,'' \emph{Sci. Data}, vol.~8, no.~1, p.~34, 2021.

\bibitem{de2023airogs}
C.~De~Vente \emph{et~al.}, ``Airogs: artificial intelligence for robust glaucoma screening challenge,'' \emph{IEEE transactions on medical imaging}, 2023.

\bibitem{9740985}
M.~Subramanian \emph{et~al.}, ``Classification of retinal oct images using deep learning,'' in \emph{2022 International Conference on Computer Communication and Informatics (ICCCI)}, 2022, pp. 1--7.

\bibitem{ultrasound-nerve-segmentation}
\BIBentryALTinterwordspacing
A.~Montoya \emph{et~al.}, ``Ultrasound nerve segmentation,'' 2016. [Online]. Available: \url{https://kaggle.com/competitions/ultrasound-nerve-segmentation}
\BIBentrySTDinterwordspacing

\bibitem{Fetal}
X.~P. Burgos-Artizzu \emph{et~al.}, ``Evaluation of deep convolutional neural networks for automatic classification of common maternal fetal ultrasound planes,'' \emph{Sci. Rep.}, vol.~10, no.~1, p. 10200, 2020.

\bibitem{ouyang2020video}
D.~Ouyang \emph{et~al.}, ``Video-based ai for beat-to-beat assessment of cardiac function,'' \emph{Nature}, vol. 580, no. 7802, pp. 252--256, 2020.

\bibitem{polat2022improving}
G.~Polat \emph{et~al.}, ``Improving the computer-aided estimation of ulcerative colitis severity according to mayo endoscopic score by using regression-based deep learning,'' \emph{Nes. Nutr. Ws.}, p. izac226, 2022.

\bibitem{misawa2021development}
M.~Misawa \emph{et~al.}, ``Development of a computer-aided detection system for colonoscopy and a publicly accessible large colonoscopy video database (with video),'' \emph{Gastrointestinal endoscopy}, vol.~93, no.~4, pp. 960--967, 2021.

\bibitem{smedsrud2021kvasir}
P.~H. Smedsrud \emph{et~al.}, ``Kvasir-capsule, a video capsule endoscopy dataset,'' \emph{Sci. Data}, vol.~8, no.~1, p. 142, 2021.

\bibitem{ozyoruk2020endoslam}
K.~B. Ozyoruk \emph{et~al.}, ``Endoslam dataset and an unsupervised monocular visual odometry and depth estimation approach for endoscopic videos,'' \emph{Med. Image Anal.}, vol.~71, p. 102058, 2021.

\bibitem{HyperKvasir}
H.~Borgli \emph{et~al.}, ``Hyperkvasir, a comprehensive multi-class image and video dataset for gastrointestinal endoscopy,'' \emph{Sci. Data}, vol.~7, no.~1, pp. 1--14, 2020.

\bibitem{nwoye2022data}
C.~I. Nwoye and N.~Padoy, ``Data splits and metrics for method benchmarking on surgical action triplet datasets,'' \emph{arXiv preprint arXiv:2204.05235}, 2022.

\bibitem{ma2021ldpolypvideo}
Y.~Ma \emph{et~al.}, ``Ldpolypvideo benchmark: a large-scale colonoscopy video dataset of diverse polyps,'' in \emph{Proc. Int. Conf. Med. Image Comput. Comput.-Assisted Intervention}.\hskip 1em plus 0.5em minus 0.4em\relax Springer, 2021, pp. 387--396.

\bibitem{yan2018deeplesion}
K.~Yan \emph{et~al.}, ``Deeplesion: automated mining of large-scale lesion annotations and universal lesion detection with deep learning,'' \emph{Journal of medical imaging}, vol.~5, no.~3, pp. 036\,501--036\,501, 2018.

\bibitem{armato2011lung}
S.~G. Armato~III \emph{et~al.}, ``The lung image database consortium (lidc) and image database resource initiative (idri): a completed reference database of lung nodules on ct scans,'' \emph{Medical physics}, vol.~38, no.~2, pp. 915--931, 2011.

\bibitem{liew2022large}
S.-L. Liew \emph{et~al.}, ``A large, curated, open-source stroke neuroimaging dataset to improve lesion segmentation algorithms,'' \emph{Sci. Data}, vol.~9, no.~1, p. 320, 2022.

\bibitem{saha2023artificial}
A.~Saha \emph{et~al.}, ``Artificial intelligence and radiologists at prostate cancer detection in mri—the pi-cai challenge,'' in \emph{Medical Imaging with Deep Learning, short paper track}, 2023.

\bibitem{bien2018deep}
N.~Bien \emph{et~al.}, ``Deep-learning-assisted diagnosis for knee magnetic resonance imaging: development and retrospective validation of mrnet,'' \emph{PLoS medicine}, vol.~15, no.~11, p. e1002699, 2018.

\bibitem{duffy2022high}
G.~Duffy \emph{et~al.}, ``High-throughput precision phenotyping of left ventricular hypertrophy with cardiovascular deep learning,'' \emph{JAMA cardiology}, vol.~7, no.~4, pp. 386--395, 2022.

\bibitem{ghahremani2022deep}
P.~Ghahremani \emph{et~al.}, ``Deep learning-inferred multiplex immunofluorescence for immunohistochemical image quantification,'' \emph{Nature machine intelligence}, vol.~4, no.~4, pp. 401--412, 2022.

\bibitem{national2011national}
N.~L. S. T.~R. Team, ``The national lung screening trial: overview and study design,'' \emph{Radiology}, vol. 258, no.~1, pp. 243--253, 2011.

\bibitem{SNOW}
K.~Ding \emph{et~al.}, ``A large-scale synthetic pathological dataset for deep learning-enabled segmentation of breast cancer,'' \emph{Sci. Data}, vol.~10, no.~1, p. 231, 2023.

\bibitem{cellxgene}
C.~S.-C. Biology \emph{et~al.}, ``Cz cell{\texttimes}gene discover: A single-cell data platform for scalable exploration, analysis and modeling of aggregated data,'' \emph{bioRxiv}, pp. 2023--10, 2023.

\bibitem{genbank}
\BIBentryALTinterwordspacing
D.~A. Benson \emph{et~al.}, ``{GenBank},'' \emph{Nucleic Acids Res.}, vol.~41, no.~D1, pp. D36--D42, 11 2012. [Online]. Available: \url{https://doi.org/10.1093/nar/gks1195}
\BIBentrySTDinterwordspacing

\bibitem{tarhan2023single}
L.~Tarhan \emph{et~al.}, ``Single cell portal: an interactive home for single-cell genomics data,'' \emph{bioRxiv}, 2023.

\bibitem{gencode}
\BIBentryALTinterwordspacing
A.~Frankish \emph{et~al.}, ``{GENCODE reference annotation for the human and mouse genomes},'' \emph{Nucleic Acids Res.}, vol.~47, no.~D1, pp. D766--D773, 10 2018. [Online]. Available: \url{https://doi.org/10.1093/nar/gky955}
\BIBentrySTDinterwordspacing

\bibitem{HumanCellAtlas}
\BIBentryALTinterwordspacing
A.~Regev \emph{et~al.}, ``Science forum: The human cell atlas,'' \emph{eLife}, vol.~6, p. e27041, dec 2017. [Online]. Available: \url{https://doi.org/10.7554/eLife.27041}
\BIBentrySTDinterwordspacing

\bibitem{UCSCgenome}
\BIBentryALTinterwordspacing
B.~J. Raney \emph{et~al.}, ``{The UCSC Genome Browser database: 2024 update},'' \emph{Nucleic Acids Res.}, vol.~52, no.~D1, pp. D1082--D1088, 11 2023. [Online]. Available: \url{https://doi.org/10.1093/nar/gkad987}
\BIBentrySTDinterwordspacing

\bibitem{doi:10.1021/pr501254j}
N.~J. Edwards \emph{et~al.}, ``The cptac data portal: A resource for cancer proteomics research,'' \emph{Journal of Proteome Research}, vol.~14, no.~6, pp. 2707--2713, 2015.

\bibitem{Ensembl}
F.~J. Martin \emph{et~al.}, ``Ensembl 2023,'' \emph{Nucleic Acids Res.}, vol.~51, no.~D1, pp. D933--D941, 2023.

\bibitem{RNAcentral}
\BIBentryALTinterwordspacing
The RNAcentral Consortium, ``{RNAcentral: a hub of information for non-coding RNA sequences},'' \emph{Nucleic Acids Res.}, vol.~47, no.~D1, pp. D221--D229, 11 2018. [Online]. Available: \url{https://doi.org/10.1093/nar/gky1034}
\BIBentrySTDinterwordspacing

\bibitem{Armstrong2020pdbe}
\BIBentryALTinterwordspacing
D.~R. Armstrong \emph{et~al.}, ``Pdbe: improved findability of macromolecular structure data in the pdb,'' \emph{Nucleic acids research}, vol.~48, p. D335—D343, 1 2020. [Online]. Available: \url{https://europepmc.org/articles/PMC7145656}
\BIBentrySTDinterwordspacing

\bibitem{Consortium2023uniprot}
\BIBentryALTinterwordspacing
T.~U. Consortium, ``Uniprot: the universal protein knowledgebase in 2023,'' \emph{Nucleic Acids Research}, vol.~51, pp. D523--D531, 1 2023. [Online]. Available: \url{https://doi.org/10.1093/nar/gkac1052}
\BIBentrySTDinterwordspacing

\bibitem{lincs1000}
\BIBentryALTinterwordspacing
I.~NeuroLINCS (University~of California, ``imn (exp 2) - als, sma and control (unaffected) imn cell lines differentiated from ips cell lines using a long differentiation protocol - rna-seq,'' 2017. [Online]. Available: \url{http://lincsportal.ccs.miami.edu/datasets/#/view/LDS-1398}
\BIBentrySTDinterwordspacing

\bibitem{gdsc}
W.~Yang \emph{et~al.}, ``{Genomics of Drug Sensitivity in Cancer (GDSC): a resource for therapeutic biomarker discovery in cancer cells},'' \emph{Nucleic Acids Research}, vol.~41, no.~D1, pp. D955--D961, 11 2012.

\bibitem{CCLE}
\BIBentryALTinterwordspacing
M.~Ghandi \emph{et~al.}, ``Next-generation characterization of the cancer cell line encyclopedia,'' \emph{Nature}, vol. 569, pp. 503--508, 2019. [Online]. Available: \url{https://doi.org/10.1038/s41586-019-1186-3}
\BIBentrySTDinterwordspacing

\bibitem{bycroft2018uk}
C.~Bycroft \emph{et~al.}, ``The uk biobank resource with deep phenotyping and genomic data,'' \emph{Nature}, vol. 562, no. 7726, pp. 203--209, 2018.

\bibitem{Zhao2021cgga}
Z.~Zhao \emph{et~al.}, ``Chinese glioma genome atlas (cgga): A comprehensive resource with functional genomic data from chinese glioma patients,'' \emph{Genomics, Proteomics \& Bioinformatics}, vol.~19, pp. 1--12, 2021.

\bibitem{johnson2019mimic}
A.~E. Johnson \emph{et~al.}, ``Mimic-cxr, a de-identified publicly available database of chest radiographs with free-text reports,'' \emph{Sci. Data}, vol.~6, no.~1, p. 317, 2019.

\bibitem{bustos2020padchest}
A.~Bustos \emph{et~al.}, ``Padchest: A large chest x-ray image dataset with multi-label annotated reports,'' \emph{Med. Image Anal.}, vol.~66, p. 101797, 2020.

\bibitem{irvin2019chexpert}
J.~Irvin \emph{et~al.}, ``Chexpert: A large chest radiograph dataset with uncertainty labels and expert comparison,'' in \emph{Proc. AAAI Conf. Artif. Intell.}, vol.~33, no.~01, 2019, pp. 590--597.

\bibitem{garcia2018overview}
A.~Garc{\'\i}a Seco~de Herrera \emph{et~al.}, ``Overview of the imageclef 2018 caption prediction tasks,'' in \emph{Working Notes of CLEF 2018-Conference and Labs of the Evaluation Forum (CLEF 2018), Avignon, France, September 10-14, 2018.}, vol. 2125.\hskip 1em plus 0.5em minus 0.4em\relax CEUR Workshop Proceedings, 2018.

\bibitem{he2020pathvqa}
X.~He, Y.~Zhang, L.~Mou, E.~Xing, and P.~Xie, ``Pathvqa: 30000+ questions for medical visual question answering,'' \emph{arXiv preprint arXiv:2003.10286}, 2020.

\bibitem{tsuneki2022inference}
M.~Tsuneki and F.~Kanavati, ``Inference of captions from histopathological patches,'' in \emph{Proc. Int. Conf. Medical Imaging Deep Learn.}\hskip 1em plus 0.5em minus 0.4em\relax PMLR, 2022, pp. 1235--1250.

\bibitem{wagner2020ptb}
P.~Wagner \emph{et~al.}, ``Ptb-xl, a large publicly available electrocardiography dataset,'' \emph{Sci. Data}, vol.~7, no.~1, p. 154, 2020.

\bibitem{pelka2018radiology}
O.~Pelka \emph{et~al.}, ``Radiology objects in context (roco): a multimodal image dataset,'' in \emph{Intravascular Imaging and Computer Assisted Stenting and Large-Scale Annotation of Biomedical Data and Expert Label Synthesis: 7th Joint International Workshop, CVII-STENT 2018 and Third International Workshop, LABELS 2018, Held in Conjunction with MICCAI 2018, Granada, Spain, September 16, 2018, Proceedings 3}.\hskip 1em plus 0.5em minus 0.4em\relax Springer, 2018, pp. 180--189.

\bibitem{subramanian2020medicat}
S.~Subramanian \emph{et~al.}, ``Medicat: A dataset of medical images, captions, and textual references,'' in \emph{Findings of the Association for Computational Linguistics, ACL 2020: EMNLP 2020}.\hskip 1em plus 0.5em minus 0.4em\relax Association for Computational Linguistics (ACL), 2020, pp. 2112--2120.

\bibitem{zhang2023pmc}
X.~Zhang \emph{et~al.}, ``Pmc-vqa: Visual instruction tuning for medical visual question answering,'' \emph{arXiv preprint arXiv:2305.10415}, 2023.

\bibitem{saha2018machine}
A.~Saha \emph{et~al.}, ``A machine learning approach to radiogenomics of breast cancer: a study of 922 subjects and 529 dce-mri features,'' \emph{British journal of cancer}, vol. 119, no.~4, pp. 508--516, 2018.

\bibitem{li2022ispy2}
\BIBentryALTinterwordspacing
W.~Li \emph{et~al.}, ``{I-SPY 2 Breast Dynamic Contrast Enhanced MRI Trial (ISPY2)}.'' [Online]. Available: \url{https://doi.org/10.7937/TCIA.D8Z0-9T85}
\BIBentrySTDinterwordspacing

\bibitem{Gamper_2021_CVPR}
J.~Gamper and N.~Rajpoot, ``Multiple instance captioning: Learning representations from histopathology textbooks and articles,'' in \emph{Proc. IEEE Conf. Comput. Vis. Pattern Recognit.}, June 2021, pp. 16\,549--16\,559.

\bibitem{clark2013cancer}
K.~Clark \emph{et~al.}, ``The cancer imaging archive (tcia): maintaining and operating a public information repository,'' \emph{Journal of digital imaging}, vol.~26, pp. 1045--1057, 2013.

\bibitem{balogh2015improving}
E.~P. Balogh \emph{et~al.}, \emph{Improving diagnosis in health care}.\hskip 1em plus 0.5em minus 0.4em\relax National Academies Press (US), 2015.

\bibitem{ueda2023diagnostic}
D.~Ueda \emph{et~al.}, ``Diagnostic performance of chatgpt from patient history and imaging findings on the diagnosis please quizzes,'' \emph{Radiology}, vol. 308, no.~1, p. e231040, 2023.

\bibitem{wu2024collaborative}
S.-H. Wu \emph{et~al.}, ``Collaborative enhancement of consistency and accuracy in us diagnosis of thyroid nodules using large language models,'' \emph{Radiology}, vol. 310, no.~3, p. e232255, 2024.

\bibitem{ali2023using}
S.~R. Ali \emph{et~al.}, ``Using chatgpt to write patient clinic letters,'' \emph{The Lancet Digital Health}, vol.~5, no.~4, pp. e179--e181, 2023.

\bibitem{abd2023large}
A.~Abd-Alrazaq \emph{et~al.}, ``Large language models in medical education: Opportunities, challenges, and future directions,'' \emph{JMIR Medical Education}, vol.~9, no.~1, p. e48291, 2023.

\bibitem{karabacak2023advent}
M.~Karabacak \emph{et~al.}, ``The advent of generative language models in medical education,'' \emph{JMIR Medical Education}, vol.~9, p. e48163, 2023.

\bibitem{kung2023performance}
T.~H. Kung \emph{et~al.}, ``Performance of chatgpt on usmle: Potential for ai-assisted medical education using large language models,'' \emph{PLoS Digital Health}, vol.~2, no.~2, p. e0000198, 2023.

\bibitem{cocskun2024integration}
A.~B. Co{\c{s}}kun \emph{et~al.}, ``Integration of chatgpt and e-health literacy: Opportunities, challenges, and a look towards the future,'' \emph{Journal of Health Reports and Technology}, vol.~10, no.~1, 2024.

\bibitem{lee2023benefits}
P.~Lee \emph{et~al.}, ``Benefits, limits, and risks of gpt-4 as an ai chatbot for medicine,'' \emph{New Engl. J. Med.}, vol. 388, no.~13, pp. 1233--1239, 2023.

\bibitem{chen2023soulchat}
Y.~Chen \emph{et~al.}, ``Soulchat: Improving llms’ empathy, listening, and comfort abilities through fine-tuning with multi-turn empathy conversations,'' in \emph{Findings of the Association for Computational Linguistics: EMNLP 2023}, 2023, pp. 1170--1183.

\bibitem{luo2023biomedgpt}
Y.~Luo \emph{et~al.}, ``Biomedgpt: Open multimodal generative pre-trained transformer for biomedicine,'' \emph{arXiv preprint arXiv:2308.09442}, 2023.

\bibitem{huawei2022general}
L.~Huawei Technologies~Co., ``A general introduction to artificial intelligence,'' in \emph{Artificial Intelligence Technology}.\hskip 1em plus 0.5em minus 0.4em\relax Springer, 2022, pp. 1--41.

\bibitem{larson2020ethics}
D.~B. Larson \emph{et~al.}, ``Ethics of using and sharing clinical imaging data for artificial intelligence: a proposed framework,'' \emph{Radiology}, vol. 295, no.~3, pp. 675--682, 2020.

\bibitem{salerno2019overdiagnosis}
S.~Salerno \emph{et~al.}, ``Overdiagnosis and overimaging: an ethical issue for radiological protection,'' \emph{La radiologia medica}, vol. 124, pp. 714--720, 2019.

\bibitem{kaur2022trustworthy}
D.~Kaur \emph{et~al.}, ``Trustworthy artificial intelligence: a review,'' \emph{ACM Computing Surveys (CSUR)}, vol.~55, no.~2, pp. 1--38, 2022.

\bibitem{haendel2020many}
M.~Haendel \emph{et~al.}, ``How many rare diseases are there?'' \emph{Nat. Rev. Drug Discov.}, vol.~19, no.~2, pp. 77--78, 2020.

\bibitem{guan2021domain}
H.~Guan and M.~Liu, ``Domain adaptation for medical image analysis: a survey,'' \emph{IEEE Trans. Biomed. Eng.}, vol.~69, no.~3, pp. 1173--1185, 2021.

\bibitem{liu2024decade}
Z.~Liu and K.~He, ``A decade's battle on dataset bias: Are we there yet?'' \emph{arXiv preprint arXiv:2403.08632}, 2024.

\bibitem{cassidy2007lung}
A.~Cassidy \emph{et~al.}, ``Lung cancer risk prediction: a tool for early detection,'' \emph{Int. J. Cancer}, vol. 120, no.~1, pp. 1--6, 2007.

\bibitem{gama2014survey}
J.~Gama \emph{et~al.}, ``A survey on concept drift adaptation,'' \emph{ACM computing surveys (CSUR)}, vol.~46, no.~4, pp. 1--37, 2014.

\bibitem{wang2021annotation}
S.~Wang \emph{et~al.}, ``Annotation-efficient deep learning for automatic medical image segmentation,'' \emph{Nat. Commun.}, vol.~12, no.~1, p. 5915, 2021.

\bibitem{tajbakhsh2021guest}
N.~Tajbakhsh \emph{et~al.}, ``Guest editorial annotation-efficient deep learning: the holy grail of medical imaging,'' \emph{IEEE Trans. Med. Imaging}, vol.~40, no.~10, pp. 2526--2533, 2021.

\bibitem{sun2024trustllm}
L.~Sun \emph{et~al.}, ``Trustllm: Trustworthiness in large language models,'' \emph{arXiv preprint arXiv:2401.05561}, 2024.

\bibitem{sokol2020one}
K.~Sokol and P.~Flach, ``One explanation does not fit all: The promise of interactive explanations for machine learning transparency,'' \emph{KI-K{\"u}nstliche Intelligenz}, vol.~34, no.~2, pp. 235--250, 2020.

\bibitem{bommasani2023foundation}
R.~Bommasani \emph{et~al.}, ``The foundation model transparency index,'' \emph{arXiv preprint arXiv:2310.12941}, 2023.

\bibitem{chen2023algorithmic}
R.~J. Chen \emph{et~al.}, ``Algorithmic fairness in artificial intelligence for medicine and healthcare,'' \emph{Nat. Biomed. Eng.}, vol.~7, no.~6, pp. 719--742, 2023.

\bibitem{motoki2023more}
F.~Motoki \emph{et~al.}, ``More human than human: Measuring chatgpt political bias,'' \emph{Available at SSRN 4372349}, 2023.

\bibitem{felkner2023winoqueer}
V.~Felkner \emph{et~al.}, ``Winoqueer: A community-in-the-loop benchmark for anti-lgbtq+ bias in large language models,'' in \emph{Proceedings of the 61st Annual Meeting of the Association for Computational Linguistics (Volume 1: Long Papers)}, 2023, pp. 9126--9140.

\bibitem{gehman2020realtoxicityprompts}
S.~Gehman \emph{et~al.}, ``Realtoxicityprompts: Evaluating neural toxic degeneration in language models,'' in \emph{Findings of the Association for Computational Linguistics: EMNLP 2020}, 2020, pp. 3356--3369.

\bibitem{wei2024jailbroken}
A.~Wei \emph{et~al.}, ``Jailbroken: How does llm safety training fail?'' \emph{Advances in Neural Information Processing Systems}, vol.~36, 2024.

\bibitem{baeroe2020achieve}
K.~B{\ae}r{\o}e \emph{et~al.}, ``How to achieve trustworthy artificial intelligence for health,'' \emph{Bull. World Health Organ.}, vol.~98, no.~4, p. 257, 2020.

\bibitem{chen2023trustworthy}
P.-Y. Chen and C.~Xiao, ``Trustworthy ai in the era of foundation models,'' in \emph{Proc. IEEE Conf. Comput. Vis. Pattern Recognit.}, 2023.

\bibitem{dwyer2023high}
M.~Dwyer-White \emph{et~al.}, ``High reliability in healthcare,'' in \emph{Patient Safety: A Case-based Innovative Playbook for Safer Care}.\hskip 1em plus 0.5em minus 0.4em\relax Springer, 2023, pp. 3--13.

\bibitem{rawte2023survey}
V.~Rawte, A.~Sheth, and A.~Das, ``A survey of hallucination in large foundation models,'' \emph{arXiv preprint arXiv:2309.05922}, 2023.

\bibitem{li2023task}
C.~Li and J.~Flanigan, ``Task contamination: Language models may not be few-shot anymore,'' in \emph{Proceedings of the AAAI Conference on Artificial Intelligence}, vol.~38, no.~16, 2024, pp. 18\,471--18\,480.

\bibitem{yao2023editing}
Y.~Yao \emph{et~al.}, ``Editing large language models: Problems, methods, and opportunities,'' in \emph{Proceedings of the 2023 Conference on Empirical Methods in Natural Language Processing}, 2023, pp. 10\,222--10\,240.

\bibitem{hoelscher2023detecting}
J.~Hoelscher-Obermaier \emph{et~al.}, ``Detecting edit failures in large language models: An improved specificity benchmark,'' in \emph{Findings of the Association for Computational Linguistics: ACL 2023}, 2023, pp. 11\,548--11\,559.

\bibitem{raghu2017expressive}
M.~Raghu \emph{et~al.}, ``On the expressive power of deep neural networks,'' in \emph{Proc. Int. Conf. Mach. Learn.}\hskip 1em plus 0.5em minus 0.4em\relax PMLR, 2017, pp. 2847--2854.

\bibitem{dosovitskiy2020image}
A.~Dosovitskiy \emph{et~al.}, ``An image is worth 16x16 words: Transformers for image recognition at scale,'' in \emph{Proc. Int. Conf. Learn. Represent.}, 2020.

\bibitem{liu2021swin}
Z.~Liu \emph{et~al.}, ``Swin transformer: Hierarchical vision transformer using shifted windows,'' in \emph{Proc. IEEE Int. Conf. Comput. Vis.}, 2021, pp. 10\,012--10\,022.

\bibitem{zhao2022elements}
S.~Zhao \emph{et~al.}, ``Elements of chronic disease management service system: an empirical study from large hospitals in china,'' \emph{Sci. Rep.}, vol.~12, no.~1, p. 5693, 2022.

\bibitem{chen2020deep}
C.~Chen \emph{et~al.}, ``Deep learning on computational-resource-limited platforms: a survey,'' \emph{Mob. Inf. Syst.}, vol. 2020, pp. 1--19, 2020.

\bibitem{deng2020model}
L.~Deng \emph{et~al.}, ``Model compression and hardware acceleration for neural networks: A comprehensive survey,'' \emph{Proc. IEEE}, vol. 108, no.~4, pp. 485--532, 2020.

\bibitem{ding2023parameter}
N.~Ding \emph{et~al.}, ``Parameter-efficient fine-tuning of large-scale pre-trained language models,'' \emph{Nat. Mach. Intell}, vol.~5, no.~3, pp. 220--235, 2023.

\bibitem{griffith2023desperate}
E.~Griffith, ``The desperate hunt for the ai boom's most indispensable prize.'' \emph{International New York Times}, pp. NA--NA, 2023.

\bibitem{gupta2021chasing}
U.~Gupta \emph{et~al.}, ``Chasing carbon: The elusive environmental footprint of computing,'' in \emph{2021 IEEE International Symposium on High-Performance Computer Architecture (HPCA)}.\hskip 1em plus 0.5em minus 0.4em\relax IEEE, 2021, pp. 854--867.

\bibitem{henderson2020towards}
P.~Henderson \emph{et~al.}, ``Towards the systematic reporting of the energy and carbon footprints of machine learning,'' \emph{J. Mach. Learn. Res.}, vol.~21, no.~1, pp. 10\,039--10\,081, 2020.

\bibitem{park2019deep}
A.~Park \emph{et~al.}, ``Deep learning--assisted diagnosis of cerebral aneurysms using the headxnet model,'' \emph{JAMA network open}, vol.~2, no.~6, pp. e195\,600--e195\,600, 2019.

\bibitem{steiner2018impact}
D.~F. Steiner \emph{et~al.}, ``Impact of deep learning assistance on the histopathologic review of lymph nodes for metastatic breast cancer,'' \emph{Am. J. Surg. Pathol.}, vol.~42, no.~12, p. 1636, 2018.

\bibitem{kim2020changes}
H.-E. Kim \emph{et~al.}, ``Changes in cancer detection and false-positive recall in mammography using artificial intelligence: a retrospective, multireader study,'' \emph{The Lancet Digital Health}, vol.~2, no.~3, pp. e138--e148, 2020.

\bibitem{tschandl2020human}
P.~Tschandl \emph{et~al.}, ``Human--computer collaboration for skin cancer recognition,'' \emph{Nat. Med.}, vol.~26, no.~8, pp. 1229--1234, 2020.

\bibitem{han2021dynamic}
Y.~Han \emph{et~al.}, ``Dynamic neural networks: A survey,'' \emph{IEEE Trans. Pattern Anal. Mach. Intell.}, vol.~44, no.~11, pp. 7436--7456, 2021.

\bibitem{vaswani2017attention}
A.~Vaswani \emph{et~al.}, ``Attention is all you need,'' \emph{Proc. Adv. Neural Inf. Process. Syst.}, vol.~30, 2017.

\bibitem{shazeer2016outrageously}
N.~Shazeer \emph{et~al.}, ``Outrageously large neural networks: The sparsely-gated mixture-of-experts layer,'' in \emph{Proc. Int. Conf. Learn. Represent.}, 2016.

\bibitem{you2023implicit}
C.~You \emph{et~al.}, ``Implicit anatomical rendering for medical image segmentation with stochastic experts,'' in \emph{International Conference on Medical Image Computing and Computer-Assisted Intervention}.\hskip 1em plus 0.5em minus 0.4em\relax Springer, 2023, pp. 561--571.

\bibitem{gu2023mamba}
A.~Gu and T.~Dao, ``Mamba: Linear-time sequence modeling with selective state spaces,'' \emph{arXiv preprint arXiv:2312.00752}, 2023.

\bibitem{yi2023towards}
H.~Yi, Z.~Qin, Q.~Lao, W.~Xu, Z.~Jiang, D.~Wang, S.~Zhang, and K.~Li, ``Towards general purpose medical ai: Continual learning medical foundation model,'' \emph{arXiv preprint arXiv:2303.06580}, 2023.

\bibitem{kojima2022large}
T.~Kojima \emph{et~al.}, ``Large language models are zero-shot reasoners,'' \emph{Adv. Neur. In.}, vol.~35, pp. 22\,199--22\,213, 2022.

\bibitem{wang2022learngene}
Q.-F. Wang \emph{et~al.}, ``Learngene: From open-world to your learning task,'' in \emph{Proc. AAAI Conf. Artif. Intell.}, vol.~36, no.~8, 2022, pp. 8557--8565.

\bibitem{tan2022federated}
Y.~Tan \emph{et~al.}, ``Federated learning from pre-trained models: A contrastive learning approach,'' \emph{Adv. Neur. In.}, vol.~35, pp. 19\,332--19\,344, 2022.

\bibitem{zhuang2023foundation}
W.~Zhuang \emph{et~al.}, ``When foundation model meets federated learning: Motivations, challenges, and future directions,'' \emph{arXiv preprint arXiv:2306.15546}, 2023.

\bibitem{zhu2022uni}
J.~Zhu \emph{et~al.}, ``Uni-perceiver-moe: Learning sparse generalist models with conditional moes,'' \emph{Adv. Neur. In.}, vol.~35, pp. 2664--2678, 2022.

\bibitem{geng2020recent}
C.~Geng \emph{et~al.}, ``Recent advances in open set recognition: A survey,'' \emph{IEEE Trans. Pattern Anal. Mach. Intell.}, vol.~43, no.~10, pp. 3614--3631, 2020.

\bibitem{li2023scaling}
Y.~Li \emph{et~al.}, ``Scaling language-image pre-training via masking,'' in \emph{Proc. IEEE Conf. Comput. Vis. Pattern Recognit.}, 2023, pp. 23\,390--23\,400.

\bibitem{ma2022multimodal}
M.~Ma \emph{et~al.}, ``Are multimodal transformers robust to missing modality?'' in \emph{Proc. IEEE Conf. Comput. Vis. Pattern Recognit.}, 2022, pp. 18\,177--18\,186.

\bibitem{mohri2018foundations}
M.~Mohri \emph{et~al.}, \emph{Foundations of machine learning}.\hskip 1em plus 0.5em minus 0.4em\relax MIT press, 2018.

\bibitem{yuan2023power}
Y.~Yuan, ``On the power of foundation models,'' in \emph{Proc. Int. Conf. Mach. Learn.}\hskip 1em plus 0.5em minus 0.4em\relax PMLR, 2023, pp. 40\,519--40\,530.

\bibitem{jimenez2020drug}
J.~Jim{\'e}nez-Luna \emph{et~al.}, ``Drug discovery with explainable artificial intelligence,'' \emph{Nat. Mach. Intell}, vol.~2, no.~10, pp. 573--584, 2020.

\bibitem{qayyum2020secure}
A.~Qayyum \emph{et~al.}, ``Secure and robust machine learning for healthcare: A survey,'' \emph{IEEE Rev. Biomed. Eng.}, vol.~14, pp. 156--180, 2020.

\bibitem{schlarmann2023adversarial}
C.~Schlarmann and M.~Hein, ``On the adversarial robustness of multi-modal foundation models,'' in \emph{Proc. IEEE Int. Conf. Comput. Vis.}, 2023, pp. 3677--3685.

\bibitem{habli2020artificial}
I.~Habli \emph{et~al.}, ``Artificial intelligence in health care: accountability and safety,'' \emph{Bull. World Health Organ.}, vol.~98, no.~4, p. 251, 2020.

\bibitem{vinuesa2020role}
R.~Vinuesa \emph{et~al.}, ``The role of artificial intelligence in achieving the sustainable development goals,'' \emph{Nat. Commun.}, vol.~11, no.~1, pp. 1--10, 2020.

\bibitem{kaack2022aligning}
L.~H. Kaack \emph{et~al.}, ``Aligning artificial intelligence with climate change mitigation,'' \emph{Nat. Clim. Change}, vol.~12, no.~6, pp. 518--527, 2022.

\bibitem{menghani2023efficient}
G.~Menghani, ``Efficient deep learning: A survey on making deep learning models smaller, faster, and better,'' \emph{ACM Computing Surveys}, vol.~55, no.~12, pp. 1--37, 2023.

\end{thebibliography}

\end{document}